\documentclass[a4paper,12pt,twoside,oldfontcommands]{memoir} 
\usepackage{textcomp} 
\usepackage{amsmath} 

\usepackage[round]{natbib} 

\usepackage{graphics}

\usepackage{citeref} 
\setcitestyle{aysep={}} 

\citeindextrue

\usepackage{makeidx} 
\makeindex

\newcommand{\A}{$A$}
\newcommand{\B}{$B$}
\newcommand{\C}{$C$}
\renewcommand{\a}{a}
\newcommand{\f}{f}
\newcommand{\h}{h}

\newcommand{\M}{M}
\newcommand{\m}{m}

\newcommand{\n}{n}
\newcommand{\p}{p}
\newcommand{\bp}{\mathbf{p}}
\newcommand{\bq}{\mathbf{q}}

\newcommand{\R}{R}
\newcommand{\br}{\mathbf{r}}
\newcommand{\s}{\textrm{s}}
\renewcommand{\bs}{\mathbf{s}}

\newcommand{\V}{V}

\newcommand{\w}{w}
\newcommand{\x}{\textrm{x}}
\newcommand{\bx}{\mathbf{x}}
\newcommand{\z}{z}

\newcommand{\raw}{\textrm{raw}}
\newcommand{\cooked}{\textrm{cooked}}
\newcommand{\Tr}{\textrm{Tr}}

\renewcommand{\thesubsection}{\arabic{chapter}.\arabic{section}.\alph{subsection}} 

\newcommand{\LeftDoubleAngle}{\left\langle \! \left\langle}
\newcommand{\RightDoubleAngle}{\right\rangle \! \right\rangle}

\setsecnumdepth{subsection}
\settocdepth{subsection}

\begin{document}
\frontmatter


\pagestyle{empty}
\begin{flushright}
\vspace*{1in}
\large
\textsc{The Computable Universe:} \\
\huge
From Prespace Metaphysics to Discrete Quantum Mechanics

\vspace{1in}
\large
Martin Leckey\\
B.Sc. (Hons.), M.A.\\

\vspace{1in}
Submitted for the degree of Doctor of Philosophy\\
Philosophy Department, Monash University\\
October 1997
\end{flushright}

%

\cleardoublepage



\pagestyle{ruled}
\tableofcontents

\clearpage
\listoffigures

\chapter{Abstract}

The central motivating idea behind the development of this work is the concept of prespace, a
hypothetical structure that is postulated by some physicists to underlie the fabric of space or
space-time. I consider how such a structure could relate to space and space-time, and the rest
of reality as we know it, and the implications of the existence of this structure for quantum
theory. Understanding how this structure could relate to space and to the rest of reality
requires, I believe, that we consider how space itself relates to reality, and how other so-called
``spaces'' used in physics relate to reality. In chapter \ref{space}, I compare space and space-time
to other spaces used in physics, such as configuration space, phase space and Hilbert space. I
support what is known as the ``property view'' of space, opposing both the traditional views of
space and space-time, substantivalism and relationism. I argue that all these spaces are
property spaces. After examining the relationships of these spaces to causality, I argue that
configuration space has, due to its role in quantum mechanics, a special status in the
microscopic world similar to the status of position space in the macroscopic world. In
chapter \ref{prespace}, prespace itself is considered. One way of approaching this structure is through
the comparison of the prespace structure with a computational system, in particular to a
cellular automaton, in which space or space-time and all other physical quantities are broken
down into discrete units. I suggest that one way open for a prespace metaphysics can be
found if physics is made fully discrete in this way. I suggest as a heuristic principle that the
physical laws of our world are such that the computational cost of implementing those laws
on an arbitrary computational system is minimized, adapting a heuristic principle of this type
proposed by \citet{Feynman1982}. In chapter \ref{qm_discrete_phys}, some of the ideas of the previous chapters
are applied in an examination of the physics and metaphysics of quantum theory. I first
discuss the ``measurement problem'' of quantum mechanics: this problem and its proposed
solution are the primary subjects of chapter \ref{qm_discrete_phys}. It turns out that considering how
quantum theory could be made fully discrete leads naturally to a suggestion of how standard
linear quantum mechanics could be modified to give rise to a solution to the measurement
problem. The computational heuristic principle reinforces the same solution.  I call the
modified quantum mechanics Critical Complexity Quantum Mechanics (CCQM). I compare
CCQM with some of the other proposed solutions to the measurement problem, in particular
the spontaneous localization model of \citet{GhirardiRiminiWeber1986}. Finally, in chapters \ref{qm_complexity} and \ref{entropy_time}, I argue
that the measure of complexity of quantum mechanical states I introduce in CCQM also
provides a new definition of entropy for quantum mechanics, and suggests a solution to the
problem of providing an objective foundation for statistical mechanics, thermodynamics, and
the arrow of time.

\chapter{Statement of Originality}

This thesis contains no material which has been accepted for the award of any other degree
and, to the best of my knowledge, contains no material previously published or written by
another person, except where due reference is made in the text of the thesis.\\\\\\\

Martin J. Leckey

\chapter{Acknowledgments}

I would like to thank my principal supervisor John Bigelow for his assistance and
encouragement during the course of my work on this thesis. His metaphysical realism \citep{Bigelow1988,BigelowPargetter1990}, particularly concerning the existence of
properties, has been very influential on me. This shows through in chapter \ref{space}, on the metaphysics of space, although he actually holds a substantivalist view of space,
in opposition to the property view of space that I support in that chapter. I would also like to
thank him for asking me to assist him with his work on laws of nature, which has resulted in a
joint paper, \citet{LeckeyBigelow1995}, as well as many joint papers at conferences and
philosophy department seminars. During the course of this collaboration, I came up with a
new view of laws of nature. I have not included this work in this thesis, but I have included,
bound at the back of the thesis, a reprint of the joint paper in which my theory of laws of
nature is put forward. As explained in chapter \ref{space}, I feel that a theory of laws of nature of
this general type is needed to support a property view of space.

I would also like to thank my co-supervisor Bill Wignall, from the University of
Melbourne School of Physics, for his help, especially with his assistance with many aspects
of quantum physics. His uncompromisingly realist approach to quantum physics and his
particular interpretation of quantum theory \citep{Wignall1993} have been stimulating in the
development of my own realist interpretation of quantum theory.

Others I would like to thank include the staff and fellow students at Monash
University philosophy department, and the staff and fellow students at La Trobe University
philosophy department, at which I began this thesis. I spent one quarter of a year at the
University of California at Davis in 1994, and I would like to thank the staff and students in
the philosophy department there, in particular Paul Teller, my supervisor during that time, and
Michael Jubien, for stimulating discussions. Many of the staff and students at the School of
Physics and the department of History and Philosophy of Science at the University of
Melbourne, and the Physics Department at Monash University, have also been of great
assistance. The librarians and secretarial staff at all these universities have been of great
assistance in doing many tasks that most students would ordinarily do for themselves, so I
owe them a good deal of gratitude. The same could be said for many of the people who have
given me special assistance at many conferences, departments and other places around the
world.

There are many people I have learned from in discussions of this work, and from
questions asked at conferences and seminars. I will not be able to name all of those who
should be thanked, but I would like to thank the following: David Albert, Erik Anderson,
Harvey Brown, Jeremy Butterfield, John Earman, Brian Ellis, Edward Fredkin, Alan Hajek,
Stephen Hawking, Andreas Kamlah, Rae Langton, David Lewis, Caroline Lierse, John
Norton, David Papineau, Michael Redhead, Tom Toffoli, Damian Verdnic, Mark Zangari.

I would also like to thank my family and friends for their support during my work on
this thesis, my attendants for their general assistance, and Julia Robinson and Alex Broch for
their proofreading.

\mainmatter
\chapter{Introduction}
\label{intro}
\pagestyle{ruled}

This thesis consists in a study in the metaphysics of currently accepted physics, and in
speculations on the metaphysical and theoretical underpinnings of a possible future physics. I
examine the metaphysics of space, time and quantum physics, and the metaphysics of
``prespace,'' a hypothetical structure that has been postulated by some physicists to underlie
the fabric of space or space-time. As part of this metaphysical study, I consider what would follow
if physics were fully discrete; that is, if all physical quantities, including space and time and
field values were not continuous but came in discrete ``chunks.'' I argue that if this discrete
physics were to hold true then it is likely that quantum mechanics would require some
significant modifications. However, it turns out that making some appropriate changes leads
to a modified version of quantum mechanics which, I argue, provides a natural solution to a
problem with quantum mechanics known as the ``measurement problem.'' This is a desirable
result, because although modifications to quantum mechanics have been suggested before to
solve the measurement problem, these modifications generally have the feel of being \textit{ad hoc},
having little independent theoretical or philosophical motivation. In this work I outline a
range of metaphysical positions that not only motivate a solution to the measurement problem
but suggest a whole program of theorising, which has potential for making contributions in
many fields.

My primary interest when I began to work on this thesis was with the notion of
prespace. I also was interested in the idea, which several people have put forward, of
comparing the universe in some way to a computational system. While this analogy between
the universe and a computational system should not be taken literally, it seems to me that the
concept of a computational system could be a valuable heuristic concept in exploring the
notion of prespace. 

Connected with my interest in prespace is an interest in the nature of space:
understanding how prespace could relate to space and to the rest of reality requires, I believe,
that we consider how space itself relates to reality, and how other so-called ``spaces'' used in
physics relate to reality. This task is taken up in chapter \ref{space}. I compare space and space-time
with other spaces used in physics, such as configuration space, phase space and Hilbert space. 
I look at questions such as what these spaces are, whether they exist or not, and how they
differ in metaphysical status. I support what is known as the ``property view'' of space,
opposing both the traditional views of space and space-time, substantivalism and relationism. 
One of the reasons I support the property view is that it is opposed to substantivalism,
according to which space is a substance which ``contains'' all matter within it: substantivalism
would seem to rule out the existence of prespace. However, I mainly argue for the property
view on other grounds, grounds of parity of reasoning, for example: in my view, the grounds
substantivalists give for the existence of a substance corresponding to position space are not
sufficiently unique to position space to pick out position space from the other spaces used in
physics. Rather than supposing that space is a substance and the other spaces are mere
abstractions, I suggest they should all be granted the same ontological status: in my view, they
are all property spaces. I consider how each space relates to causation, and suggest that
position space does have special links with causation, which I label causal locality and causal
dominance, and these links with causation explain the special relation position space has to
our senses, and why we tend to think of position space as a substance. I consider the spaces
used in quantum theory, and suggest that in the microscopic realm, another space called
``configuration space'' has the special properties of causal locality and causal dominance, so
that we should grant this multidimensional space a special status in the microscopic world
similar to the status of position space in the macroscopic world. In coming to this view, I
assume a realist understanding of quantum mechanics. Furthermore, the interpretation of
configuration space that is arrived at in this chapter assists in developing a particular realist
interpretation of quantum mechanics, which is pursued further in chapter \ref{qm_discrete_phys}.

In chapter \ref{prespace}, prespace itself is considered. In the previous chapter I consider
``causal'' theories of space, in which spatial relations are reduced to causal relations among
objects, and these objects do not exist in position space. Prespace is the structure that these
objects are ``embedded in'' or constitute. I investigate in this chapter a particular kind of
causal theory of space which I call a ``functionalist'' theory of space. One way of approaching
the prespace structure is through the metaphor of a computational system, so I look at some
of those writers who have compared the universe in some way to a computational system, in
particular to a cellular automaton, in which space and time and all other quantities are broken
down into discrete units. I suggest that one way open for a prespace metaphysics can be
found if physics is made fully discrete in this way. I suggest as a heuristic principle that the
physical laws of our world are such that the computational cost of implementing those laws
on an arbitrary computational system is minimized. I adopt a particular form of this heuristic
principle, based on a heuristic principle suggested by \citet{Feynman1982}. I also compare the
prespace metaphysics developed in this chapter with the metaphysics of Leibniz and Kant,
and discuss possible versions of prespace metaphysics, discussing their relationships to
traditional metaphysical positions such as materialism, holism, mind/body dualism and
theism.

In chapter \ref{qm_discrete_phys} some of the ideas of the previous chapters are applied in an
examination of the physics and metaphysics of quantum theory. I first discuss the
``measurement problem'' of quantum mechanics; this problem and its proposed solution are
the primary subjects of chapter \ref{qm_discrete_phys}. I discuss some of the proposed solutions to the
measurement problem, concentrating on those solutions that propose a modification to the
linear evolution of the wave function of quantum mechanics, in particular the spontaneous
localization model of Ghirardi, Rimini and Weber (GRW), and the related Continuous
Spontaneous Localization model (CSL). It turns out that considering how quantum theory
could be made fully discrete leads naturally to a suggestion of how standard linear quantum
mechanics could be modified to give rise to a solution to the measurement problem. I call the
new theory of collapse the ``critical complexity'' theory of collapse. I call quantum mechanics
modified to accommodate this collapse law Critical Complexity Quantum Mechanics
(CCQM).  

It has been suggested by \citet[598]{Leggett1984} that there may be ``corrections to linear
quantum mechanics which are functions, in some sense or other, of the \textit{degree of complexity}
of the physical system described.'' Leggett does not, however, suggest a measure of
complexity of physical systems. The CCQM model can be seen to result from taking up this
suggestion of Leggett that corrections to the deterministic evolution of the wave function
depend on the complexity of the system. The theory can be seen as proposing a new measure
of complexity of wave functions, and proposing a criterion of collapse based on this measure. 

The theory has some features in common with the GRW theory and I compare the
critical complexity theory with GRW/CSL in this chapter. I discuss the interpretation of the
wave function in CCQM, comparing it with the interpretation of the wave function given by
the supporters of GRW/CSL. I show that the interpretation of the wave function I provide is
compatible with the interpretation suggested by \citet{Bell1990} for the wave function in
GRW/CSL; namely, that the modulus square of the wave function gives the ``density of stuff''
in configuration space. The implications of CCQM for relativity and nonlocality are
discussed, and possible directions for the generalization of CCQM to quantum field theory
are put forward. 

In chapter \ref{qm_complexity} I then discuss the measure of complexity of wave functions utilized in
CCQM in terms of a physical measure of complexity and two measures of computational
complexity, showing that the measure of complexity is well motivated. As already
mentioned, \citet{Feynman1982} has proposed that, as a heuristic guide to the discovery of laws of
physics, we should investigate whether those physical laws could be simulated on finite
computers. He suggests that we should accept as physically possible only those laws that
could be exactly simulated this way. After defining a rule of simulation, he applies this
heuristic principle to quantum physics and concludes that quantum physics, as it now is,
could not be exactly simulated this way, even if it were made fully discrete. Feynman leaves
it as an open question how quantum mechanics should be altered so as to conform to this
principle. In this chapter I show that the modifications to linear quantum mechanics
introduced in CCQM are natural modifications to introduce in order to work towards
this goal. It is shown that CCQM is able to be simulated according to Feynman's rule of
simulation, and the constraints on further development of CCQM in compliance with this rule
are discussed.

In chapter \ref{entropy_time} I discuss the measure of complexity of CCQM as a new measure
of quantum mechanical \textit{entropy}. I give general arguments why it should be viewed as a
measure of entropy and show, in one example at least, that it has comparable magnitude to the
standard thermodynamic entropy. I discuss some of the philosophical and theoretical
problems with the standard ``ensemble'' definitions of entropy. While the new measure of
entropy will not always have comparable magnitude to the thermodynamic or statistical
mechanical entropy, I argue that it, and CCQM, can assist in providing an objective
foundation for an ensemble entropy, and so hopefully provide an objective microscopic
foundation for statistical mechanics, thermodynamics, and the arrow of time.

My interest in metaphysics is primarily driven by a desire to understand present
physics more deeply, and a desire to explore alternative metaphysical possibilities, with the
hope that new physical theories might develop from these metaphysical speculations. These
two goals are connected. By carefully analysing current physical theories, using philosophical
tools and concepts, one understands more deeply the strengths and weaknesses of these
theories, and the range of interpretations of these theories that are possible. Furthermore, the
logical space for alternative physical theories and metaphysical systems becomes clearer,
along with some of the potential strengths and weaknesses of these alternatives. The picture
of physical reality implicit in one or more of these metaphysical systems can then be
developed into fully-worked-out physical theories.

I follow this pattern of reasoning in this thesis. I begin with a philosophical analysis
of our current theories of space, time and matter, pointing out some of their weaknesses. I
then explore some alternative metaphysical systems, based on the notion of prespace. Finally
I develop one of these alternatives, centred on the idea that all physical quantities are discrete,
within the realm of quantum mechanics, developing the outlines of a discrete quantum
mechanics. Thus the thesis covers ground ranging from rather esoteric metaphysical
discussions within the philosophy of space, to the outline of a new theory of wave function
collapse in quantum theory. Some of the latter discussion may be regarded as speculative
theoretical physics rather than philosophy. I do not believe however that there is a sharp
boundary between metaphysics and theoretical physics, or between philosophy and science
more generally. 

In hoping that the metaphysical speculation contained in this thesis may lead to new
empirically confirmable physical theories, I am following a path that has proved fruitful in
many cases in the past. The early stages of much scientific theorising could be classified as
metaphysical or philosophical in character. One prominent example lies in the work of Albert
Einstein. In developing his relativity theories, Einstein began with philosophical
considerations concerning space, time, and motion, and philosophical/theoretical objections
to physics as it then stood. His philosophical reflections on space were built fairly directly on
those of other scientists and philosophers who preceded him. These considerations played a
leading role in his development of both special relativity and general relativity. 

When I say that I hope that the philosophy developed in this thesis might be fruitful in
stimulating the development of physical theories, I mean either by further development of the
theory I outline in chapter \ref{qm_discrete_phys} or in the development of quite different theories. These
theories can then hopefully be tested by the different empirical predictions that they make
from the currently accepted physical theories. The approach to physical theories suggested by
this philosophy represents a fairly radical departure from most current approaches, but I
believe that it has promise for the development of fields ranging from quantum measurement,
quantum field theory and statistical mechanics to discussions of the mind/body problem and
the ``origin'' of the universe.

It would be arrogant hubris to think current theories in physics are so nearly right that
they will only ever be supplemented, never corrected, that revolutions are things of the past:
all past announcements of such ``ends of history'' have always been premature, and it would
be equally premature to make any such announcement in our own times.

\chapter{Spaces, Causation, and Reality}
\label{space}

\section{Introduction }
The problem of the metaphysical status of space and space-time has received much attention
from philosophers. By ``space'' I mean the three-dimensional space that appears to surround
us and to contain all the objects of the universe within it. I will refer to this space as ``position
space.'' In mathematics, science and philosophy, use is made of other so-called ``spaces.'' 
Some examples from physics are phase space, configuration space, momentum space, Hilbert
space and Fock space. In philosophy there is logical space or ``possible world'' space. 
Comparatively little philosophical work has been done on the metaphysical status of these
other spaces. Some notable exceptions have been \citet{Strawson1959} and \citet{Reichenbach1957,
Reichenbach1991}. For the most part the status of position space has been discussed totally independently
of any discussion of the status of these other spaces. In my opinion much can be learned
about the status of position space by comparing it to these other spaces. This is particularly
so in light of a comparatively new view on the nature of space known as the ``property view.'' 
The property view of position space is the metaphysical view defended in this thesis, and in
fact I argue that all the spaces under discussion should be afforded the same ontological status---I argue that a property theory should be given for all these spaces. I claim that the apparent
differences between the spaces can be located primarily in the different relations these spaces
hold to causality in the physical world. 

A particular focus of this chapter will be the status of position space and other spaces
in the quantum domain. A previously unpublished paper by \citet{Reichenbach1991} discussing
just this question has recently appeared in \textit{Erkenntnis}. The paper is followed by a
commentary by \citet{Kamlah1991}. Kamlah explains that it was most likely written in 1926 but
was withheld from publication by Reichenbach. The view that Reichenbach discusses,
namely the possible ``reality'' of configuration space rather than position space in the quantum
domain, has similarities with the view I defend here.

\section{Substantivalism, relationism, and the property view }
The traditional debate on the nature of space or space-time has been between absolutism and
relationism. According to \textit{absolutism} or \textit{substantivalism} space (or space-time) is a substance,
something that exists independently of objects and forms a kind of substratum or container
for the objects. According to \textit{relationism} space (and space-time) do not exist--all that exist
are objects and relations between objects, including two-place spatial or space-time relations
between objects. According to relationism, as it is normally understood, any points or regions
not occupied by objects do not exist because relationism only accepts relations between actual
objects.

There is another view about the nature of space or space-time called the ``property'' or
``quantity'' view \citep{Horwich1978,Teller1987,Teller1991}. According to this view, space or space-time is not a substance but a collection of properties. According to the property view there is
no need to postulate the existence of space-time points as individuals, part of the substance
``space-time''---rather we substitute for them space-time properties, one place (monadic)
properties such as ``having position $\p$.'' According to the property view, unoccupied regions
exist in just the same sense as occupied regions---they are aggregates of position properties,
where in the case of unoccupied regions these properties are not instantiated by any objects.

With regard to the other spaces that will be considered, such as velocity space, the
substantivalists and relationists would be in agreement: both would take a relationist view of
these spaces. They would agree that there exist relations of relative velocity between objects,
but that velocity space itself does not exist---it is merely an abstraction. I argue that a property
view should also be adopted for velocity space and other spaces, so that all these spaces exist
in just the same sense as ordinary position space.

\section{Spaces and perception }
Having given a preliminary sketch of the different positions, I will now turn to the relation
between spaces and perception. I believe that one reason that position space has been
traditionally viewed as a substance and other spaces have not been is because of the special
relationship between position space and human perception. In this section I will be
comparing position space with velocity space.

\begin{figure}
\begin{center}
 \includegraphics{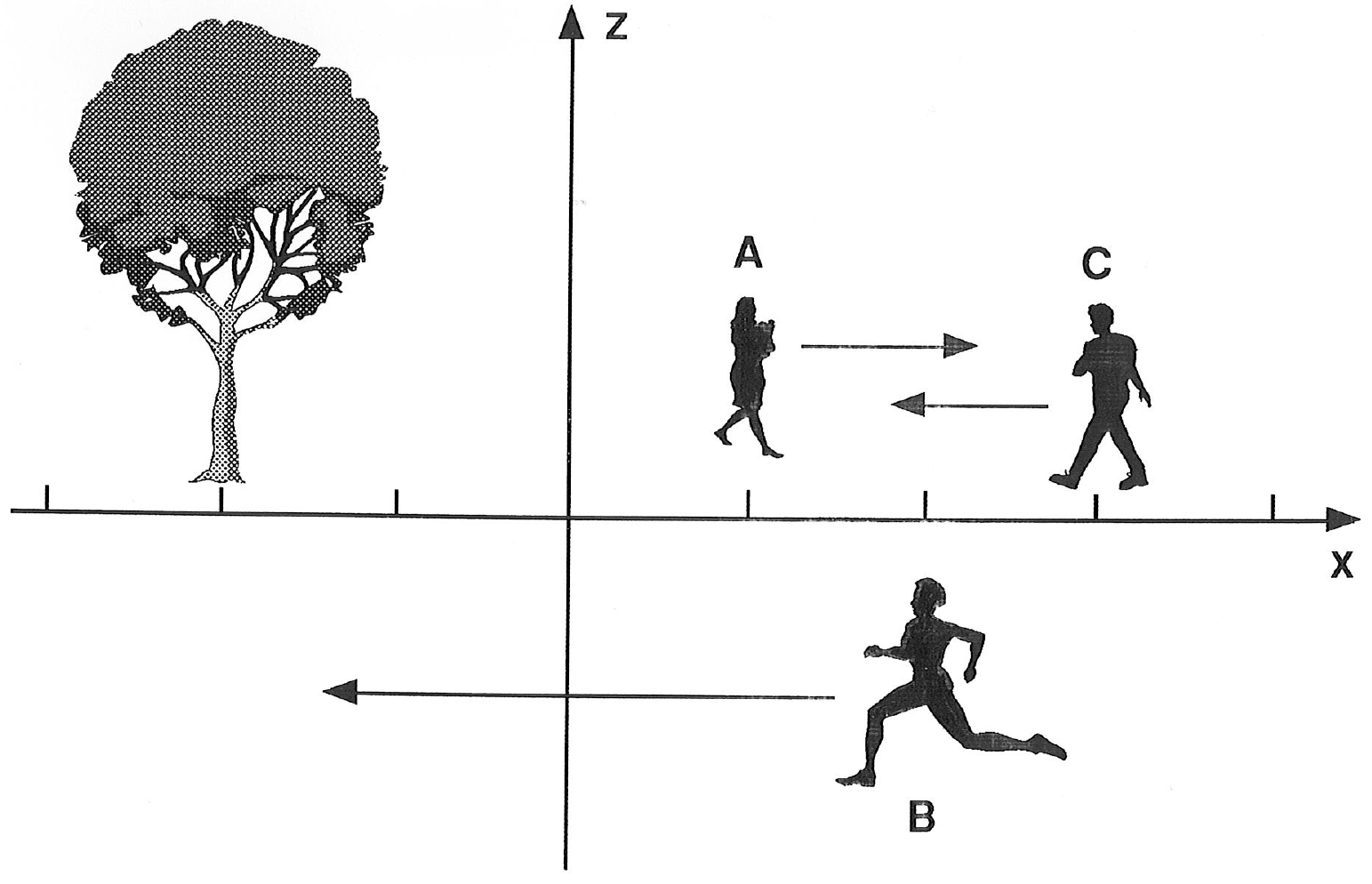} 
\end{center} 
\caption{Position space}
\label{FigurePositionSpace}
\end{figure}

\begin{figure}
 \begin{center}
 \includegraphics{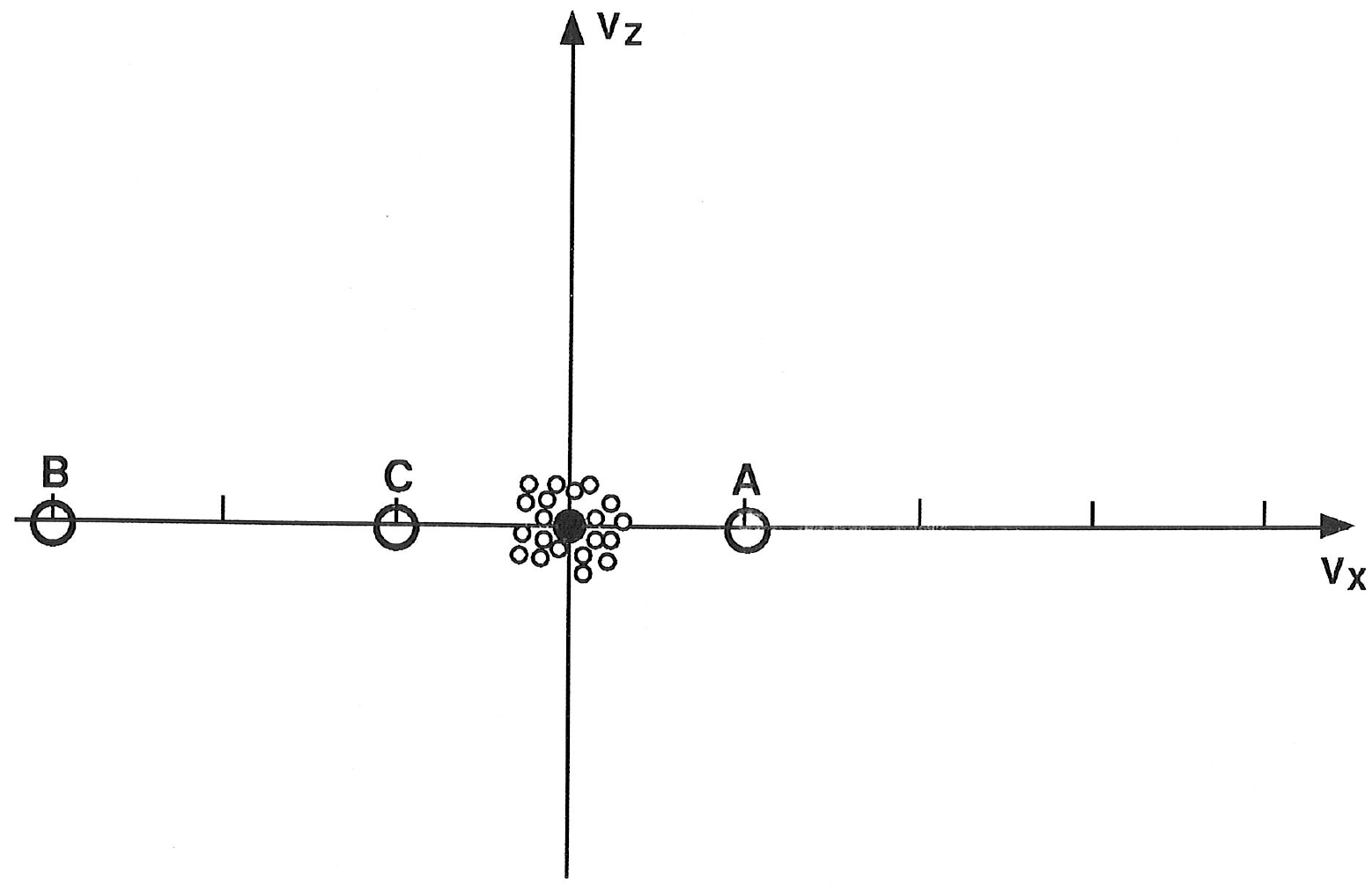} 
\end{center}
\caption{Velocity space}
\label{FigureVelocitySpace}
\end{figure}

Consider the scene represented in Figure \ref{FigurePositionSpace} (on page \pageref{FigurePositionSpace}.) We can see a tree and
people walking back and forth. Figure \ref{FigurePositionSpace} shows an ordinary position space representation of
this scene. In Figure \ref{FigureVelocitySpace} a velocity space representation of the same scene is shown. Roughly
speaking, a velocity space representation of a scene is a representation where the objects are
ordered according to their velocity rather than their position.

In order to make this comparison vivid, imagine that there were some alien beings that
could ``see'' velocity space rather than position space. I will suppose they have eyes
something like ours, but that their minds process the information it receives from their eyes in
a different way. Suppose that as a result of this processing their consciousness were like ours
except that what for us is our visual consciousness of position space, which I will call
visual-spatial consciousness, provides for them a measure of the velocity of objects in the
scene they are looking at. Then when they looked at the scene, something like what they
might see is given by the velocity space representation of the scene. Note that it is may not be
physically possible that such creatures could exist, but I am assuming that this does not matter
for the purposes of this thought experiment.

Call the horizontal direction the x-direction, and the vertical direction the z-direction. 
Then for us the distance along our visual-spatial x-axis provides a measure of the position
component in the x-direction, whereas for them it provides a measure of the velocity
component in the x-direction. Similarly for the z-axis.\footnote{Note that in the diagrams the visual-spatial consciousness of ourselves and the hypothetical aliens are represented by two-dimensional Euclidean spaces. Our visual-spatial consciousness is sometimes described as ``two-and-a-half dimensional,'' since it contains some information about depth, due to our binocular vision. Our visual space may also not be Euclidean. (See \citealt{French1987}.)}

For them the great majority of objects would be at the origin of their visual-spatial
field, presuming that most objects would be at rest. These objects would include the tree, and
other objects as rest in their field of view. Person A walking towards the right would appear
some distance to the right, and persons B and C walking leftwards would appear to the left of
the origin. The leaves of the tree rustling in the wind would oscillate around the origin. (The
leaves are shown as dots scattered around the origin.) 

A most remarkable feature regarding our visual-spatial consciousness is that our
minds simultaneously maintain in that consciousness an ordered representation of many,
many positions, whether or not those positions are currently occupied by objects. This feature
of our visual-spatial consciousness makes it easy for us to judge when objects are \textit{next} to each
other in position, and in general it is easy for us to make quick judgements about the relative
ordering of objects in position space. 

Suppose we can see two objects, one three units from the origin, the other one. Then
because our consciousness presents to us simultaneously all the positions occupied and not
occupied, it is easy to measure-off in our minds the distance ratio of three to one by
successively stepping-off a unit measure across these positions. But notice that the
representation of empty positions is essential for this to work.

In the case of velocity space, we are able to imagine if we like objects having
velocities other than the velocities that we perceive objects having, but this is something
different to the simultaneous awareness of velocities not possessed by objects. Our awareness
of properties not possessed by objects is unique to the property of position. It is not shared by
other qualities we can sense. For the alien beings we are imagining, it is their awareness of
velocity that has this unique feature.

\section{Spaces and reality }
I will now examine the attitudes that might be taken to the relationships between these spaces
and reality. The answer to the question ``Is a space real?'' will depend on what one means by
a space and what one means for something to be real. Ultimately I will argue that the
distinction between ``real'' and ``not real'' is not a useful one in comparing position space and
other spaces, but it is in terms of this distinction that I shall begin the discussion.

A natural first step in trying to make this distinction might be to assume that those
things we can see or sense directly are real, and those things we cannot or could not see are
not real. Certainly there will be a bias towards regarding those things one can see (while fully
conscious) as real and those things one cannot see as not real.

Now I have said that we can see unoccupied positions, and in fact we can see whole
aggregates or regions of positions. Thus we can virtually say that we can see position space. 
Thus using the criterion of reality as being those things we can see, then position space is real. 
On the other hand, we cannot see unoccupied velocities, or regions of velocity space, so under
this criterion velocity space is not real. 

Since we can ``see'' position space whether or not it is occupied by objects, I think that
there is also a natural bias towards supposing that position space would be a structure that
would be left over if there were no objects in the universe. Thus it would be natural to take
the view that position space is a substance, an entity existing independently of those objects,
forming a kind of substratum or container for those objects. It is not always clear what is
meant by a ``substance,'' and some of those I class as substantivalist have objected to this
term, but I take it that a substantivalist takes space to be an aggregate of individuals,
something more akin to a material object than to an aggregate of properties. On the other
hand since we cannot see velocity space, I think that it is natural to suppose that velocity
space is not a substance, but some kind of abstraction.

Our hypothetical alien friends can see unoccupied velocities perfectly, and regions of
velocity space as well, so if they were to apply the same criterion, then velocity space would
be real according to them. Also we could imagine that they might conclude velocity space is
the structure that would be left over if there were no objects in the universe, since they can see
velocity space whether or not it is occupied by objects. That is, they would assume that
velocity space is a substance, and presumably conclude that position space is a mere
abstraction.

I believe that this example shows that our definition of the reality of a space in terms
of whether it can be seen is unacceptable, because it makes reality dependent on how our
sense apparatus happens to be. I will take it as part of what it means for something to be real,
as far as the discussion of spaces go, that it exists independently of us and our minds, and
independently of how we happen to be made up. According to substantivalism and the
property view space is real in this sense, but according to the relationists space is not real in
this sense, since although spatial relations between things exist, space itself is merely a ``form
of representation,'' which is not in the world but is merely something that we use in our
theories.

Another possible definition of what it means for something to be real is that there
exists independent of us something that matches some common sense or ``folk'' notion of
what that thing is like. I argued that there is a natural bias towards regarding the space that is
``directly visible'' as a substance, although I suggest that the substantival view of position
space is not universally shared, even by the ``folk.'' So if the question ``Is space real?'' means
``Is space as many people take it to be, namely a substance?'', then I think the answer is that
space in this sense is not real, since I argue that space is not a substance. However, I argue
that rather than simply saying that space is not real, it is better to say that space is not what
many people think it is: it is a property space.

\section{Spaces as property spaces }
The position I will be supporting is that both position space and velocity space are spaces of
properties. I argue that both position properties and velocity properties exist, so both of these
spaces exist---it is just that they are not substances. Thus I argue that both we and the
hypothetical aliens can ``see'' a real space, only the spaces we can see are not substances but
collections of properties. I am not suggesting that the ``hypothetical alien'' thought
experiment demonstrates by itself that it is mistaken to regard position space as a substance,
and mistaken to assign velocity space a different ontological status to position space. I am
suggesting that this example demonstrates that we have a natural bias towards regarding
position space as a substance due to the special relationship of our senses to position space. 
Thus we must be particularly careful to check that arguments for the substantival nature of
position space are good ones, and that they do not apply equally to velocity space. I will be
arguing that there are no such good arguments to be found, and that it is therefore preferable
to assign the same ontological status to position space and velocity space. Adopting a
property view of position space and velocity space achieves this, and at the same time retains
the same advantages over the relational view that the substantival view of space enjoys.

Since there may be parts of these spaces that are not occupied by any objects,\footnote{According to current physics, all of position space is occupied by at least one field at every point, so it may be argued that the problem of uninstantiated properties does not arise for this space. However, I do not wish to assume the ultimate truth of all of current physics in this discussion. The view that fields continuously fill all of space is sometimes challenged by rival views, such as ``particle'' theories and ``discrete space'' theories, according to which there can be empty positions. In the case of position space, the actual or \textit{possible} existence of uninstantiated positions is one feature that distinguishes the property view from the relational view. The relational view rules out the existence of (mind-independent) uninstantiated positions \textit{a priori} whereas for the property view whether or not there are uninstantiated positions is a contingent matter. Furthermore, other 
property spaces we will discuss include properties that are not instantiated in the actual world, so the problem of uninstantiated properties does arise in general.} I am
supposing that those positions and velocities that do not happen to be possessed by any
objects exist as well as those that are possessed by objects. Thus we need to examine the
philosophical status of properties, particularly properties that are not instantiated.

In philosophical tradition, properties and relations are generally taken to be \textit{universals. 
}There are other views on the nature of properties, but I will not consider these. Universals are
generally thought of as those things that are or can be instantiated by more than one thing at
once. While many universals are indeed of this sort, I do not assume that all properties or
relations have to be of that kind. I allow that there may be properties which are necessarily
unique, which can be instantiated at most once. Thus I do not take multiple instantiation to
be an essential feature of universals. Rather, I take the defining characteristic of universals to
be the fact that they \textit{can} be instantiated by individuals or other universals; a universal can
either instantiate or be instantiated: an individual can instantiate but cannot be instantiated. 
Thus there can, on my view, be uninstantiated universals or universals instantiated only once. 
\citet{BigelowPargetter1990} defend this view. Others, such as \citet*{Armstrong1978}, accept the
existence of fewer universals, and then there are nominalists, who do not accept the existence
of any universals. See \citet{Field1989} for nominalist arguments against a property view of
space.

To say that a property exists, I maintain, is not to say that some entity exists in some
kind of Platonic heaven. The claim that a property exists is logically equivalent to the claim
that something could instantiate such a property. There are some ``properties'' that do not
exist---these are self-contradictory properties that nothing could instantiate. The property of
``being round and square'' does not exist, nor does the property of ``being less than two feet
wide and more than five feet wide in the very same direction.'' For each of these putative
properties, there are predicates in English, but there are no properties that correspond to these
predicates. I think that this sense of what it means for a property to exist corresponds well
with ordinary scientific usage. We might say: ``there exist certain discrete energies that are
solutions to equation H, but there exist no energies between these that are solutions of this
equation.'' 

There is one philosophical tradition of supposing that properties must be instantiated
in order to exist, but I suggest that the extensive usage of uninstantiated properties in science
is one good reason for allowing for the existence of uninstantiated properties. One can try to
paraphrase away the reference to properties in a philosophical reconstruction of the scientific
talk, as is attempted by nominalists such as \citet{Field1989}, or one can suppose that the
reference to properties in scientific equations and scientific models is reference to mere
``conceptions,'' where these conceptions are viewed as mental entities of some kind which do
not exist, other than perhaps in the mind of scientists. However, I believe it is simplest and
most natural to interpret the scientific terms at face value: when it is supposed that some
property exists as a solution to a certain equation, or is referred to in a scientific model, this
should be interpreted as a reference to a property existing independently of us and our
thoughts. 

One position that could be held is to accept the property view of space but to adopt an
antirealist view of properties. That is, one might accept that space and space-time ``consist''
of properties rather than individuals, but then suppose that properties are ``mere conceptions''
and their use does not involve any ontological commitment to something existing
independently of us. According to this view, space and space-time do not exist, since
properties do not really exist. This is certainly a possible position. My primary goal is to
support the property view over substantivalism and relationism, and it is a secondary goal to
support realism about properties. I think that the antirealism about space that follows from
this compromise position is unfortunate, as I would like to adopt a position that respects the
intuition that many people share that the position space that we can ``see'' does exist. 

I identify the property view of space (or space-time) with the view that space or space-time consists in an aggregate of spatial (or spatio-temporal) properties, such as ``having
position $\p$.'' I identify the substantival view with the claim that space (or space-time) consists
in an aggregate of individuals---spatial (or spatio-temporal) points. 

In relativity theory three-dimensional position space is combined with the single time
dimension to form four-dimensional space-time. I argue both space and space-time are
property spaces. In the case of space-time, just which space-time properties are instantiated
depends greatly on the ontology of the past, present, and future. According to those who
support \textit{presentism}, only the present exists, which means that only the individuals in the
present exist. Individuals in the past used to exist, but no longer do so, and individuals in the
future are yet to come into existence. If presentism is true, the only space-time properties that
are instantiated are the ones that are instantiated by present individuals. All space-time
properties that refer to the past or future are uninstantiated.\footnote{The notion of a unique ``present'' may presuppose some privileged reference frame, so some people see a conflict existing between presentism and special relativity. I will not discuss this issue here, although the issue of the relationship between relativity and the existence of privileged reference frames will come up in chapter \ref{qm_discrete_phys}.} These properties all exist,
however, because it is certainly logically possible that these properties could be instantiated:
in fact, many of them have been instantiated in the past, but are no longer instantiated, and
many of them will be instantiated in future. The major opposing view to presentism is the
``four-dimensional'' view of space-time, according to which the past, present and future all
exist equally. If this view is true, then individuals in the past, present and future exist, and
many space-time properties that refer to the past or the future are instantiated. Presentism is
the view that I support, although I will not be arguing in its favor, against the four-dimensional view, and I will not be assuming either view for the remainder of this chapter.

The notion of a substance has in the philosophical tradition usually been conceived of
as having further characteristics than I have identified as characterising individuals. I have
identified individuals by their role as fundamental bearers of properties---individuals
instantiate properties and are unable to be instantiated themselves. While being a bearer of
properties in this sense is one of the roles traditionally associated with being a substance,
some philosophers have supposed that to count an entity as a substance, the entity must meet
other conditions, such as being self-subsistent and being active. I will not require any of these
further conditions to be met for space to be classified as a substance---I only require that it
consists of an aggregate of individuals. 

Newton himself, who is considered to be the original champion of substantivalism
about space, argued that space, and the parts of space, are neither substances nor properties. 
He said that spatial extension ``has its own manner of existence which fits neither substances
nor accidents'' \citep[132]{HallHall1962}. Newton said that extension 
\begin{quote}
 is not a substance; on the one hand, because it is not absolute in itself, but as it
were an emanent effect of God; on the other hand, because it is not among the
proper dispositions that denote substance, namely, actions, such as thoughts in
the mind and motions in bodies. Moreover, since we can clearly conceive
extension existing without any subject, as when we may imagine spaces
outside the world or places empty of body, and we believe [extension] to exist
wherever we imagine there are no bodies, and we cannot believe that it would
perish with the body if God should annihilate a body, it follows that
[extension] does not exist as an accident inherent in some subject. And hence
it is not an accident. \citep[132]{HallHall1962}
\end{quote}
We can see from the quoted passage that Newton rejected the idea that space, or its parts, are
substances because he was considering a traditional notion of a substance, whereby a
substance must be self-subsistent and active. What if we were to consider whether Newton
considered space, and the parts of space, to be individuals or properties? The criterion to be
met to be an individual is simply to not be a property: to be a bearer of properties rather than
something that can be instantiated. Using this criterion, it seems to me that Newton
considered space, and the parts of space, to be individuals rather than properties, since he
clearly rejects the idea that extension is a property. He also says that extension ``approaches
more nearly to the nature of substance'' \citep[132]{HallHall1962}. It is true that Newton is
presupposing a different conception of properties to the one I am using. The quoted passage
makes it clear that Newton believes that properties must be instantiated in order to exist,
because he argues that extension exists even where there are no bodies, so that extension
could not be a property instantiated by bodies, and it follows from this that extension is not a
property at all.

I have noted above that I would deem space to be a substance if the parts of space
were individuals. I do not require that space have any of the other characteristics traditionally
associated with the concept of substance to be counted as a substance. On this definition of
substance, Newton should be classified as a substantivalist, despite his explicit denial that
space is a substance. I defend my definition of substantivalism on the grounds that the view
that space consists of individuals (points) is the view most directly opposed to the property
view, according to which space consists of properties. This contrast between space consisting
of properties on the one hand, and space consisting of individuals on the other hand, is the
major contrast I wish to focus on, so I have made this contrast central to my definitions of the
ontological views about space. The division between properties and individuals is a clear
dichotomy, and I argue that it enables us to capture at least one major point of disagreement
in the debate over the ontology of space or space-time. Newton supposes space to consist of
individuals rather than properties, I have argued, and this is enough to class him as a
substantivalist in my view.

I argued earlier that the fact that we can ``see'' position space plays a role in giving
plausibility to the view that space is a substance, something analogous to a material entity
existing independently of us. I think that this can be seen in a passage from Newton's
writings. In arguing for the existence of spatial extensions as individuals existing
independently of objects, Newton says that we have an ``exceptionally clear idea of extension,
extracting the dispositions and properties of a body so that there remains only the uniform and
unlimited stretching out of space in length, breadth and depth.'' It seems to me that a creature
that could ``see'' velocity space would have just as ``exceptionally clear'' a conception of
extension in velocity space stretching out in all directions in velocity space, so that this is not
a good argument for the existence of position space as a space of individuals rather than of
properties.

My definition of substance differs from that given by Earman. Earman's definition is
similar to the extent that he also characterizes substances in terms of their role as bearers of
properties. \citet[14]{Earman1989} says that in substantivalist analyses of space-time theories
``space-time points are being treated as substances in the sense of objects of predication.'' 
Thus he supposes that if space-time points are being assigned properties by the theory then
the space-time points must be substances, according to the theory, or at least they are ``being
treated as substances.'' However, according to the definition of substance I am using, just
because something instantiates properties does not mean that it is an individual. It could
also be a property. The crucial question is whether the thing in question can itself be
instantiated: if it can be then it is a property, if it cannot then it is a substance.

Similar remarks apply to Earman's discussion of Newton's substantivalism. \citet{Earman1989} also characterizes Newton as a substantivalist, but I do not fully agree with Earman's
reasons for counting Newton as a substantivalist. \citet[114]{Earman1989} says that on a
conservative reading of Newton's remarks, both bodies and space are substances in that
bodies and space points or regions are ``elements of the domains of the intended models of
Newton's theory of the physical world.'' In contrast, he says, ``bodies alone exhaust the
domains of the intended models of the relationist's world.'' Earman is supposing that the
elements of the domains of the intended models of a theory must be individuals, and he thinks
that identifying the parts of space as individuals is enough to be classed as a substantivalist,
from the point of view of the modern debate about the metaphysics of space. I agree with the
second part of Earman's view: I agree that we should class Newton as a substantivalist
because he viewed the elements of space to be individuals. However, I do not agree that if
something is an element of the domain of the intended model of a theory then it must be an
individual rather than a property. 

\citet[114]{Earman1989} says that the property view ``removes space points from the
ontology (the domains of the intended models) to the ideology (the predicates applied to the
elements of the domain).''\footnote{The term ``ontology'' is usually taken to mean ``that which exists'', and since I believe that properties exist, I take them to be part of the ontology, in this sense of the word. The terms ``ontology'' and ``ideology'' are sometimes used to refer to the categories of individuals, on the one hand, and properties (or predicates), on the other hand. A metaphysical realist, who believes in the existence of properties, must be careful not to conflate these senses of the word. } In my view, we can hold on to the property view and at the same
time suppose that Newton's theory and other theories of space have the parts of space within
the domains of the intended models. We simply hold that the elements of the domain are
properties rather than individuals. 

Earman is correct that properties are those things that can be instantiated. The
important point is not whether the elements of space-time are being assigned properties in
Newton's theory. The important point is whether the elements of space-time are
fundamentally those sorts of things that have properties but cannot be instantiated themselves.

The point can be made in terms of Quine's criterion of ontological commitment. The
argument for the existence of space-time is often stated in the following terms. Whatever is
quantified over in our best theories we should \textit{prima facie} give ontological commitment to. 
In our best theories of motion we quantify over space-time points, so we should believe that
space-time points exist. As a scientific realist, I assent to this \textit{prima facie} principle. The
principal in this form originates with \citet{Quine1961a}, who states that we can identify the
ontological commitment of a theory by translating the theory into first-order logic and seeing
what entities are quantified over in that first-order logic. However, Quine does not assume
that what is quantified over must be individuals rather than universals. He makes this clear in
his article ``Logic and the Reification of Universals'' \citep{Quine1961b}. Whether we should be
ontologically committed to universals or not must be decided by comparing the merits of
theories which quantify over universals with theories that do not quantify over universals. It
is true that Quine has certain objections to theories that quantify over properties, but this is
another matter. He does not presume \textit{a priori} that what is quantified over must be
individuals. 

Earman's assumption that the elements of the domain of the intended model of a
theory are individuals rather than properties is, in the language of model theory, equivalent to
the assumption that when the theory is translated into first-order logic, those things that are
quantified over are individuals rather than properties. One possible reason for this
assumption could be a presupposition of nominalism. It is assumed that properties do not
exist, so the ontological commitment must be to individuals rather than to properties. 
Without this presupposition of nominalism, there seems to be no good reason to assume that
the quantification is over individuals rather than properties. I am assuming a metaphysical
realism about properties, so that it must be decided on a case by case basis whether what is
being quantified over are properties or individuals. It is true that in quantifying over the
space-time manifold we assign properties to the entities that make up the manifold, and
typically individuals are those entities to which we assign properties, but we could just as well
be assigning properties to properties as properties to individuals. 

Earman specifically denies that he has a commitment to nominalism \citep[154]{Earman1989}. Nevertheless, the nominalist tendencies of Quine, which led him to eschew
quantification over properties, has led, I believe, to a tradition of nominalist predilections
among philosophers who have followed him and adopted his criterion of ontological
commitment. This tendency to nominalism may explain why the property view has not
gained wide popularity among philosophers, and substantivalism has been seen as the natural
alternative to relationism.

Earman is partly correct, in that if we claim that the elements of space are properties
then we must show that they can be instantiated by some individuals. In a meta-analysis of a
theory of space (or space-time), we can sort the elements of the ontology of that theory into
individuals and properties, based on whether the elements are the sorts of things that can be
instantiated or not, according to our best theories of the nature of those elements. However,
we need not insist that the theory itself quantify only over individuals. I claim that the
elements of space (and space-time) are properties, so I must identify individuals that
instantiate these properties or could instantiate those properties.

In the case of theories of space or space-time which refer to classical particles, the
individuals of the theory are the particles. These particles instantiate at least some of the
space-time properties that make up the manifold of space-time, and those space-time
properties that are not instantiated are the sorts of things that \textit{could be} instantiated by
particles. 

The conception of the world according to field theories takes the world not to be
occupied by solid objects or particles but by fields and potentials having values at every point
in space. One metaphysical interpretation of these theories, is to regard the individuals of the
theories as being the field values at each point in position space. These individuals instantiate
the properties of having certain positions, and the individual field values evolve in time,
having causal relations only with other field values with neighbouring positions. Another
interpretation, the one I favour, is to suppose there are objects which we refer to by such
labels as ``electromagnetic fields,'' and these co-instantiate many pairs of properties, the
property of having a certain position paired with the property of having a certain field
magnitude corresponding to that position. Alternatively, we could suppose that each part of
the field instantiates the properties of having a certain position, and this property in turn
instantiates the property of having a certain electromagnetic field magnitude. This is the most
direct translation of the space-time theories, if we suppose that we are quantifying over space-time properties.

In modern presentations of space-time theories, the theories are often referred to in
model-theoretic terms. A space-time theory is given by a model, consisting of a manifold $M$,
with fields $ O_{i} (i = 1, \ldots, N)$ defined on it, including geometric fields. There is discussion in
the literature \citep{EarmanNorton1987, Maudlin1988, Hoefer1996} about what structure
should be identified with space-time itself---either the manifold $M$ alone or the manifold
together with its geometric fields, or some other structure. For the sake of definiteness I will
be identifying space-time with the manifold plus the geometric fields. As Maudlin and
Hoefer point out, the manifold by itself has only topological structure, and the time dimension
is not even distinguished from the position space dimensions on the manifold itself. The
manifold has few of the properties one would normally ascribe to space-time, so I argue it is
more natural to identify space-time with the manifold plus the geometric fields. (Here I am in
agreement with Maudlin. By contrast, Hoefer identifies space-time with the geometric fields
(the metric field) alone.) \citet[519]{EarmanNorton1987} argue that if we include the
geometric fields as parts of space-time then they ``cannot see how we can consistently divide
between container and contained.'' However, as I reject substantivalism I reject the notion of
space as a ``container'' for space, and I suggest that even a substantivalist should not take this
metaphor very seriously. The realist claim that space-time exists independently of matter can
be expressed by the claim that the manifold and the geometric fields would still exist even if
there were no matter fields. Earman and Norton also argue that the geometric fields can carry
energy and momentum, and this is a reason to suppose that they are fields within space-time
rather than part of space-time, but this argument seems weak: I cannot see why we should
suppose that space-time cannot carry energy.

One could suppose, if one holds the property view, that this manifold and its
geometric fields consist of an aggregate of space-time properties, and relational properties of
those properties. Again the individuals can be regarded as the fields, the matter fields that is,
and these co-instantiate the elements of the model, including the manifold and the geometric
and matter field magnitudes defined on that manifold. On this view, supposing that space-time is continuous, we can suppose the matter fields co-instantiate an uncountable number of
$N+1$-tuples of properties, each $N+1$-tuple consisting of the property of having a certain
position in space-time, conjoined with the properties of having the $N$ matter field magnitudes
and geometric field magnitudes corresponding to that position. Alternatively, as before, we
can suppose the matter fields instantiate the space-time properties making up the manifold,
and these space-time properties co-instantiate $N$-tuples of properties, the properties of having
the $N$ matter field magnitudes and geometric field magnitudes corresponding to that position. 
The only ``fields'' that should be regarded as individuals are the matter fields. The geometric
``fields'' should be regarded as properties of these individuals directly, or as properties of the
space-time properties.

Many definitions of substantivalism have been offered other than the one I have given. 
According to one view, space is a substance if it exists independently of objects. However, it
also follows from the property view, as I have characterized it, that space exists independently
of objects, so this feature does not distinguish substantivalism from the property view. In
defining substance in terms of its role as a bearer of properties, Earman deviates from an
earlier definition given by \citet[521]{EarmanNorton1987}. According to this definition,
which Earman and Norton call the ``acid test of substantivalism,'' space is a substance if the
world would be different if all things in space were translated three feet East, retaining all the
relations between bodies. This will follow from the substantival view if the substantival
points have primitive identities (haecceities), so that the objects would exist at genuinely
different points if this translation were to take place. But it will also follow from the property
view that this translation would bring about a real change, if the space-time properties have
primitive identities. For example, parts of matter that instantiated a particular position $\p'$
would no longer do so were such a translation to occur. Thus I argue that this ``acid test'' also
fails to distinguish the property view from the substantival view. Incidentally, if it were the
case that space-time properties did not have primitive identities, but instead derived their
identities from their relation to some piece of matter in the universe, or in relation to some
reference frame picked out by the distribution of matter in the universe, then it would not be
possible to move all objects three feet to the East. But the same would be true if space-time
consists of individuals without primitive identities, and the identity of space-time individuals
were determined in a similar way. Thus I argue that passing the ``acid test'' is neither
necessary nor sufficient for either substantivalism or the property view.

Both the substantival view and the property view of position space gain certain
advantages over the relational because both views accept the existence of the full spatial and
space-time manifold, irrespective of whether the whole manifold is occupied by objects. For
example, substantivalists and those who hold the property view can explain why it is the full
position space-time manifold is useful in our physical theories---it is useful because it exists. 
See \citet{Earman1989} for an account of the advantages the substantival view and property view
share over the relational view. I will take these advantages over the relational view to be
decisive, so I will take the contest to be between the property view and the substantival view.

I have stated my support for the position that all properties that are able to be
instantiated exist. I will call this view property realism, which is opposed to various views of
a more nominalist nature. I believe that property realism should be accepted on grounds that
go beyond the issues discussed in this chapter. However, it follows from property realism that
the property spaces position space and position space-time exist, including any parts of these
manifolds that are not instantiated by any objects. Thus as long as one is a property realist,
one can gain the advantages of substantivalism over the relational view of position space-time
``for free'' without invoking the existence of a substance ``position space-time.'' Therefore I
think we need not and should not believe in the existence of such a substance, and instead
support the property view of position space-time.

Some of the same arguments that have been used for the existence of position space or
position space-time can be used to argue for the existence of other spaces, such as Hilbert
space. For instance, Hilbert space is also extremely useful in our physical theories, so this
provides support for the existence of Hilbert space. And in fundamental quantum theories we
often quantify over momentum space rather than position space. No substantivalist argues
that all these other spaces used in physics consist of substantival points, so substantivalists are
left with the problem of explaining why their arguments lead them to infer that position
space-time exists as a substance, but similar arguments do not lead them to infer that these
other spaces exist as substances. Those who hold the property view have no such problem -
the existence of all these spaces can be embraced, as spaces of properties.

As I have stated it, one way that the property view of space-time can be distinguished
from the relational view is by the fact that the property view accepts the existence of empty
regions of space whereas the relational view accepts only spatial relations between actual
objects. \citet{Teller1991} proposes what he calls ``liberalised relationism'' which allows all 
\textit{possible} spatial relations between objects, so covering all of space. In this paper he prefers
liberalised relationism to his earlier property view of space due to considerations of the so-called ``hole argument.'' I will not be considering this argument here. One problem with his
``liberalised relationism'' theory is that it is not clear what is meant by ``possible relations'':
does Teller mean uninstantiated relational properties or is he introducing some kind of
``possible'' entities? The introduction of ``possible'' entities into the ontology of a
philosophically explanatory theory is generally frowned upon, so I will assume the former
interpretation. As I will be understanding it, liberalised relationism interpreted this way is
just a version of the property view of space. While the property view of space is most
conveniently stated by using monadic, non-relational, space-time properties, I wish to leave it
as an open question whether these properties might in turn be defined in terms of more
primitive space-time relations. I will take the defining characteristic of the property view to
be that it accepts the existence of the entire spatial or space-time manifold, whether or not
every part of the manifold is instantiated, and that this manifold consists of properties rather
than individuals. Whether this manifold is best described as being made up of a collection of
monadic properties or of relational properties I will consider to be a secondary question.\footnote{See \citet{BigelowPargetter1990} for an account of quantifiable properties, which include properties such as position and momentum, referred to there as ``quantities.'' In this case, non-relational quantities are defined in terms of relational quantities, but it is left an open question which should be taken as ontologically primitive in each particular circumstance. } 
Most likely it is essential to the identity of a particular monadic space-time property that it
stand in certain relations to other space-time properties, so I suggest there is not a great
difference between the positions. Nevertheless, I prefer to express the property view in terms
of monadic properties.

\section{Spaces and causation }
Before returning to the metaphysical status of different spaces, I wish to see what are the most
significant differences between the spaces in our theories about the world. One difference is
clearly that they are spaces of different properties. But what I wish to see is whether there is
some other basis that would distinguish between the spaces in a significant and objective way.

A significant difference between position and velocity space shows up when we
consider the link between these spaces and causality. The velocity space representation of a
scene places next to each in that space objects that usually do not effect each other in any
way. For example, a leaf and a person may be represented next to each other because they are
moving at about the same velocity, but if they are separated in position space they will most
likely not significantly influence each other at all. On the other hand, objects next to each
other in position space do tend to influence each other, because the macroscopic world we
experience is governed mainly by \textit{contact} forces, or by influences that rapidly decrease in
intensity with distance. 

Those things that are exerting an influence on each other I will say are \textit{causally
related} to each other. So here a causal relation means an ``almost immediate'' causal
dependence. I will define this causal relation to hold between two parts of a physical system
at time $t$ if the state of one system-part is causally dependent on the state on the other system-part, within some short time interval $\delta t$ around $t$. That is, a causal relation will exist between
two parts of a system if it is the case that if there were a suitable change in the state of the first
part then this change would cause a change in the state of the second part, within $\delta t$. While
here I am not proposing a general analysis of causation, in this chapter I will be assuming that
causal dependence can be identified with, or at least correlates closely with, a kind of
counterfactual dependence. That is, a causal relation will exist between two parts of a system
if it is the case that if there were a suitable change in the state of the first part then there
would be a change in the state of the second part within $\delta t$, and this change would not have
occurred if the change in the first part had been different.\footnote{Thus the causal relation between system parts can be explicated in terms of counterfactual dependence, in a similar way to Lewis's analysis of ``causal dependence'' between events, with the addition of the time constraint $\delta t$. \citet{Lewis1973a} uses his notion of causal dependence to analyse causation in general. While Lewis's analysis of causation in general in terms of counterfactual dependence suffers from difficulties in certain circumstances, I contend that these difficulties will not infect the notion of a causal relation used here, at least not for causal relations in the limit of small $\delta t$, and it is mainly causal relations in this limit that will be of concern here. I will not propose how counterfactuals should be analysed, but note that according to \citet{Lewis1973b, Lewis1986}, laws of nature play a prominent role.}

Thus for a particle theory, a causal relation will exist between two particles in a
system if a force exists between them such that there is a causal dependence between the state
of one particle and the state of the other particle within $\delta t$. For ``field'' or ``wave'' theories, a
causal relation will exist between two parts of that field or wave if, according to the
formalism of the theory, there is a causal dependence between the magnitude of one part of
the field and the magnitude of the other part of the field, within $\delta t$. 

I noted that our macroscopic world of everyday experience is dominated by influences
that drop off rapidly with distance in position space. When we consider the description given
by science of the macroscopic world, which is given by modern classical physics, the \textit{local}
nature of influences in position space is even more strongly emphasized. In classical physics,
reality is described by a number of particles, fields and potentials, the fields and potentials
having values at every point in three dimensional space. The behaviour of the particles,
potentials and fields at a particular point depends immediately only on the magnitude of the
potentials and fields at that point and at immediately neighbouring points.

With the causal relation defined above we can say that for modern classical physics
causal relations are \textit{local} in position space---in the limit of small $\delta t$, objects only have causal
relations with other objects immediately neighbouring them in position space.

The relationship of locality between causal relations and the metric on position space
is highly invariant---it holds for all types of bodies and all regions of position space. On the
other hand, not only are causal relations not local in velocity space, there is \textit{no} invariant
relationship between the velocity difference between things and their causal relations. For
velocity space, it is \textit{more likely} that objects will influence each other if they are close together
in this space, because they could be travelling alongside each other, as part of a single larger
object for example. But this relationship is not at all an invariant one---there are many things
that share the same velocity but do not causally relate, and there are many things that interact
that do not have the same velocity. The degree to which this relationship holds can be
explained by the locality of influences in position space. 

The existence of an invariant relationship between relations of position and causal
relations is an additional difference between position space and other spaces. This could be
expressed by saying that position space has greater \textit{causal significance}. More contentiously
one might infer that the invariant relationship between position metrical relations and causal
relations is explained by causal relations being largely \textit{determined by} position differences. In
other words the existence of causal relations, and their strength, are highly \textit{dependent} on
position differences. By contrast, the existence of causal relations is taken in our physical
theories to be largely independent of velocity differences, although in the case of velocity
dependent forces the velocity differences will modify the strength of causal relations. 
Whether a causal relation exists at all is highly dependent on their position differences,
however. Whether or not one makes this further inference, it is clear that the invariant
relation between causal relations and position relations means that position space is the most
causally significant space. It could be labelled the \textit{causally dominant} space.

Incidentally, we should not insist that objects \textit{next} to each other causally influence
each other for the space to have great causal significance---as long as there is \textit{some} invariant
relation between the metric of the space and pattern of causal relations that exist. For
example, the relationship could be such that those objects separated by 100 units influence
each other instantaneously and most strongly. Thus it is possible that a space could be
causally dominant without being causally local, although in our world position space is both
causally dominant and causally local.

This locality feature explains a lot of the features that position space has. For
instance, since the forces centred on objects prevent objects getting too close together, it is
generally the case that no two things can be in the same place in position space. On the other
hand with velocity space, not having this link with causation, many things can share the same
velocity, and similarly for other properties---many things can have the same colour, or the
same odour and so on. Thus position space individuates entities very efficiently, more
efficiently than other spaces. \citet{Strawson1959} emphasizes this aspect of position space. 
However, this property of position space can be seen as arising from the fundamental link
between space and causation.

In order to survive, we human beings are vitally interested in those things that causally
relate to us, or potentially causally relate to us. If we wish to causally relate to food, for
example, then we have found that we must move towards it in position space. Thus we must
be vitally concerned with the position of the things in our environment relative to us. 
Judgements of relative position are much easier to make with a position space visual-spatial
consciousness than a velocity space one, so in evolutionary terms it is not surprising that we
have ended up with the position space consciousness we have.

Not only do I think that the connection between position space and causation helps to
explain why we ``see'' position space and not other spaces, but it is one of the factors that can
be seen to explain how one could be led to suppose that space has a different ontological
status to other spaces---regarding space as an existing entity or substance and other spaces as
mere abstractions. I suggest that it is best just to note the connection with causality, and to
make the metaphysical distinction is unnecessary and unwarranted.

\section{Space-time and causation }
The theories of classical physics are local not only in space but also in time---that is the
particles and field are directly influenced only by the particles and fields at the immediately
preceding time. For the relativistic versions of the classical field theories we would say that
the classical theories are causally local in space-time. Comparing position space-time with
another property space combined with time, such as velocity space-time, the same reasoning
as before could be used to argue that space-time is causally dominant. 

For general relativity, the metric in position space not only determines what the
gravitational causal relations between objects will be, but the distribution of objects and
energy helps determine the metric of space-time. This dependence of the metric on the
distribution of matter shows that in general relativity the intimate connection we have already
noted between position and causation is even deeper than in Newtonian physics, since not
only do locations in position space largely determine what the causal relations between
objects will be, but also the causal relations between objects help determine the metrical
structure of space-time itself. Although the connection between causation and the metric on
space-time is two-way rather than one way, I do not think that general relativity serves to
provide any significantly new factors in favour of substantivalism about space-time over the
property view of space-time.

\section{Properties and laws }
The relation between the spaces and causation is given by the laws of physics. According to
the view of universals I am supporting, any logically consistent properties or combination of
properties exists. Thus there will exist a countless number of space-time geometries---all the
logically possible geometries. However, our world has a particular space-time geometry,
which I will call the \textit{actual} space-time. The actual space-time differs from the other space
times because it constrains the behavior of bodies in the universe. According to the
substantivalist, the actual space-time is the only one that exists. The actual space-time is the
collection of space-time points that are either occupied by things in our world or could be
occupied by infinitesimally small elements of matter (of vanishingly small mass-energy) in
our universe, given the actual distribution of matter and energy that obtains, and given the
laws of physics. This space-time can be uniquely characterized by the laws of physics,
boundary conditions and any other constraints, and the actual distribution of matter. On the
property view, this actual space-time can be picked out in this fashion from other space-times,
all of which exist, in just the same way as, on the substantivalist's view, the actual substantial
manifold of space-time points can be picked out from other possible space-time manifolds that
do not exist. 

Here I have introduced a distinction between actual properties and non-actual
properties, in addition to the distinction between those properties that exist and those that do
not. The properties that exist are those that could logically possibly be instantiated, and those
that do not exist are necessarily not instantiated. The non-actual properties exist, so there is a
logical possibility that things could instantiate them, but there is a law of nature that excludes
things in our world from taking on these properties. One can express this using a kind of
necessity weaker than logical necessity that is called \textit{natural} necessity. For the properties that
are non-natural there is no \textit{natural} possibility that these properties could be instantiated.

There will also exist an uncountably infinite number of velocity spaces and velocity
space-times. The actual velocity space for our world is a sphere with the radius of the speed
of light---this is the maximum velocity that objects in our world could have while obeying the
laws of physics (assuming that there are no tachyons, which are hypothetical particles that
travel faster than light). The actual velocity space-time is, in some particular frame of
reference, the tangent space for the actual space-time for that frame of reference. Thus the
actual velocity space and velocity space-time will be picked out from other possible spaces in
a similar way to that for the position spaces. 

\citet{Earman1989} points out that it is possible for there to be ``holes'' in the space-time
manifold, which contain no space-time points at all, and he uses this as an argument against
Teller's original property view of space, which seems not to allow for this possibility . On
the property view argued for here there can be holes in the geometry of \textit{actual} space-time, by
means of boundary conditions or other constraints which have the effect of excluding objects
in our universe from those regions. Recall that the actual space or space-time points are those
points that \textit{can} be taken on by objects in our world, given the laws of physics and other
boundary conditions and constraints. According to the substantivalist no space-time points
will exist in the hole. According to the property view many space-time properties will exist
corresponding to the hole, but no actual space-time properties will exist there, since no
objects in our universe will be permitted to take on any space-time properties corresponding
to this region.

The laws of physics and the boundary conditions and other constraints of our world
can thus be viewed as the metaphysical grounding of the distinction between the actual space-time and other space-times that also exist. Thus the question of the metaphysical basis of
space is transferred to a certain extent to the question of the metaphysical basis of the laws of
physics and the boundary conditions and other constraints. This question is beyond the scope
of this thesis, but I have developed an account of laws of nature that I claim would be
sufficient to provide the required grounding. This theory is given in \citet{Leckeyforthcoming}
and \citet{LeckeyBigelow1995}. A reprint of this latter paper is bound at the back of this
thesis. Other accounts of laws of nature may also be adequate for the task, such as Tooley's
\citep{Tooley1987}. I suggest the substantivalist will also require the laws of physics to ground the
metric of space, although the topology of space could be a brute fact rather than being based
in boundary conditions or other constraints, as seems to be required for the property view. 
This would seem to be an advantage of the substantival view. However, we could also
suppose on the property view that it is a primitive fact which space-time properties are actual
and which are non-actual, and the constraint that bodies in our world cannot enter a certain
hole follows from the primitive fact that there are no actual space-time properties within the
hole. This approach is consistent with my account of laws of nature, as presented in Leckey
and Bigelow (1995).

\section{Spaces and reduction }
A case can be made that position space is ``more fundamental'' than many other spaces, such
as ``odour'' space for example. To the extent that odours are objective and in the world, they
correspond to certain molecular shapes. The odour of an object will thus supervene on, and
perhaps reduce to, the positions of the constituents of those objects. If this reduction holds,
then we can define a sense in which position is a more fundamental property than odour and
position space is more fundamental than odour-space---the distribution of objects in odour
space at a given time reduces to the distribution of objects in position space at that same time,
and the reverse does not hold. (If this is too controversial to regard odour as objective, and
reducing to position properties, it is at least plausible that molecular shape reduces to position
properties so position space is more fundamental than ``molecular shape space.'') 

Now it might also be claimed that position space or position space-time is more
fundamental than velocity space because the velocity reduces to a collection of space-time
properties. Thus taking the space-time view of macroscopic particles, thinking of particles as
world-lines in space-time, the velocity is the tangent to the world line at each point, and one
might claim that this is determined by the series of space-time properties. However, this
depends on us doing our physics in position space-time rather than velocity space-time. We
could also represent particles by world-lines in velocity space-time, and then
change-of-position would be the integral of these velocities over time. Now it is true that to
find the actual position we also need to supply the initial positions of the particles, whereas
the velocities are fully given by the position-space world lines. Thus there is some
asymmetry in the ease in which position and velocity are defined in terms of each other. Does
this difference explain why we do our physics in position space-time rather than velocity
space-time? I do not think so---I maintain that the differing relations between the spaces and
causality is much more significant, and in fact it is because of this difference that we do our
physics in space or space-time, not velocity space or velocity space-time.\footnote{Note that we are currently discussing classical physics. When we come to discuss the physics of microscopic objects---quantum physics, we will be comparing position space, or more generally configuration space, with momentum space. In this case the position space representation and the momentum space representation will be seen to be \textit{completely} interdefinable, and the only significant difference between them will be seen to be their relations to causality.}

Nevertheless it might be supposed that space-time is the more fundamental space,
although this view is not universally held. Under the presentist interpretation of space-time,
in which only objects in the present exist, the position and the velocity properties of the
objects in the present both determine the behaviour of those objects in the future. So on the
presentist interpretation, the properties of position and velocity are both metaphysically
fundamental, and space-time is generated as time passes.

It might be suggested that because position space and position space-time are causally
local, causally dominant, and because position space-time may be more fundamental, then we
should infer the existence of a substance position space or space-time underlying the position
space-time properties, whereas velocity space and velocity space-time remain as spaces of
properties only. However, if we begin regarding both as spaces of properties then the causal
locality and causal dominance of position space enables us to explain why position space
plays such an important role in our physics, and also why we ``see'' position space. To infer
the existence of a substance space-time does not seem justifiable because this entity plays no
further explanatory role. As I indicated earlier, due to the fact that we ``see'' position space,
there is natural bias towards regarding space as a substance. I argue that we must beware of
this bias and be sure that we have good reasons for supposing that space is actually a
substance. I argue that no such reasons are to be found.

Some physicists have speculated that there may be a more fundamental structure than
space or space-time. This structure is sometimes known as prespace or pregeometry \citep{MisnerThorneWheeler1973}. The properties of objects that appear to exist in space, including
their positions, would then supervene on the properties of objects in prespace. The possible
existence of such a structure is one more reason that space should not be assumed to be a
substance, because if it is taken to be a substance it is hard to see how it could be taken as not
being fundamental, since space being a substance is usually taken to mean being an ultimate
substratum for all objects. To assume there is a substance space-time would seem to be ruling
out a whole branch of physical inquiry. Prespace is discussed in the next chapter.
\vfill

\section{Configuration space, phase space and logical space }
Another space of interest is known as \textit{configuration space}. This space will be seen to have
great importance in quantum physics. However, I will first discuss classical or macroscopic
configuration space. When we represent two objects, we normally represent them by their
separate positions in three dimensional position space. In a configuration space
representation of two objects, the objects are represented by a single point in a six
dimensional space, where the six dimensions consist of the three spatial dimensions for each
object. Similarly the configuration space for three particles has nine dimensions and so on. 

The utility of configuration space is that it enables the state of an entire system to be
represented by a single point in the space. The other points in the space are then possible
states for the system to be in. This space is a ``state space,'' or in philosophical language, it is
a type of ``possible world'' space or logical space as defined by Lewis. There is no question of
any causal interaction at all between objects at different points in configuration space, for
there are clearly no causal interactions between possible states of systems. Even \citet{Lewis1986}, who supposes that the possible worlds exist as concrete individuals, agrees that there
will be no causal interaction between them. Thus by our causal criterion, configuration space
in classical physics is not causally significant, and the same is true of other ``possible world''
spaces. 

Now the configuration space is the space that is used when using the Lagrangian
formulation of classical physics, a formulation that is logically equivalent to Newtonian
mechanics, but formulated in a more general way. Another logically equivalent formulation
is the Hamiltonian formulation, which is formulated in another state space called phase space. 
For a system of $\n$ particles, this space usually has $6\n$ dimensions---the $3\n$ spatial dimensions,
plus another $3\n$ dimensions representing the momentum of each particle. 

This space is of interest because the quantities position and momentum are treated on
an equal footing. In fact co-ordinate transformations are allowed among all the position and
momentum co-ordinates, creating co-ordinates that are mixtures of the two types. 
Momentum is of course closely related to velocity---if one chooses Cartesian co-ordinates, the
momentum it is just the velocity times the mass of the object. The utility of the Hamiltonian
formulation suggests that it would be advisable to choose a metaphysics where position and
momentum, and velocity as well, have the same type of ontological status. Thus I suggest
that a metaphysics where all the quantities are regarded as properties of the particles is
preferable to one where position is singled out as referring to a substance, and the other
quantities as referring to properties.

In his \textit{Philosophy of Space and Time} \citeyearpar{Reichenbach1957}, Reichenbach compares configuration
space to position space by making use of the notion of causal locality, in a somewhat similar
treatment of the comparison made between these spaces to that put forward in this chapter,
although he does not compare position space in this way to velocity space or any other space
other than configuration space. The assumption that causation must be local Reichenbach
calls the principle of action by contact. He says that given that this principle is adopted,
position space is picked out over configuration space. As \citet{Kamlah1991} points out, his
discussion of causation in configuration space seems to be confused, but the main point to
note is that he does point out that it is only in position space that causation is local. He says
that if we take the principle of action by contact as given, then this holds only for three-dimensional space (which I have called position space) and for this reason position space can
be said to be the ``real space.'' (Reichenbach places the phrase ``real space'' in scare quotes, as
is shown here.) By saying that this space is ``real,'' he certainly does not mean by this that
position space is a substance, since Reichenbach held a relational view of space, in opposition
to substantivalism. Rather Reichenbach appears to be simply using the label ``real'' to signify
the fact that this space satisfies the principle of action by contact, and this principle is a
desirable feature to retain in a physical theory. His use of the term ``real,'' even if only in
scare quotes, only serves to confuse rather than clarify the issue. For while retaining the
principle of action by contact is a significant feature of position space, this does not in itself
provide an argument for its existence, or for the non-existence of other spaces, and the term
``real,'' I suggest, is best reserved to distinguish those things that exist from those things that
do not. 

Reichenbach rejected the substantivalism of Newton. He refers to Newton's absolute
space as a ``mystical philosophical superstructure'' \citep[53]{Reichenbach1958}. Reichenbach
says that he supports the relationism of Leibniz. \citet[6]{Earman1989} points out that
Reichenbach was mistaken in the degree to which modern space-time theory supports
relationism. Reichenbach believed that relativity supports the view that all motion is relative,
so that Newton's absolute space is unnecessary for describing the motion of objects. 
However, there is absolute motion in both special and general relativity, namely absolute
acceleration, so that these theories do not support relationism in this way. 

As I have characterized it so far, relationism can be characterized by the fact that it
only accepts the existence of space-time relations between actual objects, so it cannot allow
for the existence of empty regions of space. However, \citet[62]{Friedman1983} points out that
Reichenbach was not concerned to rule out empty regions of space. Friedman calls the type
of relationism I have been discussing so far ``Leibnizian relationism.'' Friedman points out
that Reichenbach was only concerned to reduce space-time relations to causal relations. 
Friedman calls this ``Reichenbachian relationism.'' I am sympathetic to the idea that space-time relations could be reducible to causal relations, or at least supervene on causal relations,
as long as the category of causal relations is inclusive enough to include possible causal
relations, so that empty regions of space are not ruled out \textit{a priori}, and the entire structure of
space-time can be reduced to causal relations. I will not be examining Reichenbach's attempt
to reduce spatial relations to causal relations: it is sufficient to note here that Reichenbach
does allow possible causal relations, and does admit empty regions of space.\footnote{See \citet{Fraassen1970} for a short account of Reichenbach's causal theory of space, and of subsequent attempts by other authors to develop more satisfactory causal theories. These authors generally attempt to reduce spatial relations to \textit{observable} causal properties and relations, such as connections by light rays. This is due to the fact that these authors generally have empiricist motivations. My motivation for a causal theory of space is not empiricist, rather it is to allow for the possibility that space can be reduced to the (causal) relations among (unobservable) objects in prespace. I make no requirement that these causal relations are observable. Furthermore, as will be seen in the next chapter, I admit not only possible causal relations between objects or events, but also possible causal relations between possible objects.} However, I do
not think that it follows that if spatial properties are reduced to causal relations this means
that the spatial properties do not exist, or that space itself does not exist. My disagreements
with Reichenbach are that I am promoting the property view, and I argue that all the spaces
exist, whereas, as I interpret his writings, he argues that none of the spaces exist. However,
this should not obscure the fact that the property view of space is compatible with
Reichenbachian relationism, and in fact I will be investigating a type of Reichenbachian
relationism in the next chapter.

\section{Quantum mechanics }
Now I wish to turn from the physics of macroscopic objects to the physics of microscopic
objects like atoms and molecules, which are described by quantum physics. In quantum
mechanics we describe a group of interacting particles by a field in configuration space called
a wave function. In this case, we cannot give a description of the particles in position space at
all. So while the use of configuration space is simply an optional convenience in classical
physics, it is essential in quantum mechanics. 

If we consider $N$ interacting particles then we use a configuration space of $3N$
dimensions, three for each particle. The wave function is written as a function of these $3N$
variables and time. The wave function for a single system has a magnitude at every point in
configuration space. Thus in quantum mechanics configuration space is not a possible world
space or state space. In quantum mechanics, the state space is the space formed by all the
possible wave functions of the system. This is called Hilbert space.

The evolution of the wave function with time at each point in configuration space
depends only on the values of the wave function at immediately neighbouring points. Thus
whereas the evolution of a field in classical physics is local in position space, the evolution of
the wave function in quantum mechanics is local in configuration space. Now we used the
locality of causal influences in position space to argue for the causal dominance of that space
at the macroscopic level, so equally here should see the locality of influences in configuration
space at the microscopic level as evidence for the causal dominance of configuration space at
that level. Furthermore, since we cannot give a local position space representation of systems
at the microscopic level, position space would seem to lose some of its claim to causal
dominance at that level.

The causal dominance of configuration spaces at the micro-level is not as strong as,
and has a different character to, the causal dominance of position space at the macro-level. 
For every separate system of particles there will be a separate configuration space, with the
dimensionality of the space depending on the number of particles in the system. That
particular configuration space will have causal significance only for that system, and only that
part of the space where the wave function is non-zero---the metrical relations outside this
region will have no causal significance. Furthermore individual configuration spaces may
lose causal significance whenever a group of interacting particles breaks up, although in some
cases, such as in Bell-type experiments \citep{Bell1964}, it appears the configuration space wave
function retains causal significance long after the particles have ceased interacting. 

I do not wish to claim that position space loses all its claim to causal significance at
the microscopic level. Although particles do not have well defined positions at that level, we
retain the idea that the behaviour of the particles will depend on those particles ``nearby,''
where in this case ``nearby'' means ``having strongly overlapping position probability
distributions.'' By ``behaviour of particles'' I mean the evolution of their probability
distributions. Therefore, position space will retain some claim to being causally relatively
significant.

It is worth remarking that quantum mechanics appears to allow that position-space
non-local effects occur when measurements are made. When measurements occur, according
to the usual interpretation of quantum mechanics, the wave function collapses so that the
particles measured become, instantaneously, relatively localized in position space to a
position that depends on the position of the macroscopic measuring instruments. How does
this non-local behaviour affect the claim that causation is local in position space? There is
some debate about whether these non-local effects are truly ``causal,'' but assume for the
moment they are causal. It is agreed that these non-local effects could not be used to transmit
information, and one could argue that space-time theories are primarily concerned with causal
influences that can be used to transmit information, such as light rays. This means that even
if the principle of locality of causation in position space cannot be strictly maintained, then it
can most likely be retained in a modified form. The modified form of the principle could take
one of two forms, either stating that causation is local in position space except when wave
function collapses occur, or stating that causation is local for that sub-class of causal
influences that can be used to send information.

The quantum mechanical state can be written not only as a wave function in
configuration space but in many other mathematically equivalent ways as well. The
configuration space wave function is known as the expression for the quantum state in the
``position basis.'' Another common basis is the momentum basis, which gives rise to a
momentum space wave function. As I noted earlier, the state space of possible wave
functions is called Hilbert space. Hilbert space is a vector space, and the state of the system
can be represented by a vector in Hilbert space (the state vector), the configuration space wave
function being the expansion coefficient of the state vector in the position basis. The
momentum space wave function is the expansion coefficient of the state vector in the
momentum basis. Since Hilbert space is a state space, it is not itself causally significant. 

The configuration space wave function can be singled out over all the other bases due
to its connection with causality---the time-evolution of the wave function is local in
configuration space, but it is highly non-local in momentum space and in other bases. This
locality in configuration space is completely invariant and accords strictly with action by
contact. On the other hand the type of non-local dependence differs for different systems. 
Any mathematical transformation of the configuration space wave function will give some
kind of non-local dependence, and in no case does a simple relation between the metric and
causal relations emerge. The relation between the metric and causation is remarkably simple
in configuration space. I think that it is reasonable to conclude the configuration space
metrical relations actually determine what the causal relations will be, so that configuration
space should be seen as causally more significant than momentum space and the other basis
spaces, and it could be labelled the causally dominant space for microscopic many-particle
systems.

Take, for example, the Schr\"{o}dinger equation for an atom of $\n$-electrons. In the
position basis the approximate Schr\"{o}dinger equation for this system, neglecting spin, is

\begin{align}
i\hbar\frac{\partial \psi (\bx)}{\partial t}  = - \left(\frac{\nabla_{N}^{2}}{2M}  + \sum_{j=1}^{n}\frac{\nabla_{j}^{2} }{2m}\right) \psi (\bx) + V(\bx)\psi (\bx)
\end{align}
where $\psi(\bx)$ is short for $\psi(\bx_{1}, \bx_{2},\ldots,\bx_{\n}, \bx_{N}, t)$, $\bx_{N}$ is the position of the nucleus, and the Laplacian
operators $\nabla_{j}^{2}$ act only on $\bx_{j}$. The $V(\bx)\psi(\bx)$ term is simply a multiple of $\psi(\bx)$. The potential
$\V(\bx)$ is a sum of terms like $\frac{Ze^{2}}{|\bx_{N} - \bx_{j}|}$ for the Coulomb interactions between the electrons
and the nucleus, and terms like $\frac{e^{2}}{|\bx_{i} - \bx_{j}|}$ for the electrons' interactions with each other. 

In the momentum basis, the equation becomes, if we look only for solutions where
there is no resultant translatory motion for the whole system, and we neglect the motion of the
nucleus \citep{McWeenyCoulson1949}:

\begin{align}
i\hbar{\frac{\partial}{\partial t}} \Phi (\bp_{1}, \ldots,\bp_{n})
= {} & \sum_{j=1}^{n}\frac{\bp_{j}^{2}}{2m} \Phi (\bp_{1}, \ldots,\bp_{n}) \nonumber\\
& + {\frac{Z}{\pi ^{2} \hbar }} \sum_{j=1}^{n} \int\frac{d\bp}{| \bp| ^{2}} \Phi (\bp_{1}, \ldots,\bp_{j} -\bp,\ldots,\bp_{n}) \\
 & -{\frac{1}{\pi ^{2} \hbar }} \sum_{j<k}^{n} \int\frac{d\bp}{| \bp| ^{2}} \Phi (\bp_{1}, \ldots,\bp_{j} -\bp,\ldots,\bp_{k} +\bp,\ldots,\bp_{n}).\nonumber
\end{align}

We see that the rate of change of the position wave function at a point depends only
on the values of the wave function at that point and at points in the immediate
neighbourhood. The dependence on the wave function at that point is given by the product
$\V(\bx)\psi(\bx)$ of the potential $\V(\bx)$ at that point with the wave function at the same point. In other
words, the potential is \textit{local}. Up until now, we have never needed to use anything but local
potentials in conventional quantum mechanics. The dependence on the wave function at
neighbouring points is given by the $\nabla_{j}^{2}$ terms.

Thus the behaviour of the wave function at a particular point in configuration space
depends instantaneously only on neighbouring points in configuration space. This is a
general result, true not only for the system considered but for all systems. This means that the
theory is \textit{local} in configuration space. If we refer to the position wave function as a \textit{field} in
configuration space, then we could call quantum theory expressed in the position basis a
local field theory. Here we are referring only to the time evolution as described by the
Schr\"{o}dinger equation; we are not considering the non-local features that seem to crop up
when we consider what happens at measurement.

Consider now the Schr\"{o}dinger equation for the momentum-space wave function. The
rate of change with time of this wave function at a point in momentum space, on the other
hand, depends on the values of the wave function at \textit{all} points of momentum space---in this
sense, the dependence is radically \textit{non-local}. In the example of the $\n$-electron atom, there is
a series of terms of the type $\frac{1}{2\m}\bp_{j}^{2}\Phi(\bp)$, which are simple multiples of the wave function
at the same point. Then there are the integral terms, which depend on the values of the wave
function at all points of momentum space, although the presence of the $\frac{1}{|\bp|}$ factor in the
integrals means that the contributions will tend to be greatest from nearby points of
momentum space. (Classically this corresponds to the fact that among particles orbiting a
central point, those particles that are travelling at around the same speed are likely to affect
each other the most. Thus this result can be explained by the locality of causation in
configuration space.)

\section{The wave function and reality }
I have provided an argument for the causal dominance of configuration space at the
microscopic level. It is important to note that this argument relies on the assumption that the
wave function gives us some sort of representation of reality. Many physicists and
philosophers would reject this view. For example, Heisenberg at one time said that the wave
function is ``completely abstract and incomprehensible'' and contains ``no real physics at all''
because it, or any of its alternative representations ``do not refer to real space'' \citep[26]{Heisenberg1958}.

While one might question the reality of the wave function on various grounds, notice
that in the quote given, Heisenberg rejects the idea that the configuration space wave function
gives us any ``real physics'' on the basis that it is not defined in ``real space,'' by which he
means position space. But here he is assuming from the start that any description of reality
must be based on position space, and I think that this is an overly restrictive attitude to take.

Among the other founders of quantum mechanics, Born was one who did believe in
the reality of the wave function in configuration space:
\begin{quote}
 I personally like to regard a probability wave, even in $3N$ dimensional space,
as a real thing, certainly as more than a tool for mathematical calculations. 
\citep[105]{Born1949}
\end{quote}
In the paper by \citet{Reichenbach1991} which was written in the 1920's and published
posthumously in \textit{Erkenntnis} in 1991, he suggests that because causation is local in
configuration space in quantum mechanics, perhaps this space is the ``real'' space in the
microscopic domain. (Recall the argument of Reichenbach's that the space in which the
principle of action by contact is maintained should be regarded as the ``real'' space.) 
Reichenbach did not publish this paper. \citet{Kamlah1991} suggests the reason that it was
withheld was that Reichenbach wrote the paper in 1926, basing his paper on Schr\"{o}dinger's
wave interpretation of quantum theory, and he believed the appearance, and subsequent
acceptance, of Born's statistical interpretation of the wave function invalidated what he had
written, because it showed that the wave function was not real, so it could not be causally
effective, so could not show that causation is local in configuration space. However, Born's
interpretation does not necessarily rule out the idea that the wave function is real and causally
efficacious---the quote above shows that, at least at one time, Born himself thought that the
wave function is real, even though it has a statistical or probabilistic interpretation. (Further
discussion of the interpretation of the wave function in quantum mechanics is undertaken in
chapter \ref{qm_discrete_phys}.)

As I argued earlier in the case of position space, I do not agree with Reichenbach that
the space where causation is local should be regarded as the real space. I do agree, however,
that the fact that causation is local in configuration space in quantum mechanics means that
configuration space should be afforded a somewhat similar status in the microscopic domain
to the status of position space in the macroscopic domain. This status I have labelled causal
dominance, as well as causal locality.\footnote{I very briefly proposed a thesis along these lines, that the status of configuration space at the microscopic level has similar status to that of position space in the macroscopic level, in my Master of Arts Thesis \citep[79]{Leckey1991}. I first presented this thesis in detail within the context of the property view of space at a Monash University philosophy department seminar in March 1991 (during my Ph.D. candidature). I also presented it in two papers read at conferences that year: Australasian Association of Philosophy Annual Conference 1991, British Society for the Philosophy of Science Annual Conference 1991. My thanks to Andreas Kamlah for pointing out to me, at a conference in Germany in 1991, that Reichenbach had written over sixty years ago an unpublished paper supporting a similar view to mine about the nature of configuration space in quantum mechanics. (As already mentioned, the paper was published in \textit{Erkenntnis} in 1991, along with a 
commentary by Kamlah.)}

It has sometimes been argued that although configuration space plays a fundamental
role in non-relativistic quantum mechanics, it does not do so in quantum field theory: it is
suggested that quantum field theory deals with fields in ordinary position space-time. One
might argue, as \citet[205]{Earman1986} does, that configuration space should therefore be
disregarded, because quantum field theory is a more fundamental theory than elementary
quantum mechanics. However, it can be shown that at least in the non-relativistic case,
quantum field theory is equivalent to elementary quantum mechanics \citep{Fraassen1991,
Robertson1973}, as long as elementary quantum mechanics is generalized to deal with cases
where there is an indeterminate number of particles. In this case, one uses a generalization of
configuration space known as Fock space, and a system is represented by a superposition of
wave functions, each representing a different number of particles. The Fock space
representation can also be used in relativistic quantum field theory. 

Although the formalism of quantum field theory has the superficial appearance of a
field theory in position space-time, the quantum ``fields'' are represented by quantum
operators rather than magnitudes in space-time, and the interpretation of these operators is not
straightforward. I concur with the interpretation of \citet{Teller1990, Teller1995}, who argues that the
states of systems in quantum field theory should not be regarded as given by fields in space-time; rather he argues for a Fock space interpretation, so that the state of a system is given by
a superposition of wave functions, each representing a different number of particles. In the
position representation, each of these wave functions is defined in its own configuration
space, each configuration space having a different number of dimensions.\footnote{\label{footnote} Recall that in non-relativistic quantum mechanics the wave function of $N$ particles is a function of $3N$ position variables and time, which I interpret as a wave in a $3N$ dimensional configuration space, which evolves in time. In relativistic quantum mechanics, there is a separate time variable for every particle, so the ``wave function'' of a system of $N$ particles will be a function of $4N$ variables. Thus one must be cautious in claiming in the relativistic case that the Fock space representation can be interpreted as giving a superposition of wave functions of $N$ particles, for various values of $N$, each wave function defined in a configuration space, with the number of dimensions of each configuration space being $3N$. This interpretation can only be made if there is some privileged reference frame for the system, in which case all the time variables can be 
identified with the time variable defined in that reference frame, and the wave function again becomes a function of $3N+1$ variables. As I discuss in chapter \ref{qm_discrete_phys}, I favor an interpretation of quantum mechanics that presupposes the existence of a privileged frame.}  The causal
relations involved in the evolution of an entity of this kind will be complicated. The
evolution may not be completely local (in the relevant configuration space) within each
separate wave function of the superposition, due to possible coupling between the
components of the superposition. Thus we could not describe the causal relations as being
local in any particular configuration space, and we may not even be able to say that there
exists a superposition of collections of causal relations, each local in a different configuration
space. But, if this interpretation is right, the causal relations involved in the evolution of the
state of a system in quantum field theory will certainly not be local in position space, because
the state of the system is not even defined in position space. Even if this interpretation is not
accurate, I argue that the elementary quantum theory is at least approximately true for many
systems having motion slow relative to the speed of light, and the configuration space picture
is at least approximately true, at least at one level of the description of reality. 

\citet[14]{Earman1989} argues against the property view of space by saying that in
quantum field theory we quantify over positions, so position is being treated as a substance. 
However, we can equally well express the fields in quantum field theory in momentum space
and so quantify over momentum, just as we can with the wave function in non-relativistic
quantum mechanics. This provides further evidence for the position I argued for earlier, that
just because we quantify over something does not mean that something is an individual. I
argue that in science we regularly quantify over properties as well as substances, so we cannot
make the distinction between properties and substances on the basis of what we are
quantifying over.

How the evolution of the wave function in a configuration space of many dimensions
transforms at the macro level to the evolution of particles and fields in three-dimensional
position space is an unsolved problem in quantum mechanics. This problem is intimately
linked with another problem with quantum mechanics known as the ``measurement problem.''
These problems will be addressed in chapter \ref{qm_discrete_phys}.

I have noted that configuration spaces cannot be based on, or reduced to, position
space, and the reverse also holds, at least not without a solution to the measurement problem
being provided. Thus we have two relatively causally dominant spaces---position space at the
macroscopic level and configuration space at the microscopic level. The existence of two
causally dominant spaces of different types is problematic---they do not fit together smoothly. 
We might hope to solve this problem by finding some other ``space'' or category or structure \textit{more} fundamental than either position space or configuration space, where this new structure
would provide a unified metaphysics. Another indication that such a structure might be
needed is the apparent existence of non-local effects that happen when ``measurements''
occur. These non-local effects do not sit well with either the position space or configuration
space based metaphysics we have described, because if we do grant these two spaces the
special status of the ``causally dominant'' spaces it is because of the preponderance of local
causal relations in these spaces.

As mentioned earlier, another reason for taking the property view of space is that this
fits in most easily with the view that space is not completely fundamental. If space is a
substance, a container for things, then it is hard to imagine how it could fail to be completely
fundamental. But if positions are just properties of objects then we can more easily imagine
these objects embedded in some other more fundamental structure, and the position just being
one of the properties of these objects.

\section{Conclusion }
Now I will go back and reconsider the two spaces we started with, position space and velocity
space, and the views of the world possessed by us and our hypothetical friends from outer
space. The question asked by both was: is there something in the world corresponding to the
space we see? I think that the answer depends on what it is you believe you are seeing when
you look into space. 

If you believe that you are seeing a substance, a kind of ultimate container for all
things, then I think that we have no good reason to suppose there is anything in the world
corresponding to that concept---it is most likely that space in that sense does not exist. If you
are a little more flexible in interpreting what you see, so that you could be persuaded that
what you see is a network of position properties, then I believe that there is something in the
world corresponding to this---space in this sense does exist. Furthermore it is a causally
significant space---the metrical relations you observe between objects play a dominant role in
determining the causal connections which govern the behaviour of those objects. 

Although you cannot see them there are also configuration spaces associated with
each microscopic system, and these spaces support metrical relations that play a similarly
dominant role in governing the behaviour of these systems. 

The velocity space experienced by the aliens is also a real space of properties, but this
space is not particularly causally significant. The metrical relations they observe have only a
limited role in determining the behaviour of the objects in the world. (Thus they will not find
their perceptual ability particularly helpful in negotiating their environment.)

Due to the causal dominance of position space and configuration space these spaces
have certain other attractive features. One of these is that the entities of the world are
effectively individuated by these spaces, because physical entities are taken to be the grounds
of causal action. In the macroscopic case, the entities are either taken as particles or as fields
in position space, with the causation being local in position space. In the case of fields the
fields can be broken down into infinitesimal elements or magnitudes, each having a separate
place in position space, and each of which acts locally on the elements nearby. These
infinitesimal elements can either be thought of as the individuals of the field theory, or as
parts of the individual field, or as properties of that field. I favor the last of these options. 
There may be more than one field, so there may be more than one magnitude at each point,
but the number will be small. Due to these features of position space, it will be useful to
visualize the elements as if they are embedded in this space. This space will individuate the
elements effectively and show as next to each other those elements that causally relate to each
other---the causal relations of the elements will closely match the topological and metrical
relations of the space. 

This will equally be the case for configuration space in the microscopic case as for
position space in the macroscopic case. Thus just as we do not go too far wrong in
visualising things as embedded in position space in the macroscopic case, I suggest that we
would not go too far wrong in ``visualising'' elements of reality as embedded in configuration
spaces in the microscopic case. This means that for a microscopic system of $N$ particles, it is
useful to ``visualize'' a space of $3N$ dimensions, if this is at all possible. The ``individual
entities'' of the microscopic realm can be taken to be individual values of the $3N$ dimensional
wave function, or these values can be regarded as properties of the individual field described
by the wave function, and these wave function values act locally on other wave function
values at neighbouring points in configuration space.

These ``visual'' pictures are useful at a certain level of description in their appropriate
domains, but naturally there is a tension between these descriptions if we take them seriously
and extend the image beyond the domain they are suited to. Thus I argue that neither of these
spaces should be taken as substances. 

The non-local effects that seem to occur in quantum theory indicate further that the
descriptions of reality that take either of these spaces as basic is likely to be incomplete. It
may be that non-local connections such as these may be incorporated into other unknown
spaces, or they may not be arranged into spatial forms at all, but be arranged in a structure that
does not have well defined topology, metric or dimensionality. It remains for us to discover
and understand this structure, if it exists. 

The possible nature of this structure, and how it might relate to position space, is the
subject of the next chapter.

\chapter{Prespace and Cellular Automata }
\label{prespace}

\section{Introduction }
It is commonly held that space is some kind of great container in which all the things of the
universe are embedded. Leibniz was one philosopher who did not believe this. He thought
that ultimately the universe is made up of peculiar entities called monads which are not
embedded in space, but somehow the relations between these monads result in beings like us
perceiving objects as if they were embedded in space. There are a number of present-day
physicists who can be seen as following in Leibniz's footsteps---they too have conceived that
space or space-time is not basic, and have speculated that everything may be constructed from
entities not embedded in space.\footnote{See, for example, \citet{Bohm1980, ChewStapp1988, Finkelstein1979, Frescura1988}; \citet[1203--1217]{MisnerThorneWheeler1973}; \citet{Penrose1975, Stapp1988, Wheeler1973}.} Various names have been given to the more fundamental
realm of these entities, the structure they are embedded in or make up, including pregeometry
and prespace, but I will use the common term prespace. While early attempts have been
made to make theories of such a structure, none is yet satisfactory. 

Many of the authors mentioned here are working outside the mainstream of modern
physics, but even within mainstream physics the presumption that space-time forms a
background structure for the world is often questioned. This is particularly so with those
working on string theory and on other approaches to a quantum theory of gravity. For
example, \citet[28]{Witten1996} says ```spacetime' seems destined to turn out to be only an
approximate, derived notion.'' He suggests that ``one does not really need spacetime any
more; one just needs a two-dimensional field theory describing the propagation of strings. 
And perhaps more fatefully still, one does not have spacetime any more, except to the extent
that one can extract it from a two-dimensional field theory.''\footnote{String theory is often interpreted as merely adding extra dimensions to space-time: the fundamental strings are interpreted as propagating in 10 or 26 dimensions, and the extra dimensions are supposed to be ``curled up'' on a very small scale so that they are not observable. However, as the quote from Witten indicates, string theory can also be interpreted as challenging the fundamental status of space-time, and of being based on a more fundamental two-dimensional ``space-time'': it can be interpreted as a prespace theory. This point was drawn to my attention by Stephen Hawking.} 

I will be tackling the philosophical problem of how entities that do not exist in space
could give rise to a world like ours where there are entities that at least appear to be
embedded in space. \textit{Relationism} is one philosophical view of space or space-time that rejects
the idea that space is a substance or container in which things are embedded. Relationists in
general believe that there are spatio-temporal relations between objects but that space itself
does not exist. These relations are sometimes taken as primitive, and sometimes it is hoped
that these relations will reduce to some other kind of relations, usually causal relations.  I will be considering in this chapter a form of the latter kind of relationism, which was labelled in the last chapter
``Reichenbachian relationism.'' Theories of space and space-time that embody the ideal of
reducing spatial properties or relations to causal properties or relations are known as ``causal
theories of space.'' 

As mentioned in the previous chapter, Reichenbachian relationism is compatible with
a property view of space, so is compatible with the existence of the whole of space or space-time. Even if spatial properties can be reduced in some sense to causal properties, I do not
see this as an argument for the lack of existence of these properties. It seems to me that it is
only if not \textit{all} of actual space finds a reduction that one might doubt the existence of that part
of actual space that does not find a reduction to causal relations, and hence reject realism
about space as a whole.

I will be investigating a particular kind of causal theory of space which I will call a
 \textit{functionalist} theory of space. In a functionalist theory of space, the spatial relations can be
reduced to the actual causal relations plus the \textit{potential} causal relations between states and
potential states of objects. The label functionalist is chosen by analogy with functionalist
theories of mind. In these theories, mental states are supposed to be reduced to the
``functional roles'' of some entities, the causal relations and potential causal relations of those
entities. While it is generally thought that it is neuronal structures in the brain that fulfil these
functional roles, it is also generally acknowledged that some other structures, such as patterns
of memory states in a computer, could also perhaps fulfil these roles, so it is possible that a
computer could have mental states. I will also make use of the concept of a computer to show
how it is possible that some structure could fulfil the functional roles that are involved in a
functionalist theory of space. I suggest that we could think of a computational structure in
prespace having memory states that fulfil the functional roles that give rise to spatial
relations, and give rise to the appearance of a universe such as ours. The purpose of this is to
show that a functionalist reduction of space is conceivable. This is not to suggest that our
universe is actually generated by a computer in prespace. Just as in a functionalist theory of
mind, any structure that fits the appropriate functional roles will do, so the computer concept
can be thought of as giving some indication of the kinds of non-spatial structures that could
fulfil these functional roles. 

In discussing a computational system that could serve to exemplify a functionalist
reduction of space I will be making particular reference to the concept of cellular automata. 
This reference to cellular automata can be thought of on three levels. At one level we can
view a computational structure that can implement a cellular automaton as a model for the
prespace structures that fill the appropriate functional roles, so that objects appear to exist in
space. At another level, the cellular automaton can be divested of any suggestion of being
implemented on ``hardware,'' and can be regarded as simply a set of laws of evolution of
discrete valued quantities defined on a discrete space-time. At this level, one can take
literally a physical theory that is framed in terms of a cellular automaton as describing a
discrete physics, in which all physical quantities are discrete. At a third level, the talk of
cellular automata and computational systems can be taken purely metaphorically, and one can
use these concepts as a heuristic devices in developing physical theories. I will be making use
of all three levels in what follows.

In this thesis I will be taking time as a basic concept---I will not consider ways that
time might be reduced to something else, only space. This asymmetric treatment of space and
time may not appeal to those who interpret Einstein's theory of special relativity as showing
that space and time must be treated symmetrically as inseparable parts of a four-dimensional
space-time. In this chapter I am only trying to illustrate one way that space and space-time
could be reduced to some other structure. I am not suggesting that other approaches could not
be successful. Indeed, many of the authors I noted as discussing prespace do treat space and
time in a more symmetrical manner than I do in their attempts to reduce space-time to some
other structure. However, under any interpretation, the three dimensions of space and the one
dimension of time have many different properties, and I do not feel obliged to treat them the
same way. Of course any account of the way the world might be must be compatible with the
well established empirical underpinnings of relativity. 

It may be that space is an irreducible part of the world, and functions as a ``container''
for the universe, as it is sometimes put by substantivalists. Without going into all the reasons
some physicists and philosophers suppose it is not, I will mention a general consideration for
trying to reduce space to something else. In physics we have managed to explain most
phenomena in terms of some other more fundamental phenomena, and in particular directly
perceptual properties such as colour, odour, loudness, hotness have been reduced and perhaps
in some cases eliminated in terms of entities that do not intrinsically possess these properties. 
Spatial properties, such as lengths and positions, are fairly directly perceptual properties, so if
we could explain these properties in terms of entities that do not possess these properties, then
we will have explained every perceptual property in terms of quite different properties. It
could be argued that we can be said to have a true understanding of some property only if we
have explained it in terms of other quite different properties.\footnote{See \citet[119--120]{Hanson1965}, \citet[475--476]{Hofstadter1986}.} Thus I will adopt the working
hypothesis that the entities in our universe are ultimately made out of things with completely
different properties to the way things appear to us, and that these things are not based in
ordinary three dimensional-space.

Before introducing the functionalist theory of space, which is modelled by a
computational system, I will briefly discuss two other types of structure that demonstrate
other ways that space could conceivably be reduced to some structure more fundamental than
space. These structures will also be modelled by computational systems. These models
involve a kind of dualism of mind and matter, in contrast to the major functionalist theory,
which is a materialist one. These models will introduce some of the concepts involved and
allow us to work up to the functionalist theory of space, which can then be contrasted with the
earlier models. Although these models are described mainly as a way of setting the scene for
the materialist theory, I think they are of some interest in their own right. They are offered in
the philosophical spirit that starts with the assumption that all that we can be sure of about
objects is the way they appear to us, so that we should consider any possible ontology that
captures the appearances of things to be a way that our universe could be.

\section{Building the universe}
The task we have is to imagine how a universe such as ours could be constructed such that the
properties of the objects in the constructed universe have very little in common with the
properties of the things out of which they are made. In particular the underlying ontology
must not be embedded in the same space as the constructed objects appear to be. One way to
approach this problem is to try to imagine how we could build a universe satisfying these
criteria, then by extension we could imagine that our own universe is something like the one
that we imagined building ourselves.

\subsection{UM1---Solipsistic dualism}
First consider the scenario of the ``brain in the vat.'' In this case we imagine that a brain is
nurtured in a vat and signals are fed to the brain from a vast computer so that the brain
experiences a universe just like our own. 

Now imagine that we could construct a computer that is conscious, which I will
assume is possible, and that this computer could function just like a human mind. Then this
computer-mind could be substituted for the brain in the vat. Furthermore the computer mind
could be embedded within the computer that feeds the computer-mind the appropriate signals
so that it experiences a world just like ours. I will call this system the ``computer in the
computer.'' All the components of this system are built by us. I consider this an instance of
building a universe of the type required, because it captures the appearances of things, or at
least the experiences of one person, yet the objects that the mind experiences (such as tables
and chairs) are, more fundamentally, parts of the running computer, some pattern of electrons
flowing in circuits, which have totally different properties. Call this Universe Mark 1 (UM1). 

We could thus construct a universe with a single mind using a computer whereby the
mind is made up from one part of the computer and the matter is made up from a separate part
of the computer. I label this universe an instance of \textit{solipsistic dualism}.

\subsection{UM2---Strong dualism}
Now consider a universe consisting of many computer-minds interconnected, with all of them
being fed appropriate signals so they see a world like ours. This is UM2, the ``computers in
the computer.''\footnote{There is a science-fiction story ``Non Serviam'' by Stanislaw Lem \citeyearpar{Lem1981a} that utilizes this idea. Lem describes a future field of human study called ``personetics'' in which ``worlds'' are created in advanced digital computers. These worlds are ``inhabited'' by ``personoids.'' The personoids are disembodied minds, their minds being generated directly in separate sections of the computer. These personoids are conscious, and perceive a generated world, which may be dimensionless or have an arbitrary number of dimensions chosen by the programmer of the computer.} I regard this universe as an instance of dualism because in this universe the
ontology of the mind is distinctly different to the ontology of the objects. The computer
minds in the universe will see objects in a three dimensional space like ours. I will call these
objects the \textit{phenomenal} objects. The objects that make up the computers themselves, the
hardware and the mobile electrons moving around the hardware I will call the \textit{noumenal}
objects. I will call the space of the noumenal objects \textit{prespace} and the space of the
phenomenal objects simply \textit{space}. The objects that the computer-minds see, the phenomenal
tables, chairs and even the phenomenal electrons they might observe in bubble-chamber
experiments, say, are all more fundamentally identical with certain patterns of noumenal
electrons moving around the surrounding computer. The phenomenal objects are built out of
patterns of noumenal objects, although the noumenal objects generally have totally different
properties to the phenomenal objects they constitute. The way I am thinking of it, the
phenomenal objects exist, they are real, it is just that they are built out of patterns of
noumenal objects.

The underlying structure of this universe consists entirely of noumenal objects. Thus
for a universe to be dualist it is not necessarily the case that mind and matter are ultimately
made up of different stuff. We can have the mind made out of that stuff directly, and the
matter made of the same stuff indirectly.

In the case I have given, the space of the phenomenal objects and the space of the
noumenal objects are both three dimensional spaces. However, they are not the same space. 
It seems likely to me that we could build the universe in such a way that the computer-minds
perceive a space of a different number of dimensions to ours, either fewer, say 2, or more, say
4 or 5. Similarly a 3-dimensional universe could presumably be generated by a computer in a
universe of a different number of dimensions, or some other sort of structure other than a
computer, and the underlying ``space'' may not be arranged in a dimensional structure at all.

It could be that our own universe is something like this one. Not literally a computer,
just somewhat analogous in structure, and not necessarily constructed by anyone, but it could
otherwise be similar to the universe UM2. 

There is something unsatisfactory about this idea. According to the dualistic universe
described, which I will label an instance of \textit{strong dualism}, if the inhabitants of the
phenomenal world were to look inside each others heads, they would presumably see brains
made out of neurons, if the phenomenal world fully resembles our world, but these
phenomenal neurons would be just as much artificial constructs as the other phenomenal
objects. This means that these phenomenal brains, the wet-ware in phenomenal space, play
no role in generating consciousness or even in processing signals that are fed to the mind. It
will appear to the inhabitants of the phenomenal world as though they are, but they will not
be. If the UM2 universe were a constructed universe, then we would see this aspect of the
universe as an elaborate hoax perpetrated by the universe builders to fool the inhabitants.

Most and perhaps all dualists see the brain as playing \textit{some} role in processing
information received from the senses, but that some higher functions of the mind, such as
consciousness and perhaps memory are generated by a non-physical mind that is somehow in
communication with the brain. Call this \textit{weak dualism}. I will not discuss strong dualism any
further. I will come back to weak dualism briefly later on, but before even beginning to
consider weak dualism, it is necessary to consider materialism, which is the main focus of
this chapter.

\subsection{UM3---Materialism}
The challenge I would like to put is whether we could build a universe such as ours out of a
computer, as before, but one where we do not assign separate areas of the computer to minds
and matter. Otherwise we still require there to be a similarly large gulf between the ontology
of the computer hardware, the noumenal ontology, and the ontology of the generated world,
the phenomenal ontology---these objects must exist in different spaces. What I have in mind
is a materialist universe, with the noumenal objects generating the phenomenal particles and
fields and so on, then the phenomenal brains that are made out of the phenomenal particles
and fields generating consciousness; in this universe the minds are identical with the
phenomenal brains; the problem with UM2 is that the minds have nothing to do with the
phenomenal brains and are certainly not built from phenomenal particles---the minds are made
directly from noumenal objects instead.

Is it conceivable to build such a computer? I think it is, and in fact there are some
people who can be thought of as investigating one way of doing just that.

\section{Cellular automata and discrete physics } \label{cellaut}
These investigators utilize the theory of what are known as cellular automata. Cellular
automata work on the assumption that space-time is discrete. First a network of points or
cells is defined on a three dimensional lattice. The state of each particular cell is represented
by a discrete magnitude or a collection of magnitudes, and these may be stored in computer
memories corresponding to each point. The state of each cell is updated at a certain discrete
time interval. How the values change with time is determined by some rules which depend
on the state of the cell and other cells---usually only the immediately neighbouring cells. 

A simple example of a cellular automaton, defined on a two dimensional lattice, is
Conway's ``Game of Life'' \citep{Gardner1970}. The simple rules of this game generate fairly
complex behaviour that depends greatly on the initial conditions. There are patterns that
disappear, stable patterns, cyclically repeating patterns, complex irregular behaviour, and
stable patterns that move, looking like moving objects---these are called ``gliders.''

The cellular automaton is a type of parallel processor---each cell is updated
independently by a given rule for each cell, presumed to be operating at the same time. It can
however be run on a serial machine by rapidly updating each cell one by one. The cellular
automaton is thus generally run by some computing device, parallel or serial, which could be
constructed from silicon chips, for example. 

Using more complex rules, these cellular automata can be used to model various
physical processes, and it has been proposed by some, notably Edward Fredkin, that a cellular
automaton running fairly simple rules could model all of physics and thus the entire universe. 
\citet{Fredkin1990} calls the discrete physics that is generated by the cellular automaton Digital
Mechanics. 

Aspects of string theory and other quantum theories of gravity have led many authors
to suggest that there may be a minimum length scale in the universe, at approximately the
Planck length: $10^{-33}$ cm. (For example, see \citet{Witten1996}.) To many, such as \citet[152]{Hooft1997}, this suggests that space-time may be discrete rather than continuous. It is a further
assumption that all physical quantities are discrete rather than just space and time, but \citet[176--177]{Hooft1997}
does refer to Fredkin's ideas on cellular automata as indicating a
possible way forward for physics.

Fredkin suggests that rather than thinking of Digital Mechanics as simply providing
approximations to the true physics, in which space-time, fields and so on are continuous, it
could in fact be an exact model of the universe. 

It is not clear to me what Fredkin means by this: whether he means merely that space
and time are discrete, and that Nature is essentially finite, or if he means more than this. In
his major paper \citet{Fredkin1990} he speaks as though we should simply take the laws of our
universe to be the Digital Mechanical laws, in other places he speaks of the world being made
of ``pure information'' \citet{Wright1985, Wright1988}, and other people have sometimes taken him to
mean that the physical system that runs the cellular automaton should be taken as being a
model for reality \citep{Brown1990}. In this thesis I am considering both the first and the last of
these options, but in this chapter I am mainly concerned with the last of these options, taking
the structure of the system that runs the cellular automaton as being some kind of model for
the noumenal objects existing in prespace.\footnote{This idea of generating a world inside a computer in this way has been discussed by many authors, in both fiction and non-fiction. The story ``The Princess Ineffabelle'' by Stanislaw Lem \citeyearpar{Lem1981b} is one of the fictional accounts which discusses this. The physicist \citet{Tipler1994} suggests that at the end of the universe, very special conditions could exist that would enable the inhabitants of our universe to construct a computer and generate a world where all sentient beings that have ever existed could be ``reincarnated.'' See \citet{Rossler1987} for further references.} 

It is conceivable that we need not be committed to the idea that Nature is finite in
order to make use of a similar idea. If it turns out that space and time are continuous, then it
may well be that we could use some kind of analogue automaton rather than a cellular
automaton to model the prespace structure for this continuous physics. However, I will adopt
the assumption of Finite Nature in this chapter because it is known from the general theory of
digital machines that cellular automata have certain desirable properties for our purposes.

One of these properties is great flexibility, which is in the nature of digital machines
and which can be expressed in terms of the concept of \textit{universality}. It can be shown that just
about all digital automata with a certain minimum degree of complexity, including any
personal computer, share the property of universality, which means that the automata can
model any other computer, or any computable process, given access to as much time and
memory capacity that it needs \citep{Minsky1967}.

Due to this property of universality, the nature of the structure that runs the cellular
automaton, its ``physics,'' can be very different from the Digital Mechanics that it is running. 
The structure need not even exist in a space with the same number of dimensions as the
digital physics, so a cellular automaton that was running our universe could be as different
from our universe as we liked, having perhaps many dimensions or perhaps not even having a
dimensional structure at all. The formal result of universality gives us the result we want
provided we stick to digital machines. Analogue machines are less flexible by nature. This
may mean that if we insist on strictly continuous processes in our physics then there may be
severe limitations on the possibilities for reducing this to a different physics in a prespace
structure.

I will assume that the prespace computational system, constructed from noumenal
objects, is a parallel processor rather than a serial machine, since it certainly appears that
many things are happening at many places at the same time. Thus I will assume that there is
an individual processor in prespace associated with each discrete spatial point, and this
processor stores the values of the field magnitudes at that point, and calculates how those
values should be updated, most likely at certain discrete time intervals. Call this metaphysics \textit{noumenal process} metaphysics or \textit{prespace process} metaphysics.

I will assume that each individual processor is itself a serial machine, and that the
serial machine works by some simple set of rules. The modern computer utilizes what is
called von Neumann architecture, which uses a number of memory registers to store
information and to make calculations. For now I will take this as the model for the prespace
processors, and the resulting metaphysics could be labelled \textit{von Neumann noumenal process}
metaphysics. Thus I will assume that for each processor the fairly complicated laws of
physics are stored in memory states in the processor, and it updates the field magnitudes by
making use of these laws, the current field magnitudes and the field magnitudes and other
physical magnitudes of immediately neighbouring cells, which are communicated from the
processors corresponding to these cells.

It must be remembered that the idea of building a universe from a computer in our
space is meant to be an analogue for how our universe may be in prespace. In this case it may
seem that references to computing concepts such as ``memory register'' and ``calculation'' are
unacceptable, because to suppose that analogous states and processes occur in prespace would
be to attribute anthropomorphic and purposeful attributes to the universe. However, it should
be remembered that in a computer a memory register is ultimately some physical state, some
arrangement of physical parts, an arrangement of electrons in silicon, for example, and a
calculation is some dynamical process that rearranges those parts. It follows that we could
describe the operation of a computer in purely physical, non-computational terms. Thus when
the jump is made to prespace, I think that it is not too difficult to conceive that analogues of
these states and processes could be constituted in prespace, and there need be no implication
of anthropomorphism or of attributing purposefulness to the universe.

The prespace metaphysics could certainly deviate further from the analogy of a
computer. Rather than the laws of physics being stored in memory registers in prespace, it
could be that these laws are simply primitive properties of the states in prespace that
constitute the field magnitudes. The field magnitudes corresponding to each point in space
would be automatically updated according to these laws and the magnitudes of fields
corresponding to other points without the need for any ``calculation.'' According to this
model these laws would be primitive and also the properties that these laws express would be
primitive. In this case the analogy of a computer becomes less useful, and the analogy could
be dropped altogether.

\section{Functionalism }
According to Von Neumann noumenal process metaphysics, the memory registers of the
parallel processors in prespace not only store the field magnitudes for each cell, they also
store the laws of physics and take part in making calculations. One question to be faced is
how we could be sure that the memory register or registers storing the field values in prespace
would be picked out in some way and end up being constitutive of \textit{field values} at some \textit{position in space.} This seems to me to be one of the central challenges for noumenal process
metaphysics. I think that the answer must lie in the functional roles they enter into with the
memory states of the other processors; in other words, the causal roles. 

I will label the theory that matter is more fundamentally identical with parallel
processors in prespace as an instance of \textit{matter functionalism}.

According to some functionalist theories of \textit{mind}, all the intrinsic properties of mind,
such as its conscious experiences and the relation between its inputs and outputs, must be
explained by the functional roles of its parts, rather than being a basic property of the mind or
a basic property of the parts. ``Functional roles'' means something like the relations of input
and output of information between the internal parts, which in turn depend on the causal
relations between those parts. The parts will have to share certain properties in order to enter
into these relations, properties such as being able to switch between different states, (by being
able to contain a surfeit or deficiency of electrons for example), but other properties the parts
have will be irrelevant to the functioning of the mind.\footnote{I am restricting my attention here to those types of functionalism that refer only to the functional relations between states internal to the mind. There are also varieties of functionalism that consider relations between environmental stimuli and behaviour as part of the functional roles. Examples of the internal type of functionalism are the machine functionalism of Putnam and the homuncular functionalism of Lycan. See \citet{Lycan1990}.}

Thus if other entities could fill the same functional roles, having an isomorphic set of
functional relations, but differing in properties that do not contribute to these relations, then
these entities could make up the same mind instead. For example it might be supposed that a
computer made of silicon chips could support a network of electrical signals with the same
functional roles as the electrical and chemical signals in the neurons in the brain, so that the
same mind, holding the same thoughts, could be instantiated by both a brain and a computer.

According to the matter functionalism I am proposing it is likewise supposed that all
the intrinsic properties of matter in our space can be explained by the functional roles of those
things of which it is constituted, where those things need not be spatial parts, but could be
prespatial constituents instead. Again the properties of the spatial matter must not be
primitive or explained directly by the primitive properties of the parts. 

Similarly to mind functionalism, according to matter functionalism any structure that
could fill the appropriate functional roles would suffice to generate matter in space. While I
have described models for the prespace structure as computers in prespace, the assumption of
matter functionalism makes it clear that this structure need not literally be identified with
computers.

The Von Neumann noumenal process metaphysics fully satisfies the demands of
matter functionalism. In this model all the field values and the laws of physics are stored in
memory registers, and the calculations are performed using memory registers as well. The
memory registers may be constructed from some substance or substances, the detailed
properties of which are irrelevant to its operation---as before it only needs to be able to switch
between different states in response to causal interactions. It is merely the arrangement of
these states and their dynamic evolution, their functional roles, that gives rise to the matter in
space. 

We could suppose that there are merely two of these different states that make
up the memory registers---call them ``0'' and ``1''---and all information could be stored in
binary form. It would then be merely the functional roles of arrangements of these two states
that would give rise to matter in space. This calls to mind a proposition of \citet{Wheeler1989},
which he sums up in the phrase ``it from bit,'' in which he supposes that all matter is
ultimately constructed out of pure information in the form of bits. The difference between the
idea proposed here and Wheeler's is that here the bits are assumed to have a physical basis in
prespace. The metaphysics proposed here could be summed up with the phrase ``it from bit
from it.''

Now consider a single processor. I am supposing that this processor will correspond
to a single cell, a single point in space, and some register in this processor will be constitutive
of some field value at that point in space. (In fact the field magnitude may be constituted by
many registers, but I will assume that it is constituted by only one register for ease of
discussion.) I do not suppose that this register could constitute that field magnitude merely by
virtue of the functional roles that the register enters into with other registers of the same
processor, although this is certainly one aspect of its functional roles. I believe that just as
important will be the functional roles it enters into with the memory registers of processors
corresponding to other cells that will count in determining that this register constitutes a
particular field magnitude at a particular point in space.

For instance, the fact that a particular memory register will end up being constitutive
of a field value at a particular position in space will depend on functional roles such as how
the contents of that register change in response to changes in corresponding memory registers
in other processors. If it is designed so that a change in such a register in one processor will
almost immediately lead to a change in the corresponding register in another processor then
this will go part way to ensuring that these registers constitute field magnitudes at nearby
points in space.

\section{Position and causation }
I am assuming that like the other properties of matter, the position in space of some field
magnitude will be completely characterized by functional roles such as this---call this a
functionalist theory of position. These functional roles will be certain causal relations that
hold between these memory registers. Thus this functionalist theory of position will have
something in common with causal theories of space, of the kind discussed in the previous
chapter. These causal theories are normally thought of as trying to reduce spatial relations to
causal relations. I think that a theory of this kind is viable as long as a broad class of
 \textit{potential} causal relations are allowed in the reducing structure, as well as actual causal
relations. This is the case in a functionalist theory of space: I assume that functional roles are
broader than actual causal relations, including also whatever potential causal relations are
required. Thus in order that a register in a processor corresponds to a certain point in space it
does not necessarily have to be causally interacting with other processors continuously, as
long as there are ``channels'' of causal connections arranged such that certain causal
counterfactuals hold, such as ``if the processor corresponding to a neighbouring cell were to
change in a certain way, then that register in this processor would be almost immediately
causally affected.'' However it is manifested in prespace, the functionalist assumption
amounts to saying that the fact that a certain register in prespace corresponds to a certain field
magnitude at a particular position in space is fully determined by, or supervenes on, the actual
causal relations plus the potential causal relations between that register and other registers. 
The details of how the property of spatial position is manifested in terms of intrinsic
properties in prespace are immaterial as far as the behaviour of matter in space is concerned. 
This position value may be simply stored in a memory register in each processor, and be used
to determine the appropriate response of the registers of each processor to changes in other
processes, or it could be fully constituted by the manner in which that processor is ``wired up''
to other processors. 

I have not yet said much about what the functional roles must be like in order to have
a functionalist theory of position. While I do not fully concur with the aphorism ``a thing \textit{is}
where it \textit{acts},'' I think there is something right about this. To be more precise, I think that in
order for a thing in prespace to have a certain position it must have causal relations or
potential causal relations with things having positions at immediately neighbouring points in
space. This means that I believe it is incoherent to say that something has a certain position
yet there is no possibility that this thing could make its presence felt by having some effect on
an object that occupies its immediate neighbourhood, or some object that we could
counterfactually suppose to be brought into its immediate neighbourhood. I would simply
deny that the thing has the position in question.

I do not rule out the possibility of \textit{non-local} causal interactions. I only insist that an
object have \textit{some} local interactions or potential interactions with objects or potential objects
in the neighbourhood of the point it is supposed to be at. The object might in addition have
non-local interactions with objects at other points far removed.

I may be overstating the case here. It may be that in some possible worlds, in which
substantivalism is true, there could be an object that exists at a certain point in space which
has no causal interactions or potential causal interactions with other actual or potential objects
near that point. However, all I need to claim is that it is conceivable that our world is not one
of these worlds, in order to claim that a functionalist theory of space is possible for our world. 
And I think that it is highly plausible that our world is not one of these worlds.

The functionalist account of position will hold equally for the version of prespace
metaphysics that takes the laws of our three-dimensional world as primitive, although it might
be argued that not all properties of the objects in space will gain a fully functionalist
reduction, since the laws and the properties that they involve will be primitive properties of
the universe in this case. The field magnitudes and other variable quantities like spin and
perhaps charge may still be collections of binary prespace states, and the physical quantities
they represent and the positions they occupy will still be determined by their functional roles,
but if some quantities, particularly fundamental constants like the gravitational constant G,
are simply incorporated into the primitive laws of physics, then these quantities will not have
functionalist reductions in the way I am considering them.

It may be worth defining two kinds of functionalist theories, \textit{conceptual} \textit{functionalism}
and \textit{ontological} \textit{functionalism.} According to conceptual functionalism, the identity of
properties of things in the world can be reduced to the causal relations and potential causal
relations with other objects and their properties, whereby these causal relations are defined in
terms of those objects and properties themselves. Examples of conceptual functionalist
theories may be what are known as ``causal theories of properties,'' such as that of \citet{Shoemaker1984}. These theories may be necessarily true, if what it is to be a causal relation is defined
broadly enough. Ontological functionalism on the other hand, while it takes conceptual
functionalism as a starting point, postulates an underlying ontology of entities with
completely different intrinsic properties to the properties being reduced, and it is by these
underlying entities entering into these functional roles that the entities and properties being
reduced emerge from the underlying ontology. The truth of ontological functionalism is very
much a contingent matter, and it is this kind of ontological functionalism that I am addressing
in this chapter.

\section{Mechanism }
I will argue that mechanism can be regarded as an earlier functionalist theory of matter.

The attitude of the functionalist to mind, which aims to reduce mental properties to
the dynamic arrangements of entities with a few simple properties, is similar to the attitude of
the mechanists to matter, who wished to reduce all material properties to the dynamic
arrangements of point particles with a few simple properties, namely positions in space and
certain forces of attraction and repulsion acting between them. This commitment to the role
of dynamic arrangements is a common element that goes deeper than a mere common
commitment to reductionism. We could therefore label the mechanical view of matter, which
was also called the dynamical view, as a functionalist view of matter. 

Eventually the ambitions of the mechanists were thwarted, with the failure to explain
electromagnetic waves via simple forces acting in an underlying medium or system of
microscopic particles. Instead it was accepted that the electromagnetic field is a primitive
part of reality, possessing rather complicated properties at each point that would explain its
behaviour---these complicated properties could not be expressed other than by quoting
Maxwell's equations, or some equivalent. This tendency to give up mechanism and
reductionism, and adopt as primitive ever-more complicated field properties has continued
since then, despite the progress made with the unification of some fields. Take even our best
unified theory of a single fundamental force, the electroweak force, which is taken to be fully
irreducible, a simplified description of some of its general aspects (taking into account only
the first of three generations of fundamental particles) can be thought of as given by the
Lagrangian shown in figure \ref{Austern} on page \pageref{Austern} \citep{Austern1991}. This Lagrangian is not
particularly simple.

\begin{figure}
\begin{minipage}{5in}
\scriptsize
%
\def\\{\cr &}
\def\half{{\frac{1}{2}}} 
\def\fourth{{\frac{1}{4}}} 
\def\d{\partial}
\def\dot{\cdot}
\def\slash#1{\,/\!\!\!\!#1}
\def\dslash{/\!\!\!\partial}

\def\Wp{{W^+}}
\def\Wm{{W^-}}
\def\WpU#1{{W^{+#1}}}
\def\WpD#1{{W^+_#1}}
\def\WmU#1{{W^{-#1}}}
\def\WmD#1{{W^-_#1}}
\def\WPdotWM{{\left(\Wp\!\cdot\Wm\right)}}
\def\phiO{{\phi^0}}
\def\phiP{{\phi^+}}
\def\phiM{{\phi^-}}
\def\fChangedByDavid{{e \over s c}}
\def\etaP{\eta_+} \def\etaM{\eta_-}
\def\etaA{\eta_\gamma} \def\etaZ{\eta_z}
\def\etaPb{\bar\eta_+} \def\etaMb{\bar\eta_-}
\def\etaAb{\bar\eta_\gamma} \def\etaZb{\bar\eta_z}

\def\gamV{\gamma_5}
\def\PLeft{{1 - \gamV \over 2}}
\def\PRight{{1 + \gamV \over 2}}
\def\plusHC{{\, + \,\,\hbox{h.c.}}}



Conventions:
The Higgs boson is denoted $h$, and the three
unphysical Goldstone bosons are $\phiP$, $\phiM$, and $\phiO$.
The four unphysical Fadeev-Popov ghosts are $\etaP$, $\etaM$,
$\etaA$, and $\etaZ$, and their Hermitian conjugates are
denoted by $\bar\eta$.
$v^2 \equiv 2 \mu^2 / \lambda$, $s \equiv g' / \sqrt{g^2 + g'^2}$, and
$c \equiv g / \sqrt{g^2 + g'^2}$.
$e > 0$.


\begin{align*}
\mathcal{L} = {} &
%
 - \half \left(\d_\mu \WpD{\nu} - \d_\nu \WpD{\mu} \right)
    \left(\d^\mu \WmU{\nu} - \d^\nu \WmU{\mu} \right)
   - {1\over\bx i} \left(\d^\mu \WpD{\mu} \right) \left(\d^\mu \WmD{\mu} \right)
   + \fourth g^2 v^2 \Wp \dot \Wm     \\&
- \fourth \left(\d_\mu Z_\nu - \d_\nu Z_\mu \right)^2
   - {1 \over 2 \bx i} \left(\d^\mu Z_\mu \right)^2
   + {1 \over 8} \left({e \over s c} \right)^2 v^2 Z^2
- \fourth \left(\d_\mu A_\nu - \d_\nu A_\mu \right)^2
   - {1 \over 2 \bx i} \left(\d^\mu A_\mu \right)^2   \\&
%
%
+ \half \d^\mu h \d_\mu h  -  \half \lambda v^2 h^2
+ \half \d^\mu \phiO \d_\mu \phiO
    - {1 \over 8} \left(\fChangedByDavid\right)^2 \bx i v^2 \phiO^2
+ \d^\mu \phiP \d_\mu \phiM  -  \fourth g^2 \bx i v^2 \phiP \phiM    \\&
%
%
- \half g^2 \left[\WPdotWM^2 - (\Wp)^2 (\Wm)^2 \right]
- e^2 \left[A^2 \WPdotWM -
      (A \dot \Wp) (A \dot \Wm) \right]
\\&
- c^2 g^2 \left[Z^2 \WPdotWM -
        (Z \dot \Wp) (Z \dot \Wm) \right]
\\&
- e c g \left[2 (A \dot Z) \WPdotWM
       - (A \dot \Wp) (Z \dot \Wm)
       - (Z \dot \Wp) (A \dot \Wm) \right]
\\&
+ i e \left[ \d^\mu A^\nu \WmD{\mu} \WpD{\nu}
    + \d^\mu \WmU{\nu} \WpD{\mu} A_\nu
    + \d^\mu \WpU{\nu} A_\mu \WmD{\nu} \right] \plusHC
\\&
+ i c g \left[ \d^\mu Z^\nu \WmD{\mu} \WpD{\nu}
     + \d^\mu \WmU{\nu} \WpD{\mu} Z_\nu
     + \d^\mu \WpU{\nu} Z_\mu \WmD{\nu} \right] \plusHC
\\&
%
%
- {\lambda v \over 2} h^3
- {\lambda \over 8} h^4
- {\lambda v \over 2} h \phiO^2
- {\lambda \over 8} \phiO^4
- {\lambda \over 4} h^2 \phiO^2
- v \lambda h \phiP \phiM
- \half \lambda h^2 \phiP \phiM
- \half \lambda \phiO^2 \phiP \phiM    
- \half \lambda \left(\phiP\phiM\right)^2            \\&
%
%
+ \left(i e A^\mu \phiP \d_\mu \phiM \plusHC \right)
+ \left({i \over 2} {e \over s c} (1 - 2 s^2)
    Z^\mu \phiP \d_\mu \phiM \plusHC \right)
+ {e \over 2 s c} Z^\mu \left(\phiO \d_\mu h - h \d_\mu \phiO \right) \\&
+ \left[ {i \over 2} g \WpU{\mu} \left(h \d_\mu \phiM - \phiM \d_\mu h \right)
  + \half g \WpU{\mu} \left(\phiM \d_\mu \phiO - \phiO \d_\mu \phiM \right)
  \plusHC \right]                
+ \fourth \left(\fChangedByDavid \right)^2 \! v \, h Z^2
\\&
+ \half g^2 v h \Wp \dot \Wm
+ \left({e^2 \over s} {v \over 2} \phiM \Wp \dot A \plusHC \right)
- \left({e^2 \over c} {v \over 2} \phiM \Wp \dot Z \plusHC \right)
\\&
+ \fourth g^2 h^2 \Wp \dot \Wm
+ {1 \over 8} \left(\fChangedByDavid \right)^2 h^2 Z^2
+ \fourth g^2 \phiO^2 \Wp \dot \Wm
+ {1 \over 8} \left(\fChangedByDavid \right)^2 \phiO^2 Z^2
+ e^2 \phiP \phiM A^2
\\&
+ \half g^2 \phiP \phiM \Wp \dot \Wm
+ \fourth \left({e \over s c} \right)^2 (1 - 2 s^2)^2 \phiP \phiM Z^2
+ {e^2 \over s c} (1 - 2 s^2) \phiP \phiM A \dot Z
\\&
+ \left[ {e^2 \over 2 s} h \phiM \Wp \dot A
  - {e^2 \over 2 c} h \phiM \Wp \dot Z
  + i {e^2 \over 2 s} \phiO \phiM \Wp \dot A
  - i {e^2 \over 2 c} \phiO \phiM \Wp \dot Z
  \plusHC \right]    \\&
%
%
+ \d_\mu \etaPb \d^\mu \etaP - \fourth g^2 \bx i v^2 \etaPb \etaP
+ \d_\mu \etaMb \d^\mu \etaM - \fourth g^2 \bx i v^2 \etaMb \etaM
+ \d_\mu \etaZb \d^\mu \etaZ - \fourth \left(\fChangedByDavid \right)^2 \bx i v^2 \etaZb \etaZ
+ \d_\mu \etaAb \d^\mu \etaA    \\&
%
%
+ i e \left(\d_\mu \etaPb \etaP - \d_\mu \etaMb \etaM \right) A^\mu
+ e \left( i \d_\mu \etaAb \etaM \WpU{\mu} -
   i \d_\mu \etaPb \etaA \WpU{\mu} \plusHC \right)
\\&
+ i g c \left(\d_\mu \etaPb \etaP - \d_\mu \etaMb \etaM \right) Z^\mu
+ g c \left( i \d_\mu \etaZb \etaM \WpU{\mu} -
    i \d_\mu \etaPb \etaZ \WpU{\mu} \plusHC \right)        
%
%
- {g^2 \bx i v \over 4 } \left(h \etaPb \etaP + h \etaMb \etaM \right)
\\&
- \fourth\left(\fChangedByDavid\right)^2 \! \bx i v \, h \etaZb \etaZ
+ {i g^2 \bx i v \over 4 }
 \left(\phiO \etaMb \etaM - \phiO \etaPb \etaP \right)
- {e g \bx i v \over 2}
 \left(\phiP \etaPb \etaA + \phiM \etaMb \etaA \right)
\\&
- {e^2 \bx i v \over 4 s^2 c} (1 - 2 s^2)
 \left(\phiP \etaPb \etaZ + \phiM \etaMb \etaZ \right)
+ {e^2 \bx i v \over 4 s^2 c}
 \left(\phiP \etaZb \etaM + \phiM \etaZb \etaP \right)         \\&
%
%
+ \bar\nu \left(i \dslash \right) \nu 
+ \bar e \left(i \dslash - m_e \right) e
+ \bar u \left(i \dslash - m_u \right) u
+ \bar d \left(i \dslash - m_d \right) d
- {m_e \over v} \bar e e h
- {m_u \over v} \bar u u h
- {m_d \over v} \bar d d h            \\&
%
%
- i {m_e \over v} \bar e \gamV e \phiO
- i {m_d \over v} \bar d \gamV d \phiO
+ i {m_u \over v} \bar u \gamV u \phiO
- \sqrt{2} {m_e \over v} \left(\bar e \PLeft \nu \phiM \plusHC \right) \\&
- {\sqrt{2} \over v} \left[ \bar d
        \left({m_u + m_d \over 2} + {m_u - m_d \over 2} \gamV \right)
        u \phiM \plusHC \right]            
%
%
- {g \over \sqrt{2}} \left( \bar \nu \slash\Wp \PLeft e
           + \bar u \slash\Wp \PLeft d  \plusHC \right)
\\&
+ e \, \bar e \slash{A} e    
+ {1 \over 3} e \bar d \slash{A} d       
- {2 \over 3} e \bar u \slash{A} u    
- \fourth \fChangedByDavid \bar\nu \slash{Z} (1 - \gamV) \nu
+ \fourth \fChangedByDavid \bar e \slash Z \left( 1 - 4 s^2 - \gamV \right) e \\&
- \fourth \fChangedByDavid \bar u \slash Z \left( 1 - {8 \over 3} s^2 - \gamV \right) u
+ \fourth \fChangedByDavid \bar d \slash Z \left( 1 - {4 \over 3} s^2 - \gamV \right) d.
\end{align*}
\end{minipage}
\caption{``The Standard Model Lagrangian'' \citep{Austern1991}.}
\label{Austern}
\end{figure}

Noumenal process metaphysics has little difficulty reducing complex fields like this to
substances with a few simple properties in a fully functionalist way. In the von Neumann
model, whatever the laws are they are stored in memory registers in the same way. Thus this
metaphysics is able to provide functionalist accounts for all of those properties where
mechanism failed to do so, including of course position, which was taken as primitive in the
mechanist program.

The mechanist program was seen as very attractive by many for the promise it held
out for reducing all the properties of matter to a few simple substances with a few simple
properties. Noumenal process metaphysics holds out a similar promise, where mechanism
has failed, and this is one reason I think that this metaphysics is worthy of consideration.

\section{Monism }
We could imagine that the parallel processor in prespace could be constructed from a single
substance, just as we can conceive of building processors in ordinary space from a single
substance such as silicon. (Strictly speaking, semiconducting silicon is made from at least
two substances, since at least one substance other than silicon is required to ``dope'' the silicon
in appropriate places. However, it is a fair approximation to regard a doped area as consisting
of silicon with a different property in that area.) 

This idea of having only one substance as the basis for all things can be seen as a
natural tendency of matter functionalism, since the ideal of functionalism is to explain the
behaviour of a system purely by the functional roles of its constituents rather than by the
proliferation of substances. I will call the metaphysics that satisfies this ideal \textit{functionalist
monism}.

\section{Spinoza }
How does this functionalist monism compare with the monism of Spinoza? For Spinoza also
proposed that there exists in our universe only one substance, which he gave the name \textit{God}. 
This substance he said has an \textit{infinite} number of attributes, and so by these attributes Spinoza
could explain the complexity and diversity of the world of appearances. According to
functionalist monism, there is one substance with a \textit{small} number of attributes, but the
substance can be arranged in an infinite number of possible ways, and so by the dynamic
evolution of these arrangements it can explain the complexity and diversity of the world of
appearances. Functionalist monism could also be called \textit{combinatorial monism}.

Another difference from the monism of Spinoza is that according to Spinoza the
substance is static or timeless, whereas according to functionalist monism it is essentially
dynamic, a view of substance that is shared by Leibniz.

\section{Leibniz } \label{leibniz_dol-_dols}
It may be that in explaining the world, we must deviate from the ideal of functionalist
monism and introduce further substances and attributes, but it is hoped that we will not have
to go as far as Leibniz, who adopted an infinite number of substances each with an infinite
number of intrinsic attributes. This way of putting it makes it sound as though functionalist
monism differs widely from the metaphysics of Leibniz, but they actually have a good deal in
common. 

In fact the theory of cellular automata implemented on parallel processors in prespace
could be interpreted in such a way that one could regard it as a kind of \textit{Monadology} going
under the disguise of modern jargon. We could rename the processors that run each cell of
the cellular automaton as ``materialist spatial-point monads with local windows.'' Like
Leibniz's monads each processor does not exist in the space it generates and represents
internally the laws of physics. The main differences between the noumenal processors and
Leibniz's monads are that monads represent internally the whole universe rather than just the
laws of physics, and that monads are what Leibniz refers to as ``windowless,'' meaning
causally isolated from each other, whereas the automata processors are causally affected by
processors corresponding to neighbouring cells. Furthermore, I am assuming that the
``monads'' of cellular automata are material entities, whereas Leibniz considered monads to be
(at least partially) mental entities.

The functionalist monism can be seen as an addition to a materialist Leibnizian
metaphysics, explaining how the monads function, whereas Leibniz took them as primitive,
although he did say they have internal relations and referred to them in places as ``incorporeal
automata'' \citep[229, 254]{Leibniz1898}. 

Each cell may receive at every instant updated fields and potentials originating from
just about every object in the universe. Calculating how the state of the cell should be
updated, based on this information, could be difficult to carry out sufficiently quickly. One
way to improve the speed could be to provide within each processor an internal representation
of the states of neighbouring cells or neighbouring systems and their behaviour, then the
influence of those systems could be predicted approximately ahead of time and this
information used, along with the information received from other processes, to help update
the state of the cell. Furthermore, according to some theories of consciousness, to form a
representation of something is to have some consciousness of it, thus it could be argued each
processor would have some rudimentary consciousness of the atoms and cells it represents. If
this extension to the idea of cellular automata were adopted, then this computational system
would look even more like Leibniz's idea of monads, which he says represent distinctly
within themselves only those monads which are ``nearest or greatest'' \citet[250]{Leibniz1898}, and
they have some kind of primitive perceptions of those same monads.

If we were to represent the entire universe within each processor then we could drop
the connections between processors and have windowless pre-established harmony. 

I do think that Leibniz's system has its shortcomings. It would seem a huge waste of
computing power to reproduce virtually the same calculation for every cell. I will suppose
instead that nature operates so as to generate the structures of the universe with as little
``information processing'' as possible.

\section{Kant }
The idea of a materialist prespace metaphysics has most in common with the metaphysics of
``the things in themselves'' described by Kant, and I have borrowed his terminology in
describing the prespace metaphysics. The things in themselves are the noumena, and we
cannot know the intrinsic properties of these things \citep{Langton1994}. The noumena do not
exist in space. The objects that we are aware of are the phenomena, and these at least appear
to exist in space. The phenomena reflect relational properties of the noumena. Although
Kant's later, and best known, view of the nature of space is that space, and spatial relations,
are ideal rather than real, according to \citet{Friedman1992} and \citet{Langton1994}, at an earlier time
he held a causal theory of space, whereby spatial relations are reducible to causal relations
among the noumena. One place Kant sets forth this causal theory of space is in his \textit{Physical
Monadology}. According to \citet{Friedman1992}, Kant's \textit{Physical Monadology} was developed
taking Leibniz's monadology as a starting point, adjusting it to incorporate causal relations
among the monads. It would seem that this earlier version of Kant's metaphysical view on
the nature of space and of matter is closest to the prespace metaphysics described here. The
relationship between the various views of Kant on space and the views discussed in this thesis
is worthy of further investigation, but this is beyond the scope of this work.

\section{Holism }
In many physical systems, the behaviour of the whole system does not reduce to the ``sum'' of
the behaviour of the spatial parts of the system. This is particularly true in quantum
mechanics, which is thoroughly holistic in its standard formulation \citep{Teller1986}.

The behaviour of whole systems could still be calculated on parallel processors in
prespace, and in fact accepting a prespace metaphysics would seem to be the only way of
accepting holism and still having some kind of reductionism, since holism by definition
means non-reductionism to smaller things in space. This acceptance of holism within this
metaphysics would, I believe, represent a modification of the idea of cellular automata as it is
normally understood---under the standard idea, it should be possible for there to be just one
processor per cell in space, operating with purely local rules.

There is an approach to quantum field theory, known at Lattice Gauge theory, that has
some affinities with this modified cellular automata idea, incorporating holistic effects. In
this approach space-time is broken down into a lattice, just as in a cellular automaton, and the
behaviour of quantum fields defined on this lattice are studied. Lattice Gauge theory
incorporates the holism of quantum theory by utilising path-integrals over many paths, so that
the quantities derived from the calculations depend on the whole system modelled. (I think it
is unlikely that the path integral approach is the best way of modelling the way holism
actually operates in quantum systems, although I will not discuss this further in this chapter. 
In the following chapter I will be considering an approach to quantum holism via the
conception of cellular automata in configuration space rather than in position space.) While
this approach is generally taken to be a means of approximating continuous space-time field
theory, it could be considered as a model for a discrete space-time field theory. One
advantage of the lattice idea is that it avoids altogether the infinities that plague quantum field
theories in continuous space-time---this can be seen as another reason for believing that space-time is actually discrete. 

Holists such as Rupert \citet{Sheldrake1987, Sheldrake1988} have argued that holism is not confined
to quantum mechanics but exists at many levels of natural structures. Sheldrake points out
that the formation of stable shapes or forms is often difficult to explain in terms of the
properties of the parts. He introduces the notion of morphic fields which help guide the
formation of complex objects. One of the problems with this idea is that these fields
seemingly have to pop into existence whenever needed, acting across time and space, and at
other times not to be present at all. If one has an ontology based on space-time there does not
seem to be anywhere these fields can reside, and the idea seems implausible. Prespace
metaphysics has much less problem accommodating the existence of entities of this type, if it
turns out that they are needed.

\section{Relativity }
\citet{Toffoli1990} has shown that a cellular automaton in an essentially Newtonian Euclidean
space, one where absolute simultaneity is defined, can generate physics that is relativistically
correct---that is Lorentz invariant, at least approximately, and exactly in the limit of small cell
separation. Lorentz invariance cannot hold exactly in cellular automata because the cellular
distribution of field magnitudes results in a physics that will not be precisely invariant under
translations and rotations of the co-ordinate system. However, as long as the cells are
sufficiently small, this deviation from Lorentz invariance will not be observable.

Characteristic of relativity is that time slows down for moving systems. Toffoli shows
that this can arise fairly naturally in a cellular automaton, because in order to generate a
moving object, one must transport information to elsewhere in the cellular automaton. This
takes computing power, and it comes out fairly naturally that the internal processes of the
moving system will slow to compensate for this. Toffoli proposes a new principle broader
than the principle of relativity which he calls ``conservation of computing power.'' This idea
fits in well with the suggestions I made for nature acting so to economize on its ``calculations''
in \S\ref{leibniz_dol-_dols}, and we could see the observed Lorentz invariance and the finite speed of light as
evidence for finite nature's attempts to economize in its calculations.

\section{Computational heuristic principle } \label{comphp}
The universe may not be in reality in any way analogous to a computational system, but I will
adopt, as at least a heuristic principle, the requirement that our universe must be able to be
precisely simulated on a universal digital computer. And I will suppose that the laws of
nature are such that the number of computations required in order to bring about the world we
experience is minimized in some sense. \citet{Feynman1982} has proposed a specific form of this
heuristic principle, and I will adopt a similar heuristic principle in this thesis. Feynman
supposes that the universe must not only be able to be precisely simulated on a universal
digital computer, but proposes the following rule of simulation:

\begin{quote}
The rule of simulation that I would like to have is that the number of computer
elements required to simulate a large physical system is only to be proportional
to the space-time volume of the physical system\ldots{} If doubling the volume of
space and time means I'll need an \textit{exponentially} larger computer, I consider
this against the rules (I make up the rules, I'm allowed to do that). \citep[469]{Feynman1982}
\end{quote}

The heuristic principle I will adopt is that the number of computations required to
simulate a large system of $N$ particles, per unit time, should preferably be proportional to the
number of particles, and must not grow exponentially with $N$. Call this the computational
heuristic principle. This is very similar to Feynman's rule because the number of particles in
a volume of space is usually roughly proportional to the volume. Thus if the number of
computations required to simulate $N$ particles per unit time is proportional to $N$, then the
number of computations required to simulate a volume of space-time will be roughly
proportional to the size of that space-time volume.

Feynman points out that it immediately follows from this principle that physics must
be fully discrete. He also points out that even if physics were fully discrete, standard linear
quantum mechanics could not be simulated according to this principle. Thus he suggests that
the laws of standard linear quantum mechanics must be altered, but he does not suggest how. 
I will show in the next chapter how the laws of quantum mechanics could be altered to
accommodate a discrete physics, and that this modification, along with some natural
assumptions, leads to modified laws of quantum mechanics that satisfy Feynman's heuristic
principle. It turns out that this modification, which involves the ``collapse of the wave
function,'' automatically gives rise to the outline of a solution to a well known problem with
quantum mechanics, known as the ``measurement problem.'' This problem and its resolution,
and the consequences of a discrete quantum mechanics, are the subject of the next chapter.

This is in a sense the opposite strategy to that adopted by Penrose in his recent books
\citep{Penrose1990,Penrose1995}. In these books, Penrose argues that the abilities of the human mind
cannot be explained if the mind is a computational system. He therefore argues that there
must be some non-computational processes occurring in the brain, and he hopes that these
non-computational processes may occur when ``wave function collapses'' occur. 

In mathematics, ``non-computational'' problems are infinitely more difficult than
problems that merely grow exponentially more difficult with the number of parameters of the
problem: it can be proved that for these problems no solution can be found by any
computational system, no matter how large the system or how long the computation goes on. 
Penrose assumes that ordinary physical processes are computational. This is despite the fact
that he assumes that physics is continuous rather than discrete. (For continuous processes, it
is not straightforward to define what is meant for a physical process to be computational,
because the notion of a computational problem is strictly defined only for finite mathematical
problems. In order to define what is meant by ``computational'' for a continuous process, the
continuous process is approximated in some way by a discrete problem, and a limiting
process taken.) Penrose hopes that there is some non-computable process that occurs when
the wave function of quantum theory collapses. This process he assumes is not just a random
element that is introduced when wave functions collapse, but some special process that would
allow human brains to achieve understanding and feats that no computational system could
achieve.

I am not persuaded by Penrose's argument that the mind must be non-computational. 
Furthermore, I assume that the world must not only be strictly computable, but be able to be
exactly simulated in accordance with my variation of Feynman's heuristic principle. I will
show chapter\ref{qm_complexity} that adopting this heuristic principle leads to the outline of a solution to
the measurement problem of quantum mechanics. 
\vfill

\section{Weak dualism }
Now that we have seen how a materialist prespace metaphysics might work, I will turn again
to look at weak dualism. It seems to me that weak dualism is conceivable within the scheme
being considered. I have supposed that the prespace structure that underlies physical systems
could take the form of universal machines based on parallel processing. Since a universal
machine can run any computable process, it seems not too difficult to imagine that rather than
all of these processors being involved in generating the spatial world, some of these
processors could be involved in generating minds directly in prespace instead. The existence
of these minds would obviously be compatible with mind functionalism. The minds could be
totally disembodied, or perhaps be connected in some way to spatial brains; these minds may
be used to provide some higher-order processing for the spatial brains, which could be
required to bring about full consciousness or the laying down of memory, for example. 
According to \citet{Dennett1991}, consciousness is an approximately serial algorithm run on a
parallel processor. This may be so, but why not suppose that some of that parallel processing
could take place on prespace processors as well as the processors in the spatial brain? 

Even if there is nothing more to the mind than the brain, if there existed in prespace
some procedure for generating a copy of at least the functional aspects of spatial brains, then a
mind in prespace could form a receptacle for the continuation of consciousness after death.

I am not promoting any particular view here, rather I am suggesting a range of views
that seem to me to fit fairly naturally into the scheme of the prespace metaphysics I have
discussed.

\section{Theism }
Leibniz was a dualist---he supposed that there are monads corresponding to minds as well as
to physical objects. \citet[269]{Leibniz1898} supposed that some of these minds form what he
called a ``government of spirits'' which help administer the world, and that God is the absolute
Monarch over this government. God is also the creator of the whole universe. 

I will now consider some metaphysical models compatible with prespace metaphysics
where the universe we are aware of is ``created'' by some being or beings. It might be
suggested that our universe is literally generated on a computation system which was built by
some being or beings in some other universe. However, if by a computational system one is
thinking literally of a computer, something like the computers in our world, this idea seems
implausible. Not only does it raise questions of technological feasibility and proper
motivation, this idea does not take to heart the lessons of matter functionalism, which teaches
us that any entity that fills the appropriate functional roles would be sufficient; to    assume that
the prespace structure is literally a computer would be to assume without justification that the
prespace structure has laws of physics similar to the ``classical'' laws in the macroscopic
world we are aware of, and that creative beings are something like us.

The options that I would like to consider are those where the mind or minds that
``create'' the physical universe are themselves constructed directly from the same substance or
substances in prespace from which they construct the physical universe we are aware of. It
would follow that these minds are finite in their mental power. Since I am supposing that
their minds are constructed ``directly'' from the prespace structure, I am supposing these
minds are not carried by bodies that are in turn constructed from the prespace substance. 
However, if we are to assume that they could have created the universe it must be assumed
that they are somehow able to manipulate the arrangement of prespace matter separate from
the matter that makes up their minds. 

I have left it open whether there is only one creative being or many, and what role they
would play. (I will continue to use the pronoun ``they'' for the universe creating minds, it
being understood that there may be only one such mind.) Whatever the details of how they
operate, the idea I would like to retain is that the creative mind or minds are finite in mental
and physical power. This seems to follow from assuming that the minds of the gods are
constituted by arrangements in prespace---it follows from assuming a functionalist account of
mind holds even for the minds of the ``gods'' in a finite universe. 

How could these beings create the universe? One way they might proceed would be
to design the laws of our universe, design and construct ``processors'' that store and run these
laws for a cell or group of cells, then arrange for these processors to be copied over and over
in the prespace substance, forming the massively parallel processor that constitutes our
universe. 

This idea would help overcome one of the great difficulties with dualism. We can
understand how our brains might have evolved from the unstructured matter produced in the
Big Bang, but where would these non-spatial minds come from? One option is to suppose
that at least some, or one, of the minds existed in prespace all along, but during the early
history of the universe they were not connected to anything in the spatial world. Then these
minds could have actually taken part in creating our spatial world, then somehow ``linked up''
with the brains of advanced creatures such as human beings when they evolved, so that they
experience the sensations of the creatures having those brains. Just how this ``link up'' could
operate represents one of the bigger obstacles to dualism. It is difficult to imagine how these
minds that are embedded in prespace could arrange the constructed universe in such a way
that there is a causal interaction with some of the creatures in the constructed universe such
that these minds will experience themselves the sensations of these creatures. Furthermore,
assuming that the dualism is interactive in both directions, it must be arranged so that the
external minds can influence the behavior of those creatures. Both of these requirements
present formidable difficulties.

The question then arises why such minds would create such a universe as ours, if their
aim is to somehow interact with creatures with advanced nervous systems, because
presumably no advanced nervous systems evolved anywhere in our universe for the first few
billion years---why create a universe that is so ungratifying for so long? Note that as well as
setting up the laws of the universe, these minds would have to set up the initial conditions of
the universe. Assuming that these minds are finite in their speed of processing information,
then there would be a limit to the complexity of the initial conditions they could set up, as
well as limits on the complexity of the laws. This would mean that they must begin the
universe in some fairly uniform state, such as is presumed in the Big Bang model of the
beginning of the universe, and rely on the laws of the universe to eventually produce the
complexity of nervous systems that they wish to interact with. They would have no choice
but to wait for a long time, a long time in terms of the time measured in our universe at least. 

Thus we could see the apparent uniformity of the initial conditions of our universe as
evidence for the finite power of the creators of the universe we are aware of. In addition, the
idea that the god or gods who created the universe are finite would provide independent
motivation for the computational heuristic principle. Of course the origin of this god or gods
would be itself unexplained, as would the origin of the substance and structure of prespace
that constitute them and the universe, but there are unexplained elements in every
cosmological system.

While prespace metaphysics is certainly compatible with the existence of a creator or
creators of the visible universe, the description of them as ``gods'' has only limited validity, as
these minds are envisaged as being part of the prespace structure rather than being totally
external to the physical world. Similarly, the analogy of the prespace structure that these
minds are able to manipulate with ``processors'' which ``store'' and ``run'' the laws of nature
must also not be taken too literally: as I have emphasized, any structure that fills the
appropriate functional roles will do. As discussed in \S\ref{cellaut}2.3, in a computational system memory
registers are ultimately physical states, and a calculation is some dynamical process that
involves changes in these states. It follows that we could describe the operation of a
computer in non-computational terms, in terms of functional (causal) roles. The structure that
fills analogues of these functional roles in prespace may not resemble any computational
system we could even conceive of, and the ``laws'' of the behaviour of the prespace substances
will most likely be very different to the classical laws of macroscopic objects in our universe. 
Thus there need be no implication of anthropomorphism in using the computational analogy.
\vfill

\section{Conclusion }
I have laid out a range of possible ontologies, most having in common the idea that space is
not a fundamental container for all things. One could argue that all of these ontologies defy
Ockham's razor---they all add structure to the universe beyond that we now know about. One
might reply to this by saying that in return for adding extra structure these ontologies add a
degree of unity to the world, by reducing all of reality to one substance in the extreme case of
functionalist monism. 

It may be that there is no unity to be found, but it remains true that there may well be
extra structure to the world that we do not know about, and it would be as well to speculate
what that structure might be like, and this is my main argument. One is welcome to drop the
idea of reducing the universe to a single substance, or regard the automata metaphysics I have
presented as merely a metaphor for these different systems, or take a different approach
altogether. At bottom could be pure mind or pure God or some kind of mixture with matter,
and one could approach the whole range of ontologies from many different starting points.

Bertrand Russell once questioned why philosophers spend so much time studying the
metaphysical systems of previous philosophers rather than trying to understand the world
\citep[327]{Russell1978}. One reply he was given was that the systems of philosophers are more
interesting than the world! Russell was not persuaded by this reply, not surprisingly, and he
helped turn philosophy more towards the actual world as it is revealed by science. In so
turning, philosophy has lost much of the spirit of helping to speculate about those aspects of
the world \textit{not yet} revealed by science. I think that both philosophy and science have been the
poorer for this loss. Philosophy can do a real service to science and to individuals by drawing
out each possible metaphysics in as much detail as it is able, marshalling the evidence in
favour of each one. These metaphysical systems might form ``proto-sciences,'' helping to
stimulate theoretical sciences that fill out these systems in detail.

In the final analysis, deciding on the most likely metaphysics is a personal decision,
and in making this decision a person should take into account not just empirical science,
which mostly takes into account only those factors that are measurable with scientific
instruments, but also personal experience and the experiences of others, and his or her own
standards of what seems rational and plausible.

I urge that there is room in current physics for the exploration of possible prespaces of
many different types, including ones for which comparisons with cellular automata will be of
heuristic value. Recollections of the metaphysical reflections of Leibniz and Kant, too,
maybe of heuristic value in current physics; and even the comparisons of aspects of prespace
to ``gods'' might not be altogether without heuristic value. Indeed, there may even be clear and
literal truths which may be relatively easily extracted from such talk of ``creators'' in prespace,
provided the central morals of functionalism are kept clearly in mind.

I have mapped out a space for speculations; different people may choose different
regions of this space in which to invest their own energies. Even those who choose not to
spend time in the most speculative reaches of this space might profit from an occasional
reminder of their existence.

\chapter{Quantum Mechanics and Discrete Physics}
\label{qm_discrete_phys}

\section{Introduction }
Quantum mechanics is an empirically successful theory, but it contains at least one serious
flaw: two of the central postulates of the theory appear to contradict one another. According
to one postulate of quantum mechanics, the state of any quantum system, which can be
represented by a wave function (or state vector), evolves deterministically according to the
Schr\"{o}dinger equation. According to another postulate, the wave function of a quantum
system will change indeterministically when a ``measurement'' takes place, abruptly
``collapsing'' so as to give rise to a determinate measurement outcome. (In an ideal
measurement, the measurement outcome is an ``eigenvalue'' of the measured quantity and the
wave function collapses to a state corresponding to that eigenvalue known as the
``eigenstate.'') The problem is seen acutely when one tries to describe the measurement
process itself; that is, when one treats the measuring apparatus not as separate from the
measured system, but as part of a larger system, and describes the evolution of this larger
system by the Schr\"{o}dinger equation. This evolution does not give rise to a determinate
outcome for the measurement, but instead a superposition of different possible outcomes---a
superposition of macroscopically distinguishable states. These superpositions are never
observed, and the lack of a determinate measurement outcome contradicts the measurement
hypothesis, if the assumptions are made that the collapse of the wave function is an objective
occurrence, and that the measuring apparatus is able to cause such a collapse. This internal
tension within the theory is known as the ``measurement problem.''\footnote{A general account of this problem and some of the responses to it are given by \citet*{Albert1992}.} Closely connected with
this problem is the problem of how the wave function should be interpreted: the problem of
deciding on the relationship between the wave function and reality.

There are two types of response to the measurement problem. The first type of
response has it that the collapse of the wave function never occurs, and the wave function
always evolves linearly according to the Schr\"{o}dinger equation. Interpretations of quantum
mechanics that embrace this response include the ``many-worlds'' \citep{DeWittGraham1973}
and ``many minds'' \citep{Lockwood1996} interpretations of quantum mechanics. These
interpretations have the advantage of retaining the Schr\"{o}dinger equation evolution without
modification, but in return for this simplicity, they involve extra complexity in the wave
function, compared to those theories that embrace the collapse of the wave function. 
According to these interpretations, superpositions of macroscopically distinguishable states
do occur, leading to superpositions of many ``worlds'' or many ``minds'' or many ``decohering
histories'' \citep{Gell-MannHartle1989}, but we are only able to observe one of the many
components of these superpositions. The proponents of the ``decoherent histories'' approach
attempt to demonstrate that if there did exist these macroscopically distinguishable
components other than the one we observe then they could not be detected experimentally,
due to the complicated entanglement of any macroscopic system with its environment
\citep{Gell-MannHartle1989}. The Bohm interpretation \citep{Bohm1952}, also called the ``pilot
wave'' interpretation, also involves a rejection of wave function collapse. In this
interpretation of quantum mechanics, there exist point particles in addition to the wave
function and the wave function guides the trajectories of these particles. This interpretation
also involves many extra wave function components beyond those accepted in a collapse
interpretation. These extra wave function components correspond to collections of possible
particle trajectories that are not followed by the point particles. Other interpretations known
as ``modal'' interpretations are similar to the Bohm theory in that they reject the assumption of
standard quantum mechanics that a wave function must be an eigenstate of an observable in
order for that system to have a determinate value of that observable \citep{Clifton1996}. (This is
known as rejecting the eigenstate-eigenvalue link.) These ``modal'' interpretations also reject
the collapse of the wave function. As already mentioned, all of these no-collapse
interpretations involve extra complexity in the wave function, compared to those theories that
embrace the collapse of the wave function, due to the supposition that the wave function
forms superpositions of many macroscopically distinguishable states. (I will be seeking a
solution to the measurement problem that \textit{minimizes} the complexity wave functions, so from
this point of view I must reject these no-collapse interpretations.)

The second kind of response has it that the linear evolution of the wave function is not
universally applicable to all systems and processes. One way of denying the universal
validity of the Schr\"{o}dinger equation is along the lines of the ``Copenhagen interpretation'' of
Bohr, driving a wedge between the observed system and the measuring apparatus being used
by the observer by insisting that the apparatus must be described in terms of classical physics
and only the observed system in terms of quantum mechanics. This situation is rightly
viewed as unsatisfactory if we have the aim of discovering a theory that can be interpreted
realistically as a fundamental theory, allowing all of known reality to be described, including
``measurement'' processes. I assume that we are seeking a theory of this type. According to
some interpretations of the wave function, the wave function only represents a ``state of
knowledge,'' or only describes the behaviour of an ensemble of similar systems, rather than
the behaviour of a single physical system. I am seeking a solution to the measurement
problem that allows for a realist interpretation of the wave function as a real property of a
single physical system.

Another way of denying that the linear evolution of the wave function is universally
valid is to accept that the collapse of the wave function is an objective occurrence, and that
the linear evolution of the wave function is interrupted when these collapses occur. We may
interpret in this way the wave function collapses that occur according to the standard von
Neumann-Dirac quantum mechanics. However, in general no account is given in orthodox
quantum mechanics of what a ``measurement'' is, or how it brings about the collapse. This
situation is again unsatisfactory if we assume that measurements are physical interactions and
we seek a fundamental theory that can provide an account of these physical interactions. I
will assume that measurements are physical interactions, and that wave function collapses do
occur when physical systems interact with measuring instruments. (Thus I reject approaches
to the measurement problem that suppose that collapses do not occur until somehow brought
about by the consciousness of an observer.)

Accordingly, many have attempted to introduce modifications to the linear evolution
of the wave function, modifications that will eliminate the macroscopic superpositions that
are not observed, via the collapse of the wave function, but leaving the quantum mechanics of
microscopic systems largely unaltered, so that there will be no conflict with experimental
results. In this way it is hoped that references to measurements or observers can be
eliminated from the fundamental formulation of the theory, and a unified dynamics can be
provided for microscopic and macroscopic systems. It is these kinds of responses to the
measurement problem that will be considered in this chapter.

Two closely related theories of this type, involving modifications to linear quantum
mechanics that have the effect of bringing about the collapse of the wave function, have
recently been proposed with the intention of providing a solution to the measurement
problem. These are Quantum Mechanics with Spontaneous Localization \citep{GhirardiRiminiWeber1986, GhirardiRiminiWeber1988, Bell1987}, also known as the GRW theory, and the related
Continuous Spontaneous Localization theory \citep*{GhirardiPearleRimini1990}. In this
chapter, the outline of a new theory of wave function collapse is proposed, which I call the
``critical complexity'' theory of wave function collapse. I label the quantum mechanics
modified to accommodate this collapse law Critical Complexity Quantum Mechanics
(CCQM). This new theory has some features in common with the GRW theory and I will be
comparing CCQM with the GRW and CSL theories in this chapter. 

One criticism that has been levelled at GRW/CSL is that these theories are \textit{ad hoc}---little
motivation can be given for the modifications of quantum mechanics that the theories
involve, other than to produce a solution to the measurement problem. On the other hand, I
will show that CCQM not only provides a solution to the measurement problem but can be
shown to arise naturally if we consider what would follow if physics were fully discrete, that
is, what would follow if all physical quantities were discrete rather than continuous. While
the theory can be motivated this way, I suggest that a version of the theory can stand
independently of whether physics is continuous or discrete.

It has been suggested by \citet[598]{Leggett1984} that there may be ``corrections to linear
quantum mechanics which are functions, in some sense or other, of the \textit{degree of complexity}
of the physical system described.''  The theory presented in this chapter, Critical Complexity
Quantum Mechanics, can be seen to be taking up this suggestion of Leggett that corrections to
the deterministic evolution of the wave function depend on the complexity of the system. The
theory can be seen as proposing a new measure of complexity of wave functions, and
proposing a criterion of collapse based on this measure.  The full justification for taking this
measure to be a measure of complexity is given in chapter \ref{qm_complexity}.  Feynman's heuristic
criterion stemming from the possibilities of simulating physics on digital computers is also
discussed in chapter \ref{qm_complexity}. In the final chapter I make the proposal that the measure of
complexity of wave functions I propose should not only be seen as providing a measure of
complexity in quantum mechanics but also a new measure of quantum mechanical \textit{entropy}. 

I begin this chapter by giving a short account of the Ghirardi, Rimini and Weber
(GRW) theory, and the Continuous Spontaneous Localization (CSL) theory, and discuss some
implications of these theories for a realist interpretation of quantum mechanics. I then
provide an account of discrete physics as applied to quantum mechanics and set up the
motivation for the new theory. I then give some indication of what some details of the wave
function collapse model might look like, and compare the new theory further with the GRW
and CSL theories, including possible empirical tests for CCQM, and the transition from the
``quantum realm'' to the ``classical realm.''

I discuss the interpretation of the wave function in CCQM, comparing it with the
interpretation given to the wave function by the supporters of GRW/CSL. I show that the
interpretation of the wave function in CCQM I provide is compatible with the interpretation
suggested by \citet{Bell1990} for the wave function in GRW/CSL; namely, that the modulus
square of the wave function gives the ``density of stuff'' in configuration space.

Finally the implications of CCQM and GRW/CSL for relativity and nonlocality are
discussed, and possible directions for the generalization of CCQM to quantum field theory
are put forward.

\section{The GRW model } 
The general idea behind the theory of Ghirardi, Rimini and Weber (GRW) is that every
particle in the universe is subject, at random times, to approximate spatial localizations. The
effect of a localization is to cause the wave function representing this particle to collapse
instantaneously. Suppose that before collapse the particle is part of a system represented by
wave function of $N$ particles:

\begin{align}
\Psi (\bx_{1} ,\bx_{2} ,\ldots,\bx_{N}).
\end{align}

The effect on the wave function is given by multiplying the wave function by a single-particle
``jump factor'' $j(\bx -\bx_{k})$, where the argument $\bx_{k}$ is randomly chosen from the arguments of the wave function, representing the localizing particle, and $\bx$ is the position where the localization is centred. The wave function after collapse is given by

\begin{align}
\Psi_{\bx ,k} (\bx_{1} ,\bx_{2} ,\ldots,\bx_{N})={\frac{\Phi_{\bx ,k} (\bx_{1} , \bx_{2} ,\ldots\bx_{N})}{{\left\|\Phi_{\bx ,k} (\bx_{1} , \bx_{2} ,\ldots, \bx_{N}) \right\| }}}
\end{align}
where

\begin{align}
\Phi_{\bx ,k} (\bx_{1} ,\bx_{2} ,\ldots,\bx_{N}) &= j(\bx -\bx_{k})\Psi (\bx_{1} ,\bx_{2} ,\ldots,\bx_{N}).
\end{align}
We see here that the product of the wave function and the jump factor is renormalized at the
time of collapse by the division by its norm. GRW suggest that jump factor $j(\bx)$ is a
Gaussian:

\begin{align}
j(\bx) &= \left(\frac{\alpha}{\pi} \right)^{\frac{3}{4}} e^{-\alpha\bx ^{2}/2}.
\end{align}

where $\alpha$ is a new constant of nature. The effect of multiplying by the Gaussian is to
approximately localize the particle within a radius of $\frac{1}{\surd\alpha}$. The localization radius $\frac{1}{\surd\alpha}$ is
assumed to take a value of order $10^{-5}$ cm. (This value is chosen in order that the theory will
not conflict with experimental results.) The probability density of the Gaussian being centred
at point $\bx$ is taken to be

\begin{align}
\label{COL.PRB}
P_{k} (\bx) &= \left\|\Phi_{\bx ,k} (\bx_{1} ,\ldots,\bx_{N}) \right\|^{2} = \int d \bx_{1} \ldots{}d\bx_{N} |\Phi_{\bx ,k} (\bx_{1} ,\ldots,\bx_{N})|^{2} .
\end{align}

This ensures that the probability of collapse is greatest where the magnitude of the wave
function is greatest, in close agreement with the probabilistic predictions of standard quantum
mechanics. 

It is assumed that the localizations for each particle occur at random times, with mean
frequency $\lambda = 10^{-16}$ sec$^{-1}$, and that between each localization the wave function evolves
according to the Schr\"{o}dinger equation. It follows that a single particle will suffer a collapse
every $10^{9}$ years on average, and a macroscopic system of $10^{23}$ particles every $10^{-7}$ sec. Due to
certain problems with this theory, a modified version of the GRW theory has been proposed
in which the mean frequency of hittings is proportional to the mass of the particle: $\lambda_{m} =
\frac{\m}{\m_{0}}\lambda$, where $\m$ is the mass of the particle, and $\m_{0}$ is the nucleon mass.

The theory is supposed to solve the measurement problem as follows: the collapses
are so rare for microscopic systems containing few particles that their effects will not be
observable, whereas for macroscopic systems containing large numbers of particles, the
collapses will be common, and be of a type that will prevent macroscopic systems from
entering into superpositions of macroscopically separate locations.

Consider a situation of measurement, where an initially isolated system of a small
number ($L$) of particles interacts with a system of a large number ($M$) of particles, a
measuring instrument. Assume that there is a pointer on the measuring instrument that is also
initially isolated. (The assumption that the system and pointer are initially isolated systems is
clearly an idealisation.) As a result of the interaction, a pointer on the instrument may be
thrown into a superposition. The combined system of the measured system and the pointer
might then be represented by the superposition

\begin{align}
\phi_{1} (\bs_{1} , \ldots,\bs_{L})\chi_{1} (\br_{1} ,\ldots,\br_{M})+\phi_{2} (\bs_{1} , \ldots,\bs_{L})\chi_{2} (\br_{1} ,\ldots,\br_{M}) .
\end{align}
The states of the pointer $\chi_{1}$ and $\chi_{2}$ represent different positions of the pointer on the scale of
the measuring instrument, separated by a distance greater than $\frac{1}{\surd\alpha}$, and the states $\phi_{1}$ and $\phi_{2}$
represent the corresponding states of the microscopic system undergoing measurement. (Note
that if there are identical particles in the measured system and the pointer, then the
requirement of symmetry of the wave function under exchange of the identical particles
between the pointer and the measured system is not respected by this wave function as
written. To represent the standard quantum mechanics of this situation more faithfully, the
product wave functions would need to be replaced by symmetrized product wave functions.) 
The multiplication of the wave function by $j(\bx -\br_{i})$, for any of the many arguments $\br_{i}$ of the
pointer, will reduce to close to zero all but one of the terms in the superposition. Thus we see
that the superposition will very quickly be reduced and the pointer position will be thrown
into one or other of the macroscopically distinguishable states, at least approximately. 

One problem perceived with the GRW method of collapse is that when a collapse
takes place the existing symmetry of the wave function is destroyed for wave functions of
systems containing identical particles. The symmetry (or antisymmetry) of the wave function
with respect to the exchange of identical particles is a requirement of standard quantum
mechanics. The problem arises for GRW because when a collapse occurs one particle is
treated differently to the others---the localization is focused on one particle, with only some
effect on other particles represented by the wave function.

\section{The CSL model } 
In part due to the problems with the GRW model, attention has also focussed on the
Continuous Spontaneous Localization (CSL) models that were subsequently developed
\citep*{GhirardiPearleRimini1990}. I will not describe this theory in any detail, since the new
theory I introduce resembles the GRW theory more closely. In CSL, the Schr\"{o}dinger
equation is modified by the addition of a non-hermitian term:

\begin{align}
\label{CSL_dol-_dolEQ}
{\frac{d |\Psi_{w} (t) \rangle }{dt}} & =\left[ -{\frac{i}{\hbar}} H+ \sum_{i} A_{i} w_{i} (t) - \gamma \sum_{i} A_{i}^{2} \right]|\Psi_{w} (t) \rangle
\end{align}
where $H$ is the usual Hamiltonian, the $A_{i}$ are commuting operators and the $w_{i}(t)$ are white
noise functions, which satisfy the expectation values

\begin{align}
\LeftDoubleAngle w_{i} (t) \RightDoubleAngle & =0, & \LeftDoubleAngle w_{i} (t)w_{j} ({t'}) \RightDoubleAngle & =\gamma \delta_{ij} \delta (t-{t'}).
\end{align}

In order to maintain consistency with the predictions of standard quantum mechanics,
an assumption needs to be made analogous to the GRW assumption about the probability
distribution of hitting positions. The assumption made is that the probability density of the
stochastic processes $w_{i}$ is not given by the ``raw'' distribution implied by the expectation
values given for $w_{i}$, but is given by $P_{\cooked}[w_{i}]$ which depends nonlinearly on the state vector at time $t$ according to

\begin{align}
P_{\cooked} [w(t)] &=P_{\raw} [w(t)] \left\| |\Psi_{w} (t) \rangle \right\|^{2}
\end{align}
where

\begin{align}
P_{\raw} [w(t)] &=\frac{1}{N^{*}} \exp \left[ {-\frac{1}{2 \gamma } \sum_{i} \int_{0}^{t} d \tau w_{i}^{2} (\tau)} \right]
\end{align}
and $N^{*}$ is a normalization factor.

It is assumed that the state of the system at time $t$ is given by the normalized
state vector

\begin{align}
{\frac{|\Psi_{w} (t) \rangle }{\left\| |\Psi_{w} (t)\rangle \right\| .}}
\end{align}
The set of commuting operators $A_{i}$ must be chosen. The chosen operators actually have a
continuous distribution, and can be labelled by the continuous and discrete indices $(\bx , k)$. The
operators (written as $N^{(k)}(\bx))$ are given by

\begin{align}
N^{(k)} (\bx)&=\left(\frac{\alpha}{2\pi}\right)^{\frac{3}{2}} \sum_{s} \int d\bq a_{k}^{\dag}(\bq ,s)a_{k} (\bq ,s)e^{-{\frac{\alpha}{2}} (\bq -\bx)^{2}}
\end{align}
where $a_{k}(q, s)$ and $a_{k}^{\dag}(q, s)$ are the creation and annihilation operators of a particle of type $k$
(e.g., $k =$ electron, proton\ldots) at point $\bq$ with spin component $s$. The $N^{(k)}(\bx)$ are locally
averaged density operators. Their eigenvalues can be taken as representing the average
number of particles of type $k$ in a volume of order $\alpha^{-\frac{3}{2}}$ surrounding $\bx$. The parameter $\alpha$ is
assumed to take the same value as in the case of GRW, while $\gamma$ is related to the GRW
frequency $\lambda$ by $\gamma = \lambda(\frac{4\pi}{\alpha})^{-3/2}$. 

The theory does preserve the symmetry character of the wave function. It does not
suffer from the same problem as the GRW theory because the collapse process is associated
with the density of particles averaged over a certain volume rather than with an individual
particle. There have been formulated GRW-type discontinuous hitting theories that also
preserve the symmetry character of the wave function by associating the collapse with the
locally averaged particle density. These theories transform in an appropriate infinite
frequency limit into the CSL theory \citep*{GhirardiPearleRimini1990}. In fact it has been
shown that for any CSL dynamics there is a discontinuous hitting dynamics that transforms in
an appropriate infinite frequency limit into the continuous dynamics \citep{NicrosiniRimini1990}. 
From this point of view, Equation \ref{CSL_dol-_dolEQ} can be interpreted as describing a succession of tiny
spontaneous localizations which continuously strive to reduce the state vector into one of the
common eigenstates of the operators $N^{(k)}(\bx)$. 

It might be thought that the ``tiny hittings'' of the continuous reduction process would
reduce the wave function randomly, but in fact the statistical processes are coordinated so that
the outcome of the continuous process is similar to what would have been achieved by the
discontinuous process ``all in one hit.''

For the same reasons that the GRW theory was modified to make the frequency of
hitting proportional to the mass, a modification of CSL has been suggested \citep{GhirardiGrassiBenatti1995} in which the operators $N^{(k)}(\bx)$ are replaced by the mass density operators

\begin{align}
M (\bx) &= \sum_{k} m_{k} N^{(k)} (\bx)
\end{align}
where $m_{k}$ is the mass of the particles of type $k$. The evolution equation for the state vector is
altered to become

\begin{align}
{\frac{d |\Psi_{w} (t) \rangle }{dt}} &=\left[ -{\frac{i}{\hbar}} H+ \int d\bx M(\bx)w(\bx ,t) -{\frac{\gamma}{m_{0}^{2} }} \int d\bx M ^{2} (\bx) \right] |\Psi_{w} (t) \rangle
\end{align}
where $\m_{0}$ is the nucleon mass. The stochastic processes $\w$ now satisfy

\begin{align}
\LeftDoubleAngle w(\x,t) \RightDoubleAngle &=0,&  \LeftDoubleAngle w(\bx ,t)w(\bx' ,{t'}) \RightDoubleAngle &={\frac{\gamma}{m_{0}^{2} }} \delta (\bx -\bx')\delta (t-{t'}).
\end{align}

The quantum mechanical state can be represented in many mathematically equivalent
ways, the representation depending on which observables are used to define the basis states of
the representation. The GRW/CSL models give a privileged place to the position observable,
since the localizations occur in position space rather than momentum space or the spaces of
other observables.

\section{Realism and holism }
Many criticisms can be levelled at the GRW and CSL theories. The problems of these
theories will mainly be examined when a comparison with the theory I support is made later
in this chapter. I wish here to make some remarks about how the GRW and CSL theories
relate to the concepts of realism and holism.

Ghirardi, Rimini, Weber and Pearle believe that with the modifications of standard
quantum mechanics that they provide, it is possible to interpret the wave function as a real
property of a single physical system. Their theories are also supposed to allow for the
possibility that the quantum mechanical description of physical systems is complete. That is,
the GRW/CSL models do not rely on there being extra properties of quantum systems not
implicit in the wave function---GRW and Pearle claim to provide an interpretation which
allows that the wave function may give all the intrinsic properties of the system. It is also an
aim of this chapter to provide a solution to the measurement problem that allows for a realist
interpretation of the wave function as a complete representation of a single system.

\citet[319]{Ghirardi1992} says that the CSL theory ``attributes a definite wave function to
any system at any time.'' \citet[25]{GhirardiRiminiWeber1988} state that according to their
theory, as well as standard quantum mechanics, any isolated system having initially a definite
wave function maintains such a property for all times. This may be so, but it is also true that
there are no completely isolated systems in the universe. According to standard quantum
mechanics, as long as there is non-zero strength of interaction between particles they must be
represented by a single configuration space wave function, whose arguments are the positions
of all the particles. All particles, and systems of particles, have some strength of interaction
with other particles surrounding them: the universe as a whole is the only truly isolated
system. Thus it follows from standard quantum mechanics that there can only be a single
wave function for the whole universe, and there is nothing in GRW/CSL that alters this
situation. Thus I argue that it follows from GRW/CSL, unless standard quantum mechanics
is further altered, there can only exist a single wave function for the whole universe. I will
label this situation \textit{universal holism}. In later publications, in which the supporters of
GRW/CSL give careful philosophical interpretations of these theories \citep{Ghirardi1997,
GhirardiGrassiBenatti1995}, this consequence of GRW/CSL seems to be acknowledged, for they
refer only to the state vector of the entire universe.

This universal holism infects not just the GRW and CSL theories but every theory that
takes the wave function as being a real property of a system, and which does not modify
quantum mechanics in such a way as to enable weakly interacting systems (or any other
systems) to be represented by separate wave functions. The Bohm interpretation is another
such interpretation that is infected with universal holism. 

A consequence of this universal holism, plus the fact that the Gaussian ``jump factor''
multiplying the wave function is non-zero over the entire universe, is that when a collapse
occurs then, assuming it occurs instantaneously, the collapse affects instantaneously the state
of every particle in the universe, affecting in particular the position probability distribution of
every particle in the universe. I will give the name \textit{universal non-locality} to a situation where
there is some change of state that has an effect on, or is affected by, the state of every particle
in the universe, where the effects are felt more rapidly than influences travelling at the speed
of light would allow. This universal holism and non-locality also exist in the case of CSL,
the continuous collapse continuously affecting the state of every particle instantaneously. 
Furthermore, in the GRW case, as can be seen from equation \ref{COL.PRB}, because the Gaussians
extend over all space, the probability of a collapse being centred at a particular point in
configuration space at any one moment depends at that instant on the wave function
everywhere in configuration space. Thus there is a kind of universal non-locality that shows
itself not just when the collapse takes place, but in the evolution of the instantaneous
probability of the centre of collapse being located at each point in configuration space. 

This universal holism and non-locality do not necessarily result in the prediction
that superluminal signalling is possible across the breadth of the universe. In the case of
GRW and CSL, it is hoped that in relativistic versions of these theories, although such
instantaneous effects occur to the wave function, Lorentz invariance will be maintained at the
level of the statistical results of measurements. If this Lorentz invariance holds then this
would mean that these non-local effects could not be used to send signals faster than light.

It may be that universal holism and universal non-locality do exist in our world. 
However, one should be clear that their existence is what one is committed to if one accepts a
realist interpretation of the GRW/CSL theories, unless quantum mechanics is further
modified. I wish to pursue a theory of collapse that allows us to be realist about wave
functions that represent a finite number of particles, wave functions that each represent a
number of particles a great deal less than the number of particles in the universe. Having
wave functions representing a limited number of particles will not eliminate the holistic
nature of wave functions, nor will it eliminate the presence of non-local effects---these
phenomena will merely be limited in their extent. These situations could be called \textit{limited
holism} and \textit{limited non-locality}---these phenomena can only extend at a maximum over a
certain finite distance in space.

\citet[25]{GhirardiRiminiWeber1988} say that the localization processes in the GRW
model ``tend to disentangle quantum wave functions.'' However, it is my contention that these
localization processes, which are effected by multiplying the wave function by a factor,
cannot alter the number of particles represented in a single wave function. Thus if by
disentanglement it is meant the splitting of the wave function into separate wave functions,
then I claim GRW and CSL cannot disentangle wave functions. The most that could possibly
be achieved along the lines of disentanglement by the multiplication of the wave function by
such factors as those used in GRW/CSL would be for the wave function to become
approximately equal numerically to the product of two or more separate wave functions. 
Such disentanglement is only ever approximate, due in part to the fact that the Gaussian factor
never drops to zero magnitude. However, suppose hypothetically that a wave function of $N$
particles, defined in a $3N$ dimensional configuration space, became equal in magnitude to
the product of two wave functions:
\\vfill

\begin{align}
\psi_{12} (\bx_{1} ,\ldots,\bx_{N})&=\psi_{1} (\bx_{1} ,\ldots,\bx_{j}) \psi_{2} (\bx_{j+1} ,\ldots,\bx_{N}).
\end{align}
It does not follow from this that the system would then be given by two separate wave
functions. For the wave function of a system to become numerically equal to the product of
two separate wave functions is quite a different kind of evolution to the system ceasing to be
represented by a single wave function and instead becoming represented by two separate
wave functions. It should be remembered that what is sought is a realist interpretation of the
wave function, and under a realist interpretation a single wave function in a $3N$ dimensional
configuration space is quite a different entity to a pair of wave functions in lesser dimensional
configuration spaces, even if the former is equal to a product of the latter. It is true that the
product and the separate wave functions give rise to the same empirical predictions in
standard quantum mechanics, and non-interacting systems are routinely represented by
product wave functions in standard quantum mechanics. The fact remains, however, that it
would be very difficult to maintain a realist interpretation of the wave function and at the
same time regard two separate wave functions and the product of those wave functions as
representing the same state of affairs. I will assume that such wave functions represent
different states of affairs.

Let us now examine the situation of identical particles. If two sub-systems share some
identical particles, then according to standard quantum mechanics the wave function of the
overall system cannot become equal to a product of wave functions of the two sub-systems,
since this product wave function is not symmetrized. The closest the wave function can
approach such a product while preserving the symmetry requirement is for the wave function
to be equal to a proper linear combination of products of this type, so that the wave function
is antisymmetric under the exchange of identical fermions, and symmetric under the exchange
of identical bosons. This sum is not empirically equivalent to any single product, since the
symmetrized product contains quantum coherence information not contained in a simple
product, and this quantum coherence gives rise to observable effects in many circumstances. 

For example, a fully isolated system of two electrons that were not interacting could
be represented by the antisymmetric wave function:

\begin{align}
\Psi (\bx_{1} , \bx_{2})&={\frac{1}{\sqrt{2} }} \left[ \phi_{A} (\bx_{1})\phi_{B} (\bx_{2})-\phi_{B} (\bx_{1})\phi_{A} (\bx_{2}) \right].
\end{align}
This wave function might represent the state of two electrons in a singlet state, for example. 
(The wave functions referred to here are the complete wave functions, which can be written
as a product of the spin wave function and the spatial wave function. The $\bx_{\textrm{r}}$ are assumed to
include spatial co-ordinates and spin components. The overall wave function must be
antisymmetric, and in the case of the singlet state, the spin wave function is antisymmetric, so
the spatial wave function must be symmetric in that case.) The system represented by this
wave function cannot be interpreted as a system of two separate wave functions---the
combined wave function contains the quantum coherence information, and the existence of
this information is confirmed in experiments such as those conducted by Aspect \citep*{AspectDalibardRoger1982}.

As mentioned above, if two sub-systems share some identical particles, then according
to standard quantum mechanics, as it is normally presented, there must be a joint wave
function describing them and the wave function must be antisymmetric under the exchange of
identical fermions, and symmetric under the exchange of identical bosons. It is clear that if
there exists a wave function of $N$ particles, where $N$ is less than the number of particles in the
universe, then the wave function is not symmetrized under the exchange of particles within
the wave function with identical particles outside that wave function. For in this case we
have identical particles in separate wave functions, so that no such joint wave function exists. 
Furthermore, the state comprising these two wave functions will not in general be invariant
under the exchange of identical particles in different wave functions, because the exchanged
particles will in general have different probability distributions. We can either regard the
concept of symmetry or antisymmetry under the exchange of identical particles as being
inapplicable in the case where the particles are not represented in the same wave function, or
we can judge the symmetry properties under such exchanges by the construction of a product
wave function of the two wave functions concerned. This constructed product wave function
will not be symmetric or antisymmetric under the exchange of identical particles, in general,
due to the differing probability distributions of those particles.

Thus a choice must be made between realism about wave functions representing
systems smaller than the entire universe and the principle that all identical particles must be
represented in joint, symmetrized wave functions. Realism about finite-sized wave functions
is the option being explored here, so such symmetry must not be upheld under some
circumstances. Will this have observable empirical consequences? In fact the symmetry of a
joint wave function under the exchange of two identical particles has significant empirical
content only when the individual-particle wave functions \textit{overlap} significantly, in the sense
that there are regions in position space where both have appreciable magnitude \citep[570]{FrenchTaylor1978}. Thus as long as two separate wave functions do not overlap significantly
in this sense, or as long as they represent extremely weakly interacting particles of distinct
origin, then the separate wave functions are unlikely to be empirically distinguishable from a
joint, symmetrized wave function.

Many people, including \citet{RedheadTeller1991}, have argued that (fundamental)
identical particles are ``unindividuable'' entities, or simply ``non-individuals,'' so that two
hypothetical situations that differ only in the sense that the quantum properties of two
identical particles are exchanged cannot be distinct situations at all. However, the arguments
to this conclusion by Redhead and Teller rely on the fact that there appear to exist no non-symmetrized states of identical particles in nature. Thus their argument proceeds from the
lack of existence of non-symmetrized states to the ``non-individuality'' of identical particles,
rather than the other way around. It is true that the non-individuality is offered as an
explanation for the lack of existence of non-symmetrized states, so if one were inclined to
accept the explanation one would be disinclined to think that non-symmetrized states could
exist. Otherwise one would have to say that identical particles sometimes have identity, and
sometimes do not, which would be an odd position to adopt. It has been pointed out by many
other people, including \citet{Fraassen1991}, that it is also perfectly possible to regard
identical particles in symmetrized states as individuals, with holistic relations holding
between them, and the restriction to symmetric or antisymmetric states, at least under some
circumstances, being simply a primitive property of those particles' wave functions under
those circumstances. This is the view favoured here, but further discussion of this point is
beyond the scope of this thesis. I will be assuming that identical particles are individuals in
all circumstances, although in many circumstances they exist in symmetric or antisymmetric
states.\footnote{I believe that the requirement for antisymmetry under the exchange of two identical fermions is a natural one if their position probability distributions overlap, since the antisymmetry guarantees that the Pauli Exclusion Principle holds, and this principle is vital; for example, it ensures that not all electrons pile into the ground state of an atom. The requirement of antisymmetry also ensures that various conservation laws are maintained while at the same time respecting the symmetry of the situation. However, if the position probability distributions of the particles do not overlap, then the antisymmetry requirement seems superfluous.}

If we are to have wave functions of finite size, and we assume that the number of
particles represented in a wave function can vary with time, then there must be a kind of wave
function collapse mechanism that is able to bring about, under some circumstances, the
splitting of wave functions of systems of many particles into two or more separate wave
functions, each representing subsystems of the original system; and there must be another
mechanism that is able to bring about the combination of separate wave functions. Given that
a wave function is symmetric under the exchange of identical particles before the separation
takes place, and given the result that the exchange symmetry does not exist between identical
particles in the separate wave functions after the separation takes place, it follows that this
collapse mechanism must not preserve the existing exchange symmetry in general. It is put
forward as a criticism of the GRW theory that the modification to standard quantum
mechanics it involves does not preserve the symmetry of the wave function.\footnote{\citet{DoveSquires1995} have developed versions of the original GRW model that preserve the symmetry character of the wave function, but these involve modifying that model so that the wave function involved in the collapse represents every particle in the universe.} We have seen
that the GRW wave function collapse mechanism is inadequate for our purposes, because it is
unable to bring about the separation of wave functions. However, it seems that any
modification to standard quantum mechanics compatible with the existence of wave functions
of finite size must involve a collapse mechanism that does not preserve symmetry under the
exchange of identical particles, at least under some circumstances.

\section{The Critical Complexity theory of wave function collapse } 
I wish to look for a theory of wave function collapse that not only solves the measurement
problem but introduces modifications to quantum mechanics in such a way that the theory
admits wave functions of lesser extent than the wave function of the universe. Although
these wave functions may vary in the number of particles they represent with time, they must
always remain finite in size. From this point of view it is perhaps natural that the collapse of
the wave function be triggered when it reaches some critical size, or complexity, for some
measure of the complexity of the wave function. This takes up a suggestion by \citet{Leggett1984} that there may be ``corrections to linear quantum mechanics which are functions, in
some sense or other, of the \textit{degree of complexity} of the physical system described.'' Leggett
does not, however, suggest a measure of complexity of physical systems. The theory presented
in this chapter can be seen to be taking up this suggestion of Leggett by proposing a measure
of complexity of quantum mechanical systems and proposing how the corrections to the
deterministic evolution of the wave function depend on the complexity of the system.

The measure of complexity of a wave function I propose is given by its ``relative
volume'' in configuration space. The relative volume is defined by dividing configuration
space into cells of small volume. The relative volume is the number of these cells contained
within the boundary of the wave function in configuration space. The reason that the relative
volume should be regarded as a measure of complexity will emerge in the sections that
follow. This measure of complexity will emerge especially naturally for a discrete physics, in
which configuration space is in reality divided up into discrete cells. However, consider for
the moment that space and configuration space are continuous, as they are assumed to be in
standard quantum mechanics, and we wish to use its volume in configuration space as a
measure of its complexity. Configuration space has $3N$ dimensions for a system of $N$
particles so that the dimensionality of a ``volume'' in this space will vary with the number of
particles represented by the wave function. A quantity of consistent dimensionality is
required in order to be able to compare the complexity of wave functions. This requirement
can be met if configuration space is divided up into cells of small volume. This enables a
 \textit{dimensionless} quantity for each wave function to be defined; namely, the number of these
cells contained within the boundary of the wave function in configuration space. This is the
``relative volume'' of that wave function. Thus if we wish to use the volume of a wave
function in configuration space as a measure of complexity of a wave function, we are led to
divide configuration space up into discrete cells, whether or not configuration space is in
reality discrete. 

I will assume that the wave function of a quantum system will collapse when it
reaches a certain critical level of complexity. Thus I will assume that the wave function will
collapse when it grows to a certain \textit{critical relative volume} in configuration space. This is the
essence of the new theory of the collapse of the wave function, which I call the ``critical
complexity'' theory of collapse. Quantum mechanics altered by the introduction of this
critical complexity I label Critical Complexity Quantum Mechanics (CCQM).\footnote{I presented an early version of this theory at the ``Regional Workshop on the Theory of Quantum Measurement,'' Adelaide University, 1993. This paper \citep{Leckey1993} was printed in the conference proceedings (unpublished booklet). A version of the theory closer to the present one was presented in a paper ``Quantum Measurement, Complexity, and Discrete Physics'' at a conference in Vienna, ``Complexity and Constructivism in Physics and Mathematics,'' September 15-17, 1994.}

In defining the relative volume of a wave function, we need to define what is meant by
the boundary of the wave function. Consider for example a single particle, for which
configuration space is ordinary three-dimensional space, position space. In standard quantum
mechanics the wave function of a particle is taken to vary continuously over all position
space, and so never drops to zero magnitude---if this were strictly true, then the volume
occupied by the single-particle wave function would be the volume of the universe. This
problem for the case of a continuous wave function can be overcome by means of a small
reference magnitude $ \f_{0}$. The measure of the wave function volume is taken to be given by the
volume where the average wave function magnitude inside a single discrete cell volume is
greater than this magnitude (given that the whole wave function is normalized to unity).

We see that in defining the relative volume for a continuous wave function, we have
divided configuration space into \textit{discrete} cells, and disregarded that part of the wave function
having magnitude less than a certain \textit{discrete} magnitude. If it were the case that the wave
function were \textit{in reality} discrete, able to take on only a discrete range of values, and taking on
only one value per cell in a cellular configuration space, then we will see that this would
provide independent motivation for the theory of collapse presented here. We will now
examine discrete physics.
\vfill

\section{Discrete physics }
Discrete physics is characterized by the quantities representing the state of a system being
discrete valued and finite in number \citep{Feynman1982, Fredkin1990, Minsky1982}. I will
introduce the discrete physics in two stages; first of all a discreteness in space or space-time,
and then discreteness in all the magnitudes that characterize the state of the system.

\subsection{Space and time discrete}
Physical quantities representing the state of a system are taken to be defined only at points on
a lattice of space or space-time---whereby the points on the lattice are separated by some finite
distance. While this idea is currently used as a practical device in lattice quantum field
theories, allowing the divergences that appear in continuous quantum field theories to be
regulated in a natural way, some, such as T.D. Lee \citep{FriedbergLee1983} have taken this
kind of approach as providing a picture of a truly discrete physics, rather than being merely a
method of approximation. Experimental considerations suggest that this length is not greater
than $10^{-18}$m \citep{FriedbergLee1983}. Other theoretical indications that there may be a
fundamental unit of length come from string theory and other theories of quantum gravity,
which indicate that this length might be of the order of the Planck length, of order $10^{-35}$m. 
\citet{Forrest1995} has recently discussed some of the philosophical objections to a discrete space-time, arguing that the philosophical objections are not persuasive.

\subsection{All state quantities discrete valued}
In this approach all quantities are taken to be discrete, including field magnitudes. If it is
assumed that each magnitude can take on only a finite range of discrete values, then there is
only a finite amount of information within a given region of space. This is the approach taken
by those modelling physics using cellular automata \citep{Wolfram1986, ToffoliMargolus1987}. In a cellular automaton the state of a physical system is taken to be represented by a
certain finite number of discrete magnitudes, and these quantities are taken to be defined only
at points on a spatial lattice, whereby the points of the lattice are separated by some finite
distance. These magnitudes are updated at a certain discrete time interval according to some
rules, representing the fundamental laws of physics. This approach is normally taken as a
means of modelling continuous physical phenomena, but if it turned out that all state-variables currently taken to be continuous were in fact discrete, then an approach like this
could give a more realistic picture of reality than an approach based on continuous quantities.

\subsection{Discrete physics heuristic}
I will adopt in this chapter the heuristic principle: consider how a discrete physics could
model physical laws, then only accept as possible laws of nature those physical laws that
could be successfully modelled this way. The heuristic principle has clear justification if
physics is in reality discrete, but it can also be used as a means of choosing laws that apply to
continuous physics, since there are usually laws of continuous physics that correspond closely
in form to the laws expressed in terms of discrete quantities. I claim that it is an advantage of
the critical complexity theory that it seems to follow naturally merely by considering how
quantum mechanics could be made fully discrete. Accordingly I will define a discrete-valued
wave function corresponding to the continuous-valued one. However, I propose that a version
of the theory can stand independently of whether nature is in fact continuous or discrete. In
the case where the wave function is taken to be continuous, the discrete wave function can be
seen as having merely heuristic value. In that case the magnitude $ \f_{0}$ and the lengths defining
the cell size in configuration space are taken as ``purely reference'' magnitudes---magnitudes
in nature that play a role in defining a finite, dimensionless volume for a continuous wave
function. In the case of a physics that is in reality discrete, I take it that $ \f_{0}$ and the reference
lengths take on further significance.

Cellular automata are usually modelled on digital computers and their study leads to
comparisons being made between physical systems and computational systems. (See
\citealt{Wolfram1986}.) I will make use of such a comparison at some points to provide independent
motivation for the measure of complexity I propose, although the theory I propose can stand
independently of any of the particular motivations provided for it here.

\section{Discrete physics and wave function collapse }
In this section I will describe how quantum wave functions could be represented in terms of a
discrete physics and how the constraints of a discrete physics lead naturally to a new theory of
wave function collapse and a new solution to the measurement problem. Just as in the case of
GRW/CSL, I will assume that the position representation is privileged, and the discretization
of the wave function will be carried out in this representation. I will first discuss single
particle wave functions, introducing the discrete representation and defining the relative
volume in terms of it. I then discuss how the idea of a collapse of the wave function arises for
these systems. I will then discuss many-particle systems and how the collapse of the wave
function for these systems leads to a solution to the measurement problem. I finally discuss a
typical measurement process---the detection of a particle in a bubble chamber.

\subsection{Single particle systems---the discrete representation}
I will begin with single-particle systems. I suppose that particles interacting very weakly with
their environment---``free'' particles---will sometimes be described by single-particle wave
functions. 

Consider the wave function for a single free-particle $\psi(\bx)$, neglecting spin. In
standard quantum mechanics, the wave function is continuous in magnitude and defined at all
points in a continuous space. Write the wave function as a product of a magnitude and a
phase factor as follows:
\begin{align}
\psi (\bx)&=f(\bx)e^{i\theta (\bx)} .
\end{align}
The probability of finding the particle on measurement within a cubic region of volume $d\bx $
around point $\bx$ is $|\psi(\bx)|^{2}d\bx = \f(\bx)^{2}d\bx $. The real-valued function $\f$ is the magnitude of the wave
function. The wave function is assumed to be normalized to unity:

\begin{align}
\label{normeqn}
\int d\bx \left | \psi (\bx) \right |^{2} &=1 .
\end{align}

Now consider dividing the space the wave function occupies into cubic cells of side $\a$,
volume $\a^{3}$. These cells may be constant in volume or vary with the energy of the particle. 
Consider a spatially-discrete wave function that has a single value for each cell. If we begin
with a continuous wave function, this value would be the average value of the continuous
wave function in that cell (which is equal to the volume integral over that cell of the
continuous wave function, divided by the volume of the cell). For the spatially discrete wave
function, the probability of finding the particle in a particular cell of volume $\a^{3}$ at $\bx$ is given by
$|\psi(\bx)|^{2}\a^{3} = \f(\bx)^{2}\a^{3}$.\footnote{Note that the discrete wave function, as 
well as the continuous wave function, will have units of the square root of a density [$\textrm{L}^{-\frac{3}{2}}$], 
because the modulus of the wave function squared multiplied by the volume of the cell is equal to a probability, 
which is dimensionless. Alternatively, a discrete wave function magnitude for a single cell could be derived from 
the continuous wave function by the volume integral of the wave function over the cell. In this case the discrete 
wave function would be dimensionless, and the probability of finding the particle at $\bx$ would be given by $|\psi(\bx)|^{2}$. 
I prefer, however, to retain the probability density interpretation of the modulus square of the wave function, adapted to take into account the discreteness in space.}

Now consider a wave function that has discrete wave function values as well as being
spatially discrete. Take $\f$ and $\theta$ as discrete:

\begin{align}
f(\bx)& = n_{f} (\bx)f_{0} & \theta (\bx)&=n_{\theta} (\bx)\theta_{0} 
\end{align}
where $n_{f}$ and $\n_{\theta}$ are natural numbers greater than or equal to zero. $ \f_{0}$ is the \textit{base magnitude} of
the discrete wave function and $\theta_{0}$ is the \textit{base phase angle}. I will suppose that when the
magnitude of the corresponding continuous-valued (but spatially discrete) wave function for a
single cell, $|\psi|$, falls below $ \f_{0}$, the discrete wave function has magnitude 0. In general, the
discrete-valued wave function has value $nf_{0}$ when the continuous-valued wave function has
values between $nf_{0}$ and $(\n+1)\f_{0}$.

In general the particle will also have an intrinsic angular momentum or spin. For
fundamental particles of spin half, and for photons as well, the intrinsic spin can be
represented by a two component complex vector, called a spinor:

\begin{align}
\left(\begin{array}{c}U_{1} \\
U_{2} \end{array} \right).
\end{align}
The spinor can be represented by two complex numbers in some arbitrary representation. 
Suppose the representation is the $\z$-representation, where $\z$ signifies some direction in space. 
Then $U_{1}^{2}$ gives the probability that the particle will be detected ``spin up'' in the $\z$-direction
when a measurement is made, and $U_{2}^{2}$ gives the probability that the particle will be detected
``spin down'' in the $\z$-direction when a measurement is made. The state of the particle is given
by the spatial wave function multiplied by the spinor, so the full discrete wave function of a
single particle can be represented by two complex values at each cell in space.

Define the \textit{relative volume} $v$ of a single free particle system by:

\begin{align}
v ={{\frac{\int _{V} dxdydz}{\int _{0}^{a} \int _{0}^{a} \int _{0}^{a} dxdydz}}}
={\frac{V}{a^{3}}}
\end{align}
In terms of the fully discrete wave function, $\V$ is the volume of space for which the wave
function is non-zero, and the relative volume is just the number of cells for which the wave
function is non-zero. The relative volume is a dimensionless quantity. In the case of the
wave function that is continuous in space and in magnitude, the volume $\V$ is defined as the
volume where the wave function magnitude averaged over a single cell of volume $\a^{3}$ is
greater than $\f_{0}$, or $|\psi|>\f_{0}$, assuming the wave function magnitude is constant for that cell. 
The relative volume is the number of cells that satisfy this condition. 

I have noted that the cell size may be constant or may vary with the particle and its
energy. In standard quantum mechanics the wave function is continuous, and no empirical
deviation has ever been observed from the predictions derivable from a spatially continuous
wave function. Thus for the discrete wave function to maintain empirical adequacy it must
preserve the form of the continuous wave function to within some degree of approximation. 
In order to achieve this it is natural to suppose that the size of the cells should be related to
the rate of change of the wave function with distance. In general the magnitude of the wave
function varies more slowly than the phase with distance, and the rate of change of phase with
distance is characterized by the mean de Broglie wavelength of the particle concerned, so it is
a natural choice to have the length of the sides of the cell, $\a$, proportional to the mean de
Broglie wavelength of the particle. (The mean de Broglie wavelength is inversely
proportional to the mean momentum of the particle.) Thus I will suppose that the length of
the cell sides is proportional to the mean de Broglie wavelength. (Here the assumption has
been made that the cell volume for a single wave function is independent of its position in
space, but this assumption could be removed if this were theoretically desirable.)

I am supposing that the cell size is smaller for particles of greater mass and energy. 
An alternative way for a discrete wave function to preserve the form of the continuous wave
function to within some degree of approximation would be for cell size to be constant but
extremely small; so small that the form of the wave function would be closely preserved even
for processes of extremely high energy (such as those that occurred during the first moments
of the Big Bang). If the cell size were not sufficiently small, then CCQM would lead to
predictions significantly different from the predictions of standard quantum mechanics,
deviations that have not being observed. Thus we are faced with two versions of CCQM, one
with constant cell size and another with cell size dependent on the energy. I favour the latter
version, although the choice is not crucial in what follows, until we consider the connection
with the concept of entropy in \S\ref{entropy}. I will discuss the issue of the cell size further in that
section.

It was noted earlier that the wave function of a fundamental particle will be a two
component vector. I will label as the \textit{state complexity} the number of non-zero complex
numbers used to represent the discrete wave function in the position representation. In this
single particle case the state complexity will be twice the relative volume. I have suggested
that the relative volume is a measure of complexity. If we wished to maintain that the
complexity is the number of non-zero complex numbers used to represent the state of the
system, then strictly speaking the relative volume is only a measure of complexity for
particles without spin, and the actual complexity is double this for particles with spin. 
However, whether the relative volume or the state complexity is used as the measure of
complexity of CCQM, it will make little difference to the pattern of wave function collapses.
\vfill

\subsection{Single-particle wave functions and wave function collapse}
I make the assumption that a wave function will collapse when it reaches some critical
relative volume. Thus I will suppose that there exists a \textit{critical relative volume} $v_{c}$ for any
free-particle wave function and that a wave function that reaches this volume will collapse to
some fraction of the original volume. 

The existence of this critical volume may be clearly motivated in the case where the
wave function is in reality discrete valued. Consider for example the emission from an atom
of a single particle, which could be a particle with mass or one without mass, such as a
photon. Suppose that the particle begins as a spherical wave function spreading out from the
atom. Suppose also for the moment that the particle can be regarded as remaining free of
interaction with any other particles. The spherically spreading wave function will soon
become extremely attenuated in magnitude as it spreads from the source (such as a distant
star), and if we assume that the wave function can take on only a discrete range of
magnitudes, then if the wave function did not collapse the magnitude of the wave function
would eventually fall everywhere below $ \f_{0}$, the smallest magnitude that can be represented. 
The wave function would then disappear altogether! The problem of how to represent
spherical propagating waves in a discrete physics is discussed by \citet[537]{Minsky1982}, and he
leaves it as an unsolved problem. 

One way to avoid this problem would be to suppose that the base magnitude is so
small that the problem would never arise---the maximum magnitude of the wave function
would never approach $ \f_{0}$, even if the wave function covered the whole universe. This solution
could only be successful if the universe were finite in size. Furthermore, if the base
magnitude were that small then it is unlikely that the form of the laws of discrete physics
would need to deviate in any significant way from those of continuous physics, yet the point
of applying the discrete physics heuristic is to see what modifications to the laws of physics
might arise from a discrete physics, other than simply the discreteness itself. Thus I will add
to the heuristic principle that it is disallowed to choose the discrete magnitudes arbitrarily
closely spaced so that no conflict with continuous physics arises. In fact I will suppose it is a
desirable feature of a discrete physics to have the separation between adjacent discrete
quantities as large as possible while still retaining a physics that is empirically adequate. 

I suggest that a natural solution to the problem of spherically propagating waves in
discrete quantum mechanics is for the wave function to ``collapse'' to some extent when it
reaches some critical volume, before the wave function becomes too weak, instantaneously
reducing in volume by some fraction. We have supposed that the corresponding continuous
wave function will remain normalized to unity at all times, so that when the collapse occurs
the average magnitude will rise. The wave function would then resume spreading until it
again reached the critical volume, at which time it would collapse again. And so on. It seems
to me that this is a natural solution to the problem, so it seems that some degree of
spontaneous localization of free waves is a natural consequence of a discrete physics. This is
the first important result to arise from the discrete physics heuristic. 

To see how this solution could work, consider the original wave function in more
detail. Suppose the particle were a photon emitted from a star. Given that the photon is in
the visible spectrum, emitted by a state of a typical lifetime of $10^{-8}$s, its wave function will
have the form of a spherical shell, the magnitude of the wave function falling off
exponentially, with a characteristic width of approximately 3 metres \citep[158]{Messiah1961}. 
The discrete wave function magnitude will be finite within this shell, but rapidly fall to zero
outside this region. The hypothesis is that when the volume of the shell reaches a critical
volume it collapses, so that much of the shell is eliminated, leaving a single segment of the
shell almost unchanged, while eliminating the rest of the wave function. This segment of the
shell would then continue to propagate, expanding as it recedes from the source, until its
volume again reaches the critical volume and another collapse occurs. I will make two
assumptions about the form of collapse for a single-particle system. Firstly, I assume that the
collapse occurs in such a way as to as closely as possible preserve the statistics of the original
wave function. Secondly, that it collapses to a single region rather than two or more regions,
in order to ensure that the particle number is conserved. One suggestion on how this could be
achieved is made in \S\ref{singleformc}.

For this solution to the problem to be viable, it must be the case that the collapse does
not cause deviations from quantum mechanics that are in conflict with experiments that have
already been carried out. Consider a single-particle system, a ``free'' particle. Actual single-particle wave functions may be rare, but we can use them to put some constraints on the size
of the critical relative volume. The critical volume $v_{c}$ must be large in size. We know, from
interference experiments that have been conducted, that a wave function of a free particle
spreading from a coherent source can spread out over large volumes in position space before
two segments of that wave function are combined again at the detector, producing an
interference pattern. If the critical volume $v_{c}$ is not very large then such wave functions
would collapse before being combined, eliminating one of the segments of the wave function,
which would destroy the interference, leading to conflict with experimental results. 

The volume that the wave function reduces to after collapse must also be large. If the
wave function when it collapsed did reduce in size by a very large degree then this would
have observable consequences. For example, a collapse of a photon wave function to a small
volume would cause a large spreading in the distribution of its momentum components at the
point the collapse occurs, in accordance with Heisenberg's uncertainty principle, so if a
photon from a distant star were to collapse to a small volume before it reached us then this
would affect its spectrum of frequencies in an observable way. (See \S\ref{singcol}.)

It should be noted that the assumption is being made that the limitation on the relative
volume will apply to photons as well as particles with mass. There are certain difficulties
associated with assigning a wave function to a photon, but it has recently been demonstrated
that this can be done as long as the wave function is interpreted slightly differently than is
usual in elementary quantum mechanics. Two ways of defining a photon wave function are
given by \citet{Bialynicki-Birula1994} and \citet{Sipe1995}. According to these definitions, the
probability interpretation and normalisation condition for the photon wave function differ
slightly from those given here, but not in a way that significantly affects the arguments of this
chapter. The full treatment of photons is beyond the scope of CCQM: since photons are
relativistic particles, they must be treated within an extension of CCQM to relativistic
quantum theory.

Note that the type of collapse I am proposing for a free-particle system is not itself
supposed to comprise a solution to the measurement problem. I am not here discussing
localizations to small regions, such as spots forming on photographic plates, that take place
when ``measurements'' occur. These kinds of localizations will emerge for many-particle
systems, which will be discussed in the next section. 

\subsection{Many-particle systems---the discrete representation}
A many-particle system of $N$ particles can be described by a wave function in configuration
space. Consider the wave function for a system of $N$ interacting particles, neglecting spin. In
standard quantum mechanics, the wave function is continuous in magnitude and defined at all
points in configuration space. Write the wave function as a product of a magnitude and phase
factor as follows:

\begin{align}
\psi (\bx_{1} ,\ldots, \bx_{N})&=f(\bx_{1} ,\ldots, \bx_{N})e^{i\theta (\bx_{1} ,\ldots, \bx_{N})}.
\end{align}
The real-valued function \f{} is the magnitude of the wave function. The wave function is
assumed to be normalized to unity:

\begin{align}
\int | \psi (\bx_{1} ,\ldots, \bx_{N})|^{2} d\bx_{1} \ldots{} d \bx_{N} &=1.
\end{align}

Analogously to the one-particle case, consider dividing configuration space into cells
of small finite size. Consider dividing configuration space into cells of volume $\a_{1}^{3}\ldots{}a_{N}^{3}$,
where $a_{i}$ is the length characteristic of the $i$th particle represented in the wave function. Here
the possibility has been left open that the length of the sides of the cell in configuration space
may depend on the mass or energy of the particle they correspond to. If the length turns out
to be independent of the particle and its energy then the volume of a single cell will be $\a^{3N}$. 
Similarly to the single particle case, I favor the option that $a_{k}$, the length of the three sides of
the cell corresponding to a single particle $k$, is proportional to the mean de Broglie
wavelength of that particle. This is because the mean de Broglie wavelength of that particle
roughly characterizes the rate of change of the wave function with distance in those three
directions in configuration space. Further discussion of this point will arise later in this
chapter. Note, however, that the version of CCQM in which the cell size varies with the
energy of the particles is favored by the modified version of the discrete physics heuristic,
according to which it is a desirable feature of a discrete physics to have the separation
between adjacent discrete quantities as large as possible while still retaining a physics that is
empirically adequate. (As in the one particle case the assumption has been made that the cell
volume is independent of its position in configuration space, but this assumption could be
removed if this were found to be theoretically desirable.)

We can define a spatially discrete wave function that has one value per cell in
configuration space: if we begin with a continuous wave function, this value will be the
average value of the continuous wave function for that cell. Now define a wave function that
has discrete wave function values as well as being spatially discrete. Again take \f{} and $\theta$ as
discrete:

\begin{align}
f(\bx_{1} , \ldots, \bx_{N}) &= n_{f} (\bx_{1} ,\ldots, \bx_{N})f_{0}& \theta (\bx_{1} ,\ldots{}\bx_{N})&=n_{\theta} (\bx_{1} ,\ldots{}\bx_{N})\theta_{0} 
\end{align}
where as before $n_{f}$ and $\n_{\theta}$ are natural numbers greater than or equal to zero, $ \f_{0}$ is the \textit{base
magnitude} of the discrete wave function and $\theta_{0}$ is the \textit{base phase angle}.

Define the relative volume of the wave function as the $3N$-dimensional ``volume'' of
the wave function \textit{in configuration space} divided by the corresponding volume of a single cell
in configuration space:

\begin{align}
v={{\frac{\int _{V} d^{3} x_{1} \ldots{} d^{3} x_{N} }{\int _{0}^{a_{1}} \ldots{} \int _{0}^{a_{N}} d^{3} x_{1} \ldots{} d^{3} x_{N} }} }={\frac{V}{a_{1}^{3} \ldots{}a_{N}^{3} }} .
\end{align}
In terms of the discrete wave function, $\V$ is the volume in configuration space for which the
wave function is non-zero, and the relative volume is the number of cells in configuration
space occupied by the discrete wave function. The cells are of volume $\a_{1}^{3}\ldots{}a_{N}^{3}$, where $a_{i}$ is
the length characteristic of the $i$th particle represented in the wave function. In terms of the
wave function that is continuous in space and in magnitude, the volume $\V$ is defined as the
volume in configuration space where the average wave function magnitude for the cells of
volume $\a_{1}^{3}\ldots{}a_{N}^{3}$ is greater than $ \f_{0}$. Assuming that the wave function is approximately constant
over each single cell we can write this condition for each cell as $|\psi|> \f_{0}$. The relative volume
is the number of cells which satisfy this condition. 

The full wave function of $N$ spin-half particles will be given by multiplying the spatial
wave function by a spin vector of $2^{N}$ components, so that the total wave function is given by
$2^{N}$ complex values at each point in configuration space. I have defined the state complexity
of the wave function to be the number of non-zero complex values used to represent the
discrete wave function in the position representation. In this case the state complexity is $2^{N}$
times the relative volume of the wave function in configuration space. (Configuration space
corresponds to position space for a single particle system, so this definition is consistent with
the earlier one.)

\subsection{Many-particle systems and wave function collapse}
I will assume that the critical relative volume of an $N$ particle system is again $v_{c}$. When the
wave function reaches this relative volume it collapses to some fraction of this volume.

From the point of view of discrete physics, a similar argument to the single particle
case applies. The larger the relative volume of the wave function in configuration space, the
weaker the average magnitude of the wave function. The relative volume of the wave
function in configuration space determines the average wave function magnitude because the
integral over configuration space of the magnitude of the wave function squared must be
equal to unity at all times. Thus if a wave function represented the state of a large number of
particles, some of the particles having probability distributions spreading out in position space
with time, then eventually the wave function magnitude would become zero everywhere in
configuration space. Thus it is natural from the point of view of discrete physics that in order
to keep the average magnitude well above the minimum $\f_{0}$, that the same or similar limit be
placed on the volume of the wave function in configuration space as in the one-particle case,
and that any wave function with relative volume that grows beyond this critical value should
suffer a collapse, keeping the volume under this critical value at all times. 

Consider an $N$-particle system in which each of the particles interacts relatively
strongly with every other particle in the system. Suppose that the position probability
distribution of each particle in the $N$ particle interacting system covers an absolute volume of
$\V_{\s}$ in position space. Then the corresponding relative volume of the system in configuration
space will be of the order $(v_{\s})^{N}$, since the wave function will be spread over a distance of order
$\V_{\s}^{\frac{1}{3}}$ in each orthogonal direction in the $3N$ dimensional space. Thus the relative volume will
tend to increase \textit{exponentially} with $N$ as the number of interacting particles increases.\footnote{Here $v_{\s}= \frac{\V_{\s}}{\a^{3}}$: I am making the simplifying assumption that the length of the cell-side $(\a)$ is the same for each particle, which is a reasonable approximation for the purpose of working out the order of magnitude dependence of the relative volume on the number of particles represented in the wave function. Another way to see the exponential increase of the relative volume with $N$ is to note that if the position probability distribution of all $N$ particles in a wave function were spread over some relative volume $v$ in position space then, in the position representation, we would represent the state by a $3N$-dimensional matrix of size $v^{\frac{1}{3}}$ in each dimension, so there will be $v^{N}$ numbers in the matrix.} Thus
as $N$ grows the relative volume will soon reach the critical relative volume $v_{c}$, causing the
wave function to collapse. Thus for an interacting many-particle system the volume covered
in \textit{position space} by \textit{each particle} will tend to remain small, since a small spread in volume of
the position probability distribution for each particle in position space will contribute a large
amount to the relative volume of the whole system in configuration space---if the position
probability distribution of each particle spreads out beyond a small volume in position space,
then the critical relative volume will be reached, bringing about a collapse of the wave
function, restricting each particle to the same small volume in position space again. Thus
many-particle interacting systems will have particles within them that tend to remain
localized in position space, as is observed. Furthermore, superpositions of large numbers of
interacting particles spread over significant volumes in position space will be prevented from
occurring by the collapse of the wave function. In this way the collapse mechanism will
prevent the occurrence of macroscopically distinguishable superpositions that arise in the
application of the unmodified Schr\"{o}dinger equation to a typical measurement process, where
no collapse is assumed. Instead the wave function will collapse in such a way that these
superpositions do not arise, and there will be a determinate outcome of the measurement. 
Thus I claim that the critical complexity theory of the collapse of the wave function can
provide a solution to the measurement problem. A more detailed discussion of the
measurement process follows in the next section.

Thus we see that the problem presented by the representation of spherically
propagating waves in discrete physics is turned into an advantage, because it turns out that a
modification to the evolution of the wave function introduced in order to solve this problem
provides a solution to the measurement problem as well. This comes about because an
exactly analogous problem that arises for spherically spreading waves in position space arises
when representing the evolution in configuration space of a wave function of a large number
of particles, at least some of which are spreading out in position space with time. An
analogous modification to the evolution of the wave function in configuration space will
solve that spreading problem, and it turns out that this modification also provides a solution
to the measurement problem. This is the major result to be derived from the discrete physics
heuristic.\footnote{One might argue that the critical relative volume for an $N$-particle system should be $N v_{c}$ rather than simply $v_{c}$. For if we have $N$ totally independent particles, represented by separate wave functions in space, then the limit for that system will be $Nv_{c}$, and one could then argue that the limit in the relative volume should be independent of whether or not particles are interacting. In this chapter I have supposed, following the argument from discrete physics, that there is the same limit $v_{c}$ for each separate wave function. However, even if we set the limit as $Nv_{c}$ for any $N$ particle wave function, this would make little practical difference in what follows, as the exponential growth in relative volume for many-particle interacting systems would soon overtake any linear growth in the critical relative volume. Thus wave function collapses would occur in a similar pattern 
whichever of the alternative critical relative volumes is chosen for an $N$ particle wave function.}

It should be noted that in my presentation of discrete physics I have deviated from the
original conception of discrete physics as proposed by Feynman, Fredkin and Minsky. They
propose that we envisage the world as a cellular automaton in position space, with the state of
each cell influencing only immediately neighbouring cells in position space. The first
modification to this conception I have introduced is to represent systems in terms of cellular
automata in configuration space rather than position space, except of course in the case of
single-particle systems, and the ordinary evolution of the wave function being local in
configuration space rather than in position space. The collapse itself is non-local since it is
presumed to take place instantaneously, or virtually instantaneously, over the entire volume of
the wave function, which may be greatly extended in configuration space, having probability
distributions for each particle greatly extended in position space. This non-local behaviour
when collapses occur is the second deviation from the original conception of discrete physics
put forward by Feynman, Fredkin and Minsky.

\subsection{Measurement} \label{measurement} 
A measurement process is one in which a wave function of a system of one or few particles,
which is relatively free to spread out in space, and in configuration space, interacts with a
many-particle system, becoming entangled in a many-particle wave function, so that the
number of interacting particles involved in the wave function grows with time. (The
dynamics of this process will be discussed in \S\ref{numberp}.) The many-particle wave function will
rapidly reach the critical relative volume $v_{c}$ and will be forced to collapse, eventually resulting
in a wave function in which many particles become relatively localized in space. This
localization of a large number of particles in one confined region marks the conclusion of the
measurement. I claim that all measurements involve similar kinds of localizations in many-particle systems.

For the localization of the many-particle wave function to constitute a measurement,
the position of localization of the many-particle wave function must correlate approximately
with the position probability distribution of the incoming wave function. For this to happen
the particle being measured must become involved in some kind of ``cascade'' effect. In a
cascade effect one particle interacts with some other particles, and these particles interact with
more particles, and so on. Furthermore, at each stage in this process the position of the
interactions that take place is dependent on the position of the interactions that occur at the
previous stage. It is in this way that the probability distribution of the position of localization
of the many-particle wave function approximately correlates with the position probability
distribution of the incoming particle.

As an example of a measurement process consider the formation of a track of a
particle in an unstable medium, which could be for example an alpha particle in a Wilson
cloud chamber \citep{Mott1929}. I will assume that the alpha particle impinging on the cloud
chamber can be represented (in position space) as a spherical wave. Although the gas in the
chamber consists of neutral molecules, atoms can be ionized by the particle as it passes. I
will assume that initially the alpha particle can be represented by a spherical wave spreading
through the neutral medium. For any atom the wave function passes there will be some
chance that the atom is ionized by this particle, so there will form a superposition of wave
function components, each component corresponding to a different gas atom being ionized. 
Consider just one of the components of the superposition, corresponding to a particular atom
in the gas being ionized. In this system there are now (at least) three charged systems---the
alpha particle, the ionized atom and the electron; these systems will interact with one another
with some strength, so this component of the wave function must be given by its many-particle configuration space wave function representing at least these three systems. (Note
that the atom and the alpha particle themselves consist of many elementary particles.) The
evolution of this component of the wave function stimulates the ionization of a further atom,
whereby the probability of ionizing an atom is significant only for a narrow cone, spreading
out from that atom in the direction of motion of the alpha particle \citep[130--131]{Mott1929}. This
ionization will produce further charged particles, so the wave function will then include
superposition of wave function components of higher dimensions. And so forth. We see that
the relative volume will grow at a very great rate, because each component of the
superposition will add separately to the relative volume, and as time goes on further
components of the superposition will be generated, each representing increasing numbers of
interacting particles. The ionized particles in the path stimulate the condensation of droplets
along the trail, so a truly huge number of particles will potentially become involved in the one
wave function. All the many-particle components of the superposition evolve until the
relative volume in configuration space of the total wave function reaches the critical value $v_{c}$. 
The collapse then occurs, so that many of these components of the superposition will be
eliminated. As the entire wave function continues to evolve, the wave function will
continually reach the critical relative volume and collapse again, so that eventually just one
line of droplets at a particular location in the medium survives in the superposition. I will not
specify whether the collapse to a localized path occurs sometime during the initial string of
ionization events or only when a sizeable droplet forms. Just how many particles will need to
become involved in the wave function for a collapse to a well-localized region to occur will
depend on the values of the constants of the theory, which I am not addressing in this thesis.

\section{The number of particles per wave function }  \label{numberp}
In the above I have assumed that a particle initially represented by a wave function of a small
number of particles will come to be represented by a wave function of a larger number of
particles as the system comes into contact with other particles. As discussed earlier,
according to standard quantum mechanics, as long as there is a non-zero strength of
interaction between two systems, then those systems will be most accurately described by a
single wave function rather than separate wave functions, and since there are no truly isolated
systems, it would seem that standard quantum mechanics requires that there exist only one
wave function representing all particles. In the modified quantum mechanics proposed here, I
assume that the number of particles represented by a single wave function is finite and subject
to change over time. Thus the modified quantum mechanics of CCQM must involve further
modifications to linear quantum mechanics other than simply limiting the complexity of wave
functions by introducing wave function collapses which localize the wave functions in
configuration space. The CCQM model must also involve alterations to the dynamics of
linear quantum mechanics that have the effect that the number of particles represented by a
single wave function can remain finite and change with time.

A system described by a single, spreading wave function might be postulated to
evolve as follows. We might assume that as time goes on and a spreading wave function
comes into contact with another system, which is represented by a separate wave function,
there is a certain probability that these systems combine into a single wave function, and that
the probability that they combine is related to the strength of interaction between the particles
in those systems---that strength of interaction being determined by the types of particles
involved and, under most circumstances, by the expectation value of their separation in
position space. It is natural to suppose that the greater the strength of interaction, the greater
the probability that the systems combine. The simplest way for this to occur would be for the
two wave functions to be replaced by their symmetrized product wave function:

\begin{align}
\psi_{12} (\bx_{1} ,\ldots, \bx_{N})&=S \psi_{1} (\bx_{1} ,\ldots, \bx_{j}) \psi_{2} (\bx_{j+1} ,\ldots, \bx_{N}).
\end{align}
Here the symbol $S$ refers to a symmetrization operator, which ensures that the combined wave
function is symmetric with respect to the exchange of identical bosons between the wave
functions, and antisymmetric with respect to the exchange of identical fermions. The effect
of the symmetrization operator is to form a (normalized) proper linear combination of product
wave functions of this type, with the arguments permuted, such that the required symmetry
properties hold. 

Immediately after the wave functions combine into a single wave function, this wave
function would evolve according to the standard laws of linear quantum mechanics. 
Accordingly, an instant after the wave functions combine, the single wave function could no
longer be represented by a product wave function, due to the interactions among the particles
represented by the wave function producing correlations among the probability distributions
of the observables of those particles. This wave function would continue to evolve and to
combine with other wave functions until the critical relative volume was reached, at which
time the wave function would collapse in some way.

I will assume that there are two types of collapse processes that can occur when the
critical volume is reached. One of these types of collapse processes involves a localization of
the wave function in configuration space. This is the kind of collapse process already
discussed. The other kind of ``collapse'' process that can occur is for the wave function to
split into separate wave functions. An allowance must be made for the possibility that a wave
function will break up into separate wave functions, otherwise the wave function would
simply keep growing, in the sense that it would come to represent more and more particles as
time went on, and this process would continue without limit. Every system is surrounded by
other systems it interacts with, so if there were only a law that brings about the combination of
separate wave functions, and no law that results in the breaking up of wave functions into
separate wave functions, then wave functions would continue to combine until there existed
only a single wave function for the entire universe. As wave functions continued to combine,
the relative volume could theoretically be kept under the critical volume by continuing to
localize in configuration space, but eventually the particles represented in the wave function
would become so narrowly localized that this would bring about conflicts with experimental
results, by greatly boosting the energy of the particles, for example. If there is a process
whereby wave functions can combine, there must also be a process whereby wave functions
can split into two or more separate wave functions.

One might suppose that when the critical relative volume is reached, each subsystem
of the system represented by the wave function has some probability of becoming represented
by a separate wave function, and the probability of becoming separated would be greater the
more weakly the particle or particles in the subsystem interact with the particles in the rest of
the system. If the wave function does split into separate wave functions, this process will
clearly bring the relative volume of each wave function below the critical volume. If it
happens that the wave function does not break up into separate wave functions, then the wave
function must collapse by localizing in configuration space, so bringing the relative volume
below the critical volume. One possible method for this kind of collapse will be discussed in
the next section. 

Due to these laws for combining and breaking up wave functions, in regions where
many particles are strongly interacting the number of particles in a single wave function will
usually be high; in regions where there are few particles, or weakly interacting particles, the
number of particles in a single wave function will usually be low. This is just the result that
is required in order to solve the measurement problem, since we require many particles per
wave function for many-particle, strongly interacting systems to ensure that these particles
remain localized in position space. Where there are many particles to interact with, a wave
function will spend little time with few particles in the wave function, so the probability
distributions of the particles will not spread very far before collapsing. 

Consider this argument in more detail. I have supposed that the probability of a
subsystem becoming represented by a separate wave function would be greater the more
weakly the particle or particles in the subsystem interact with the particles in the rest of the
system. In general, the more particles there are in a system, the greater the average separation
between the particles in the system, and the weaker the average strength of interaction
between particles in the system. Thus as a system grows larger, in the sense that a single
wave function comes to represent a larger and larger number of particles, the greater the
probability will tend to become that this wave function will break up into separate wave
functions. Furthermore, the more particles a wave function represents, the greater the number
of subsystems contained within it, so the greater the probability that \textit{some} subsystem will
become represented by a separate wave function. Due to these factors there will be a
tendency for there to be an upper limit on the number of particles per wave function. 
However, the more strongly particles interact, and the more closely spaced the particles are in
position space, the more likely it is that the particles will become represented by a single
wave function. Thus the number of particles per wave function will be greater for these
systems than for systems of widely spaced, weakly interacting particles.

When discussing a typical measurement process in \S\ref{measurement}, I did not take into account
the possibility that the wave functions may split into separate wave functions when the critical
volume is reached. This possibility need not necessarily significantly affect that discussion. 
We can accommodate the possibility that the wave function of the alpha particle may
eventually decouple from the wave functions of the gas particles in the bubble chamber (after
passing out the other end of the chamber, for example). It must merely be assumed that their
combined wave function, representing a sufficiently large number of particles, remains
combined for long enough to give rise to a collapse, or a number of collapses, of their
combined wave function sufficient to confine the ``track'' of the alpha particle to within the
observed track widths. Ensuring that wave functions remain combined for sufficiently long to
bring about a sufficient number of collapses is one reason that I chose to adopt the rule
that wave functions can only split when the critical volume is reached, rather than there being
a probability of wave functions splitting at any time. It may be that this precaution is not
required. In any case it is clear that the laws of the dynamics of wave function combination
and separation must be carefully adjusted, along with the laws of wave function localizations,
in order for CCQM to abide by all the known experimental constraints.

Systems that consist of very large numbers of particles, such as tables and chairs, will
consist of a large number of separate wave functions. Consider two systems ``nearby'' each
other in position space that are currently represented by separate wave functions. These
systems need not be considered as non-interacting in CCQM, but their interaction must be
handled in the CCQM model in a different way than via an all-encompassing wave function
in a higher-dimensional configuration space. 

One possibility is that the interaction could take place via external potentials, or
``mean field'' potentials. In other words, each wave function will evolve according to
Hamiltonians that only include the full interaction potentials for pairs of particles within the
same wave function. These potentials will depend on the position variables of both particles. 
The Hamiltonian terms for representing the interaction with particles in separate wave
functions will be external potentials, which are also called ``mean field'' potentials. These
``mean'' potentials will depend on the position variables of only the particles represented in
the one wave function, but will be time-dependent in general, to take into account the
evolution of other wave functions.

This model of interactions between wave functions, via external potentials, is
appropriate only for non-relativistic quantum mechanics. In quantum field theory,
interactions are represented as taking place via the exchange of particles rather than via
potentials. Thus it is unlikely that a realist picture of interactions between particles
represented by separate wave functions should rely on time-dependent external potentials. 
Instead a realist picture of these interactions between particles represented by separate wave
functions would more likely involve the exchange of particles between those wave functions. 
The treatment of quantum field theory is beyond the scope of this work, however. Some
further comments on extending CCQM in order to give an account of quantum field theory
and interactions among wave functions will be made in \S\ref{qft}.

\section{The collapse of the wave function---suggested form of collapse }
I have proposed that when the wave function of a system covers a certain large relative
volume $v_{c}$, the wave will ``collapse'' in a more-or-less random way, reducing in volume to
some fraction $F$ of the original volume, where $F$ is some number between zero and one. The
actual fraction is not important, as long as the volume of a free wave remains large after the
collapse---the fraction cannot be too small. I will now make a tentative suggestion about the
form this collapse might take, before comparing the theory further with the theory of Ghirardi,
Rimini and Weber. A number of possibilities exist for the details of the way the wave
function collapses, but I will adapt a simple system of collapse from the collapse theory of
GRW.

\subsection{Single particle systems} \label{singcol} \label{singleformc} 
I will adapt the form of collapse from the GRW theory for one particle wave-functions. I will
assume that when the collapse occurs, due to reaching the critical relative volume, the original
wave function $\psi(\bx)$ is multiplied by a ``jump factor'' $j(\bx'-\bx)$ where $\bx'$ is the randomly chosen
centre of collapse. Like GRW, I will take the jump factor $j(\bx)$ to be a Gaussian:
\begin{align}
j(\bx) &= \left( \frac{\varepsilon}{\pi} \right)^{\frac{3}{4}} e^{- \varepsilon \bx^{2} / 2} .
\end{align}
The value of $\varepsilon$ in any particular case will depend on what width Gaussian would be required
to reduce the relative volume of the wave function by fraction $F$, the form of the wave
function before collapse and the values of the constants $v_{c}$, $F$, and $ \f_{0}$. I am not considering
in detail experimental constraints in this thesis, so I will not attempt to put bounds on the
constants on the theory. However, it is certainly the case that the width of the Gaussian $\frac{1}{\surd\varepsilon}$ will
be much larger than in the GRW theory for free-particle wave functions. 

A particle's wave function spreading out from a source will spread until it reaches the
volume $v_{c}$, then it will collapse to a wave function given by the original wave function
multiplied by the above Gaussian, centred randomly with statistics determined by the wave
function. I assume the probability distribution of the collapse centre and renormalization of
the wave function after collapse would be determined the same way as in the GRW theory.

One consequence of the collapse of the photon wave function would be a slight spread
in its momentum distribution in the direction perpendicular to the direction of propagation,
which could lead to observable results in diffraction experiments and the like. This may be
the clearest testable empirical prediction of the theory presented here. This is one reason that
the width of the Gaussian must be large for single particle wave functions: the fraction $F$
cannot be too small, otherwise there would be conflicts with currently known experimental
results. There would also be a very small violation of the law of conservation of energy, just
as there is in the GRW theory.

To illustrate these consequences of wave function collapse in CCQM, consider again
the case of a photon emitted from a distant star. I will assume that the photon can be
described by a single-particle wave function. This is most likely unrealistic, but it will serve
to illustrate some aspects of the theory. Given that the photon is in the visible spectrum,
emitted by a state of a typical lifetime of $10^{-8}$s, its wave function will have the form of a
spherical shell, the magnitude of the wave function falling off exponentially, with a
characteristic width of 3 metres \citep[158]{Messiah1961}. The discrete wave function magnitude
will be finite within this shell, but rapidly fall to zero outside this region. The hypothesis is
that when the volume of the shell reaches the critical volume it collapses, being multiplied by
a Gaussian, so that much of the shell is eliminated, leaving a single segment of the shell
largely unchanged, while eliminating the rest of the wave function. Consider a point on the
shell before the collapse takes place, a great distance from the source of the photon. The
momentum vector at this point will point in the radial direction, away from the source, and
there will be a spread in the momentum magnitude in that direction, due to the finite width of
the shell. The momentum components perpendicular to this direction will be almost zero in
magnitude. After the wave function collapses, the wave function will become more localized,
and this will cause a slight spreading in the magnitude of the momentum components,
particularly in the directions perpendicular to the direction of propagation. Thus the
expectation value of the momentum magnitude in the components perpendicular to the
direction of propagation will change from near-zero to some positive number, comparable in
magnitude to the spread in momentum. The expectation value of the momentum magnitude
in the direction of propagation will remain unchanged, since the only effect of the localization
will be to increase the spread slightly around the original value. The energy of non-relativistic particles with mass is $\frac{p^{2}}{2m}$ $(\frac{|\bp|}{c}$ for photons). Thus the expectation value of the
energy will increase spontaneously when the collapse occurs, due to the increase in the
expectation value of the magnitude of the momentum components perpendicular to the
direction of propagation. This means the law of conservation of energy will be violated: the
localization results in a small increase in the energy of the system.

This increase in energy produced by the spontaneous localizations has been noted by
\citet{GhirardiRiminiWeber1986} as a general feature of their theory. (\citealp*[See also][]{Ballentine1991}.) The increase in
energy is too small to be detected experimentally \citep{GhirardiRiminiWeber1986}. The CCQM model of
collapse will also violate the law of energy conservation, although the violation will be even
less than predicted by the GRW in the case of the collapse of a one-particle wave function,
since the degree of localization predicted is less according to CCQM than according to GRW.

Another consequence of the collapse of a single-particle wave function is that, due to
the spread in the magnitude of the momentum components perpendicular to the direction of
propagation, there would be certain effects on the evolution of wave functions that could be
observable. The increased breadth of the momentum distribution of the wave function in the
perpendicular direction would result in an increased dispersion in that direction after collapse. 
The extra dispersion of the position probability distributions of free particles could possibly
lead to observable effects. Suppose an experimental arrangement were set up that confined
wave functions of free particles to spreading in a cone of a certain solid angle, by passing a
wave function from a point source through a circular hole in a screen, for example. Then if
the wave functions were free to spread to a region where their volumes grew in size beyond
the critical volume, so collapsing, then the position probability distribution of the free
particles in an ensemble of similar wave functions would extend beyond the boundaries of
this cone from the collapse region onwards. Thus according to CCQM there is some chance
that after collapse a free particle could be detected in regions outside that allowed by the
original wave function, had no collapse occurred. The detection of particles in areas such as
this that are forbidden by standard linear quantum mechanics would provide some empirical
confirmation of CCQM.

\subsection{Many particle systems}
For many particle systems, the situation is more complex, because, as discussed earlier, there
must be some probability that the wave function breaks up into separate wave functions when
the critical volume is reached. I will suppose that the probability that a particular sub-system
becomes represented by a separate wave function is greater the less strongly it interacts with
the particles in the rest of the system. In order to produce empirical agreement with what is
observed, we will need to suppose that the probability weighting of the possible systems after
separation is such that the least possible modifications are made to the joint probability
distributions of the observables of the particles. In this way the statistical predictions
derivable from the combined system will be maintained as closely at possible, including many
of the correlations among the particles' observables, such as the conservation of total
momentum and total angular momentum. 

A break-up into separate wave functions would take the relative volume below the
critical value, but there is some probability that no such break-up will occur at the time the
critical volume is reached. I will suppose that if on a particular occasion the wave function
does not split into separate wave functions when the critical volume is reached, then the wave
function will localize in configuration space in such a way that the relative volume reduces by
fraction $F$.

The straightforward generalisation of the single particle collapse mechanism is for the
wave function to be multiplied by a jump factor, the jump factor being a many-particle
generalisation of the Gaussian in position space, namely a Gaussian in configuration space. 
Thus the jump factor is $j(x'-x)$, where

\begin{align}
j(x)&=\left(\frac{\epsilon}{\pi} \right)^{\frac{3N}{4}} e^{-\epsilon x^{2}/2}.
\end{align}
In this case $j(x)$ stands for $j(\bx_{1}, \bx_{2}, \ldots{} \bx_{N})$, $x'$ is the point in configuration space where the localization is centred, and

\begin{align}
x^{2} &=\bx_{1}^{2} + \bx_{2}^{2} + \ldots{} + \bx_{N}^{2} .
\end{align}
Note that j(x) is a product of single-particle jump factors:

\begin{align}
j(\bx_{1} , \bx_{2} , \ldots, \bx_{N})&=j(\bx_{1})j(\bx_{2})\ldots{}j(\bx_{N}) .
\end{align}
The Gaussian in configuration space is a product of $N$ one-particle Gaussians in position
space. Again the size of $\varepsilon$ will be determined by the need to reduce the relative volume by
some fraction $F$. Clearly $\varepsilon$ will be much smaller than in the one particle case, since the
relative volume in configuration space can reach the critical value when each particle in the
wave function remains fairly localized in position space. Again I assume the probability
distribution of the collapse centre and renormalization of the wave function after collapse
would be determined the same way as in the GRW theory, with the new jump factor
substituted in place of the GRW jump factor.

As in the one particle case, there will be a small deviation from the conservation of
energy. In the case of a wave function of identical particles, the symmetry of the original
wave function would not be retained, just as in the GRW theory. It is true that all the
particles in the wave function are treated the same way, in the sense all the one-particle
Gaussians multiplying the wave function are the same size, so all the particles are localized to
the same degree. This contrasts with the GRW theory. However, the centre of localization of
each one-particle Gaussian will differ for each particle, depending on the centre of
localization of the Gaussian in configuration space, so that the resulting wave function will
not be symmetric or antisymmetric. This result may be seen as unsurprising, or even
desirable, since, as we have discussed, we require our collapse theory to break the symmetry
under the exchange of identical particles, at least to the extent of sometimes breaking up wave
functions into smaller, separate wave functions.

One might argue that a collapse theory should, however, preserve the symmetry of
exchange of identical particles within each separate wave function. However, it should be
noted that it is a feature of symmetric or antisymmetric wave functions that every identical
particle they represent must have the same position probability distribution. On the other
hand, after the break-up of this wave function, some of these identical particles will be
represented in separate wave functions having differing probability distributions in position
space. It seems to me that it is a natural transition step between these two situations for the
initially symmetric wave function to localize in configuration space, and in so doing
approximately localizing the identical particles in different regions in position space. This
breaks the symmetry of the wave function, but paves the way for the break-up of the wave
function into separate wave functions having differing position probability distributions for
the identical particles they represent.

Consider for example the fate of a singlet state of identical particles. It will be a
consequence of the critical complexity theory of collapse that the wave function of two
particles initially coupled and spreading apart, such as two particles in a singlet state in
experiments such as \citet*{AspectDalibardRoger1982}, will eventually decouple into
separate wave functions, as their joint configuration space wave function reaches the critical
relative volume. The quantum coherence between the particle's direction of spin would then
be lost, and experiments such as Aspect's would detect deviations from the quantum
mechanical predictions, in accordance with the Bell inequalities \citep{Bell1964}, if the
experiments were conducted over these extremely large separation distances.

Rather than always immediately decoupling into separate wave functions when the
critical volume is reached, there will also be some probability that the wave function will
initially remain intact, but localize in configuration space. When this happens one particle
will become approximately localized in one region and, due to the symmetry of the original
spatial wave function, the other particle will be approximately localized in a region
diametrically opposite with respect to the source of the particles. The symmetry of the
original wave function will thus be broken. The next time the wave function reaches the
critical relative volume the wave function may split into separate wave functions and the
result will be that the two identical particles will be represented in separate wave functions
having probability distributions for those particles in separate regions. This is a desirable
result, the initial localization in configuration space which approximately localizes the
particles being a stepping stone towards the situation where the particles are represented by
separate wave functions in separate locations.

Thus I argue that it may be a desirable result that the rule localizing the wave function
in configuration space breaks the symmetry under the exchange of identical particles, rather
than the symmetry only being broken by the rule that governs the separation of the wave
function into separate smaller wave functions. However, it is certainly possible that some
other rules of wave function localization and separation can be devised that would maintain
the symmetry of exchange of identical particles within each separate wave function, and
confine the lack of exchange symmetry to identical particles represented in separate wave
functions. The rules of wave function collapse and separation proposed here are suggestions
only.

One way that this could be done would be to alter the jump factor for the many-
particle system by using the symmetrization operator:

\begin{align}
j(\bx_{1} , \bx_{2} , \ldots, \bx_{N})&=S j(\bx_{1})j(\bx_{2})\ldots{}j(\bx_{N}) .
\end{align}
Again the symbol $S$ refers to the symmetrization operator, which ensures that the combined
wave function is symmetric with respect to the exchange of identical bosons between the
wave functions, and antisymmetric with respect to the exchange of identical fermions. The
effect of the symmetrization operator is to form a (normalized) proper linear combination of
product wave functions of the type it acts on, with the arguments permuted, such that the
required symmetry properties hold.

Multiplying this symmetric jump factor by the original symmetric wave function
would give rise to a symmetric wave function after collapse, which is the desired result. With
this rule of collapse all wave functions would at all times remain symmetric or antisymmetric
under exchange of identical particles within the same wave function.

\section{Transition from the ``quantum realm'' to the ``classical realm'' }
CCQM will result in a very rapid transition from the ``quantum realm,'' where the wave
functions of particles are free to spread out widely, to the ``classical realm'' where the
distributions of particles in position space remain closely confined.

On the GRW/CSL model, superpositions of macroscopic numbers of particles that
spread further apart than $10^{-5}$ cm are rapidly suppressed, although there will exist the ``tails''
of the Gaussians outside this region. \citet[316]{Ghirardi1992} considers that the transition from
the quantum to the classical realm takes place where collections of particles have gathered
that contain sufficient numbers to be rapidly prevented from entering into superpositions of
states separated by more than $10^{-5}$ cm. This transition will take place in the macroscopic
realm, in the sense that the number of particles required for the localization to be sufficiently
rapid will be of order $10^{20}$ or so. The transition between the quantum realm and the classical
realm will be a linear one, in the sense that the rate of localization depends linearly on the
number of particles. If we consider some number of particles $\M$ that are considered to
localize sufficiently quickly in the above sense, then a system of half this number of particles
will localize half as quickly. We will see that the transition from the ``quantum realm'' to the
``classical realm'' will be much more rapid on the CCQM model.

\citet[35--36]{Shimony1989} has criticized the GRW theory on the grounds that the collapse
takes place at this macroscopic level. He says that his strongest objection to the theory is that
``the theory permits the formation for a short time of monstrous states of macroscopic objects
- states in which the pointer needle has an indefinite position, or Schr\"{o}dinger's cat is neither
dead nor alive---and then it rapidly aborts the monstrosity. It would be much better to have a
stochastic theory which provides contraception against the formation of such a monstrosity,
by destroying unwanted superpositions in the earliest stages of the interaction of the
microscopic object with the macroscopic apparatus.''

Shimony suggests that the suppression of superpositions may take place at the level of
large molecules. He points out that pragmatically, molecular biology pays no attention to the
possibility of superpositions of large molecules over significant distances, which may indicate
that these superpositions never occur. I suggest that CCQM may allow quantum mechanics
to be modified in such a way that the collapse of the wave function takes place mainly at the
level of large molecules, leading to the suppression of these superpositions. Thus the
transition from the ``quantum realm'' to the ``classical realm'' can take place in CCQM in the
microscopic rather than the macroscopic realm.

Consider a wave function of fairly closely interacting particles that sits on the
boundary between the quantum and the classical, according to the CCQM model. I will
suppose that this wave function represents the maximum number of particles $\M$ that can be
represented by a single wave function, considering the type of particles involved and the
structure of interactions these particles are involved in. I will also suppose that the relative
volume of the wave function is just under the critical value. Suppose that the values of $\M$ and
 $v_{c}$ are such that all the particles in this system remain localized to within $10^{-5}$ cm. If the wave
function grew in such a way that the probability distribution of any of the particles spread
more widely than this, then either the whole wave function would collapse, re-localizing the
particles to within that region, or some particles would break away from the system.

Suppose, for the purpose of illustration, that the size of the cells in configuration space
is independent of the particles involved and is characterized by a length of $10^{-15}$ cm. That is,
each side of the cell has a length of $10^{-15}$ cm. Thus $10^{10}$ of these lengths fit along the side of a
cube of $10^{-5}$ cm on a side, and the number of cells inside this cube will be $10^{30}$. Thus if there
were just a single particle confined precisely to this cubic region then the relative volume of
that wave function would be $10^{30}$. I will make the simplifying assumption that each of the $\M$
particles is precisely confined to this cubic region. Then the relative volume of the $\M$-particle
wave function will be $10^{30\M}$. I am assuming that the relative volume of this system will be
close to the critical value; thus the critical relative volume $v_{c} = 10^{30\M}$. If we supposed that
$\M=10,000$ say (the number of particles in a moderate-sized molecule, for example) then $v_{c} =
10^{300,000}$.

Consider now a wave function of half the number of particles, say a molecule half the
size. The question I would like to investigate is how freely the particles represented by this
wave function could spread out in space before the critical relative volume would be reached. 
The answer is that because the relative volume increases exponentially with the number of
particles distributed over the same volume, the volume that each particle can spread over in a
system of half the number of particles, while still just remaining under the critical relative
volume, is the \textit{square} of the relative volume in position space that each particle is confined to
in the larger system. Thus the volume that each particle can spread out in the smaller system
is equal to the volume of a cube $10^{10}$ times as large on each side. Thus each particle can
spread out a distance of $10^{5}$ cm, or one kilometre, before the critical relative volume would be
reached and a collapse would occur. For a wave function representing one third the number
of particles $\left(\frac{\M}{3}\right)$, each particle could spread out a distance of $10^{10}$ km before any collapse
would occur: it can spread out a \textit{cube} of the relative volume each particle is confined to in the
system of $\M$ particles. Thus wave functions representing the numbers of particles just a few
less than $\M$ would be very free to spread out and enter into superpositions. While the rate
of transition from the ``quantum realm'' to the ``classical realm'' is linear in the number of
particles for GRW/CSL, the rate of transition is exponential in the number of particles for
CCQM.

Wave functions representing fewer particles than $\M$ would rarely spread far enough to
collapse. What would tend to happen instead is that the number of particles represented in a
wave function would increase as that wave function came into contact with other wave
functions, until the wave function represented a number of particles close to $\M$, at which point
some collapse may occur, depending on the spatial spread of the wave function and the
number of superposed components it consists of. Weakly interacting particles would be able
to spread a long way before enough particles became represented by the one wave function to
bring about a collapse. 

As noted earlier, the law governing the splitting of wave functions must prevent wave
functions gaining many more particles than $\M$, otherwise the wave functions would
sometimes be forced to localize in such an extreme manner that high energy particles would
be spontaneously produced, and other deviations from quantum mechanics would be
expected. These deviations are not seen, so a number of $\M$ or thereabouts would be a rough
limit for the number of particles per wave function. Thus for a closely interacting system, the
system will be separated into many different wave functions of roughly $\M$ particles each, and
collapses of these wave functions will keep most particles in these systems well localized,
although relatively weakly interacting sub-systems will continually separate from, and rejoin,
each many-particle wave function.

Thus we see a rapid transition from the ``quantum realm'' to this ``classical realm,''
although of course the physics is at all times governed by the modified quantum mechanics of
CCQM. Note that the numbers chosen in this section for some of the parameters of the
theory are only guesses chosen for the purpose of illustration: it may be that these parameters
must be changed by several orders of magnitude in response to experimental constraints.

\section{Comparison with other spontaneous localization theories }
The theory outlined here, which I label the ``critical complexity'' theory of wave function
collapse, has similarities with the spontaneous localization theory of Ghirardi, Rimini and
Weber (GRW). GRW attempt to explain the process of measurement in terms of the
spontaneous localization of individual particles in the measuring apparatus causing the whole
apparatus to localize, or at least a macroscopic part of the apparatus such as a pointer. In the
critical complexity theory, measurement is not explained in terms of the spontaneous
localization of individual particles, but the localization of individual (free) particles and of
many-particle systems have a common explanation---the limitation of the relative volume of
the system.

As we have discussed, the GRW/CSL theories cannot evade the universal holism of
standard quantum theory, and I take it as an advantage of the critical complexity theory that
wave functions represent a limited number of particles at all times. As I have mentioned, one
criticism that can be levelled at the GRW/CSL theories is that they are \textit{ad hoc}---little
motivation for the modification of quantum mechanics that the theories propose can be given
other than to produce a solution to the measurement problem, although attempts have been
made to motivate these theories by linking wave function collapse to gravitational phenomena
(\citet{Diosi1989}, \citet*{GhirardiGrassiRimini1990}, \citet{PearleSquires1996}). On the other hand,
CCQM can be independently motivated in a direct way, by considering what follows from
making quantum mechanics discrete, so this is an advantage CCQM has over GRW/CSL. 
The full mathematical details of the critical complexity theory have not been worked out, but
there are further theoretical advantages over the GRW/CSL theories that give the theory
promise, and I will now discuss some of those advantages.

An attractive feature of the critical complexity theory of collapse is that there will be
no collapses at all in the case of particles in energy eigenstates, since the relative volume of
the wave function of an energy eigenstate will remain constant with time. The GRW/CSL
theories do predict collapses for bound states, since the probability of collapse depends
merely on the number of particles in the system. After collapse the particles will no longer be
in energy eigenstates. If it is assumed that the collapses occur to electrons, then electrons in
atoms will spontaneously become excited, and the atoms will spontaneously emit photons as
the electrons return to the ground state. If it is assumed that the spontaneous collapses also
occur to quarks, then, for example, the quarks in protons will become excited and protons
should spontaneously radiate pions. These effects are potentially observable. There are
experiments that place very low limits on the lifetime of nucleons, and these results place
constraints on the spontaneous localizing theories. \citet{PearleSquires1994} argue that these
results rule out the original GRW theory, if it is assumed that the collapses apply to quarks. 
They argue that the CSL theory in its original form is also close to being ruled out by these
results if it is assumed to apply to quarks, but suggest that altered CSL theories in which the
rate of collapse is proportional to the mass of the particle are compatible with these results. 
The relative volume model of collapse does not suffer from these difficulties at all.

One problem that the GRW and CSL theories have is how to deal with photons. It has
been suggested that the localization of photons to the small volume given by the hitting
Gaussian would produce a measurable effect on the spectrum of the cosmic background
radiation \citep{Squires1992}. Squires argues that this problem can be avoided if GRW or CSL is
altered by making the rate of hitting proportional to the mass of the particle, so that
photons are never subject to direct hitting events. However, this adds further complication to
the theory. I suggest that the critical volume criterion for collapse can be applied equally to
all particles, including free photons. This need not produce any measurable effect on the
spectrum of the cosmic background radiation, because the assumption has been made that the
hitting Gaussian would be extremely large for one-particle systems such as photons or
neutrinos, so altering the momentum distribution very little. There would be a small effect
predicted on the spectrum, but there is no difficulty in principle of the critical volume being
large enough to be compatible with current experimental evidence. This ability to deal
adequately with photon collapse, while treating all particles equally, is a nice feature of
CCQM.

The critical complexity theory can give rise to measurements being completed at an
earlier stage than the GRW/CSL theory. In the GRW/CSL theory, in order for a measurement
to be completed the interaction of the microscopic system with the measuring apparatus must
give rise to a superposition of a macroscopic numbers of particles in sufficiently separate
locations in space. This superposition is then broken up by the collapse mechanism, so
completing the measurement. It is not clear that such superpositions will be formed for all
processes we would normally think of as measurements---for example it would seem that a
photon interacting with a photographic plate, or the retina, would not cause such a
superposition, so it would seem that these are not measurements for GRW \citep[100--101]{Albert1992}. Thus there are types of interactions that would normally be regarded as measurements
which do not seem to bring about wave function collapse on the GRW theory, at least not
until further interactions take place. I claim that they could produce collapse on the theory
presented here---a collapse of the requisite type to complete a measurement.

Consider the example discussed by \citet*{AlbertVaidman1989}. They describe a
Stern-Gerlach arrangement for measuring the z-spin component of a spin-half particle. The
Stern-Gerlach apparatus is arranged to split the wave function into spatially separate
components, one corresponding to spin-up in the z-direction and the other spin-down in the z-direction. These components then come into contact with a fluorescent screen at two distinct
points \A{} and \B{}. There the component wave functions excite many atoms, so that atomic
electrons around these atoms take on excited states, moving into higher orbitals. A short time
later, the electrons return to the ground state, emitting photons in the process, which can be
observed by the naked eye. The observer sees a luminous dot at point \A{} or \B{}, not at both
points. Albert and Vaidman say that ``according to conventional wisdom about
measurements'' the measurement is already over by the time the wave function reaches the
eye, so that one would hope that a collapse theory such as the GRW theory would predict that
a collapse of the wave function would occur before reaching the observer's eye, so that only
the \A{} or \B{} component would reach the eye. The problem is that for the GRW mechanism to
bring about a collapse involving a large number of particles we require a macroscopic number
of particles ($\sim 10^{20}$) to be in superposition of macroscopically different locations, and here this
is not the case; only one particle, the incident particle is in a superposition of macroscopically
separate positions---the particles at \A{} and \B{} are in superpositions merely of excited and non-excited states. \citet*[114]{AicardiBorsellinoGhirardiGrassi1991} respond to this problem by admitting that the
measuring apparatus described will not bring about a collapse, but they argue that a collapse
will occur in the nervous system of the observer, and this is sufficient to preserve
appearances.

On the other hand, in CCQM, as long as there are sufficient numbers of fluorescing
atoms involved, the relative volume will reach the critical value, and a collapse will occur
before the photons reach the observer. It does not matter greatly that the superposition of
states of each electron does not spread over a large volume in position space, as long as a
large number of particles is involved, the exponential growth in the relative volume with the
number of particles will ensure that the total relative volume will reach the critical relative
volume. When the collapse occurs, the wave function is multiplied by the Gaussian in
configuration space---every particle in the wave function is multiplied by the same Gaussian
in position space. Multiplying the original wave function by a product of single-particle
Gaussians of radius less than the distance between the points \A{} and \B{} will eliminate the wave
function component corresponding to one of those points, and the relative volume will be
reduced by at least a factor of two. At this stage the measurement of spin is completed, since
only one of the wave function components corresponding to spin up or spin down survives. 
Thus a single collapse which reduces the relative volume by a factor of two or more would be
sufficient to complete a measurement of the spin, and in all the specific models of the
collapse theory I have proposed I assume the factor by which the relative volume reduces
on collapse will be greater than two. 

Although it is an advantage of CCQM that a collapse can occur without a macroscopic
number of particles involved, the number involved must not be too small or no collapse of the
wave function will occur. \citet*[114]{AicardiBorsellinoGhirardiGrassi1991} note that just six fluorescent photons
reaching the eye from the screen may be sufficient for producing a spot that is visible to the
naked eye. With this small number of photons being emitted, the number of particles involved
in the wave function may be insufficient to cause a collapse in CCQM, in which case the
critical complexity theory would not predict a collapse until the photons reached the eye of
the observer, in agreement with GRW. However, as \citet*{AicardiBorsellinoGhirardiGrassi1991} point out, in general
one would not expect significant deviations from linear quantum mechanics when there are
only a small number of particles involved in a wave function, otherwise one would expect
readily observable deviations from the predictions of linear quantum mechanics that have not
been observed. \citet*{AicardiBorsellinoGhirardiGrassi1991} show that it follows from the GRW model that the
electrical signal generated in the optic nerve by these few photons striking the retina
eventually brings about a sufficiently large macroscopic superposition, leading to a collapse
of the wave function in the nervous system of the observer. For CCQM, the collapse would
also occur after the photons strike the retina and further particles became involved, although
the collapse would occur earlier than it would according to the GRW model.

The measurement of spin of a particle comes down to a measurement of position of
that particle. More generally, it is evident that according to the theory of measurement
proposed here, for there to be a measurement only the position of one particle must be in a
superposition of macroscopically distinct position states---the particle that triggers the
superposition of states of many particles at those distinct positions. Thus as long as the state
of a large number of particles depends on the (macroscopic) position of one particle, then a
``measurement'' of the position of that particle will take place; in other words, a wave function
collapse will occur such that the probability distribution of the centre of collapse of a large
number of particles will be determined by the position probability distribution of the original
particle. This result is desirable, because it is commonly held to be the case that all
measurements are based on such position measurements.

\section{Superconductivity }
According to the Bardeen-Cooper-Schrieffer (BCS) theory of superconductivity \citep{Tinkham1975}, in a superconductor
electrons form into pairs by the exchange of phonons in the lattice which produces an
attractive force between them. In the ground state the total momentum of the pair of electrons
(Cooper pair) should be zero, so the electrons must have equal and opposite momenta. 
Taking into account the antisymmetry requirement for the total wave function with respect to
exchange of the two electrons, and that the singlet state will have lower energy than the triplet
state, the spin of the electrons will also be opposite. BCS make the approximation that the
pairs only influence each other via a ``mean-field,'' in which the occupancy of each state $\mathbf{k}$ is
taken to depend only on the average occupancy of other states. Thus they took the Cooper
pairs to be effectively non-interacting with other pairs. The pairs effectively form bosons,
which means that many of them will occupy the same state. This gives rise to coherent states
that can display ``macroscopic'' quantum behaviour, such as the quantization of electric
currents around a loop.

Consider the case of a simple superconducting circuit. Following \citet{Rae1990}, the
BCS wave function of the superconducting state can be written as

\begin{align}
\psi =\psi_{1} \psi_{2} \ldots{}\psi_{N} \exp [iS(\br)]
\end{align}
where $\psi_{j}$ represents the wave function of a Cooper pair composed of electrons with
wave vectors plus and minus $\mathbf{k}_{j}$ and $S(\br)$ is the macroscopic phase associated with the
supercurrent. In standard quantum mechanics, this wave function should be symmetrized, so
that the product should be preceded by a symmetrization operator if we wish to represent the
wave function according to standard quantum mechanics.

As noted, in BCS theory, each pair is in effect represented as non-interacting with
other pairs, except via a mean field. This approximation is assumed to be a good one because
the pairs interact very weakly with each other. Thus in CCQM, each pair will, with high
probability, be represented by a separate wave function. There will be some probability that
two Cooper-pair wave functions will combine to form a symmetrized product wave function,
but even if a wave function containing these pairs were to grow beyond the critical relative
volume, the wave function would most likely break up in such a way that the Cooper-pair
wave functions would separate out again, due to the weak interaction strength of the pairs with each
other. The breakup would be unlikely to separate the individuals within the pairs, since the
electrons within a pair are assumed to interact more strongly than the pairs interact with each
other.

In BCS theory, the probability distribution of each electron in the pair is assumed to
be spread over the entire conductor. This means that the volume of each electron's position probability
distribution is less than or equal to the volume of the conductor. The relative volume in
configuration space of each Cooper-pair wave function will therefore be well under the critical value. 
Thus no collapse of Cooper-pair wave functions would be expected in CCQM. This means
that very little deviation from the predictions of standard quantum mechanics would be
predicted for this system, as far as the predictions derived from BCS theory are concerned. 

\citet{Leggett1984} has envisaged that experiments could be carried out on superpositions
of superconducting states, in particular a two-state superposition of a many-particle
superconducting system. Leggett refers to this as a ``Schr\"{o}dinger's-cat'' state, since it is a
two-state superposition of a system containing a large number of particles. However, from the
point of view of CCQM, the formation of this superposition would merely mean that each
Cooper-pair wave function would form a superposition of two spatially separated
components, thus increasing the relative volume of each Cooper-pair wave function by a
factor of approximately two. The relative volume of each wave function would remain well
below the maximum volume, and no further deviation from quantum mechanics would be
predicted. 

According to GRW/CSL, there will be many collapses of Cooper-pair wave functions;
in the GRW cases the number of collapses depends merely on the number 
of particles in the system. Thus there is a significant difference in predictions between
GRW/CSL and CCQM. However, the deviations from standard quantum mechanics predicted
by GRW/CSL are well below that which is currently observable \citep{Rae1990}.

\section{The interpretation of the wave function---the ontology of CCQM }
In this thesis I am assuming a realist interpretation of the wave function. Here I wish to spell
out this interpretation in more detail, and compare this interpretation to the interpretation
given to the wave function by the supporters of the spontaneous localization theories.
\citet{Ghirardi1997} begins his article with the following quote from \citet[29]{Bell1990}:
\begin{quotation}
\noindent In the beginning, Schr\"{o}dinger tried to interpret his wavefunction as giving
somehow the density of the stuff of which the world is made. He tried to think
of an electron as represented by a wavepacket---a wavefunction appreciably
different from zero only over a small region in space. The extension of that
region he thought of as the actual size of the electron---his electron was a little
bit fuzzy. At first he thought that small wavepackets, evolving according to
the Schr\"{o}dinger equation, would remain small. But that was wrong. 
Wavepackets diffuse, and with the passage of time become indefinitely
extended, according to the Schr\"{o}dinger equation. But however far the
wavefunction has extended, the reaction of a detector to an electron remains
spotty. So Schr\"{o}dinger's ``realistic'' interpretation of his wavefunction did not
survive.

Then came the Born interpretation. The wavefunction gives not the
density of \textbf{stuff}, but gives (on squaring its modulus) the density of probability. 
Probability of what, exactly? Not of the electron \textbf{being} there, but of the
electron being \textbf{found} there, if its position is ``measured.''

Why this aversion to ``being'' and insistence on ``finding''? The
founding fathers were unable to form a clear picture of things on the remote
atomic scale. They became very aware of the intervening apparatus, and of the
need for a ``classical'' base from which to intervene on the quantum system. 
And so the shifty split.
\end{quotation}
 
\citet[36--37]{GhirardiGrassiBenatti1995} note that \citet[30--31]{Bell1990} suggests that the spontaneous
reduction models allow one to take a ``density of stuff'' rather than a probability density
interpretation of the modulus square of the wave function. As Bell points out, this density,
the density of stuff of which the world is made, is a density in the $3N$-dimensional
configuration space rather than in position space. \citet{GhirardiGrassiBenatti1995} reject this
interpretation of Bell, and instead adopt an interpretation that involves the average mass
density in three-dimensional position space. (The title of Ghirardi 1997 is ``Replacing
Probability Densities with Densities in Real Space.'') By contrast, I adopt an interpretation of
the wave function in CCQM compatible with Bell's ``density of stuff'' interpretation. I will
first present the interpretation of GRW/CSL provided by \citet{Ghirardi1997} and \citet{GhirardiGrassiBenatti1995}.

\citet{Ghirardi1997} offers what he calls the ``average mass interpretation of CSL.'' In his
exposition of this interpretation, \citet[354]{Ghirardi1997} says that he wishes to ``give a precise
criterion to identify properties which could be legitimately claimed to be possessed by
individual physical systems.'' The criterion he gives is as follows:
\begin{quote}
 DEFINITION. We will claim that a property corresponding to a value (a range
of values) of a certain variable in a given theory is \textit{objectively possessed or
accessible} when, according to the predictions of that theory, any experiment
(or physical process) yielding reliable information about the variable would, if
performed (or taking place), give an \textit{outcome} corresponding to the claimed
value. Thus, the crucial feature characterizing \textit{accessibility} (as far as instances
about individual systems are concerned) is the matching of the claims and the
outcomes of physical processes testing the claims.
\end{quote}
Central to the average mass interpretation is the average mass density operator $\M$ defined
earlier:

\begin{align}
M (\bx) &= \sum_{k} m_{k} N^{(k)} (\bx)
\end{align}
where $m_{k}$ is the mass of the particles of type $k$ and $N^{(k)}(\bx)$ is the number density operators
giving the average density of particles of type $k$ in a volume of about $10^{-15}$ cm$^{3}$ around the
point $\bx$. The mean value of this operator for a state vector is given by

\begin{align}
\mathbf{M} (\bx) &= \left< \psi (t) | M (\bx) | \psi (t) \right>.
\end{align}
\citet{Ghirardi1997} assumes that the state vector is the state vector of the universe. The
distribution of values $\mathbf{M}(\bx)$ is referred to as the average mass density distribution. 

\citet{Ghirardi1997} first considers the question of whether in standard quantum
mechanics $\mathbf{M}(\bx)$ represents an objectively possessed (accessible) property of microscopic or
macroscopic systems. He answers this question in the negative. In standard quantum
mechanics, macroscopic systems such as pointer needles in measuring instruments can enter
into superpositions of states separated by macroscopic distances. In this case the average
mass density distribution will have two or more ``peaks,'' corresponding to the positions of the
pointer in the components of the superposition. However, in our observation of pointers, we
only ever observe the pointer in one position. This shows, in accordance with the definition
given above, that the average mass density distribution is not an accessible property of the
pointer. In the case of microscopic systems, such as a single electron, these can also enter
into superpositions with components in widely separated regions, so the average mass density
distribution will not be an accessible property of microscopic systems either.

In the case of CSL (and GRW), the dynamics of the theory suppress macroscopic
superpositions of the kind described for a pointer, rapidly eliminating all but one component
of such a superposition if it should form. Thus \citet{Ghirardi1997} claims that the mass density
distribution will be an accessible property of macroscopic systems at all times. He says that
the average mass density will be accessible at time $t$ if the variance in the mass density is very
much less than the average mass density. When there is a space region such that the mass
density is accessible for all its points, all reliable tests aimed to ascertain the mass density
value give outcomes corresponding to $\mathbf{M}(\bx)$. He says \citep[360]{Ghirardi1997} that he wishes to
``limit property attribution to $\mathbf{M}(\bx)$ itself. In doing so one retains, at the macroscopic level
alone, a space-time description of physical processes: the unique dynamical law governing all
conceivable systems just strives to make this quantity accessible.'' He also says \citep[361]{Ghirardi1997} that ``the possible nonaccessibility of the mass density distribution does not entail
that it is not real. Actually even regions in which the mass density is non accessible
according to our criterion may play an important physical role. But what matters for our
accounting of the world as we perceive it derives from the fundamental feature of the model
to make accessible precisely the mass density function.''

On the other hand, microscopic systems evolve for most of the time in the same way
as in linear quantum mechanics, forming superpositions of widely separated states, so for
microsystems the average mass density will be nonaccessible at almost all times. 
This corresponds to the fact that ``the microworld is the realm of \textit{potentialities} and the
quantum level of reality is characterized by an objective undefiniteness of properties.''

Under this interpretation, there are, \citet[363]{Ghirardi1997} says, ```two levels of `reality',
one, the microscopic one which represents the `world of potentialities' and does not admit a
spatio-temporal description, the other, related to the world of our experience, which admits a
spatio-temporal description within three-dimensional ordinary space plus the time
characterizing a nonrelativistic theory.'' 

Ghirardi says that he wishes to provide an account of what exists ``in such a way that
the resulting physical picture is compatible with our experience of the world.'' He calls this
the task of providing an account of ``psychophysical parallelism,'' and of ``closing the circle.'' 
So while he agrees with \citet{Bell1987} that the ``state vector is everything,'' he says that the
``basic elements'' of the theory are the accessible values of the average mass density
distribution. These accessible values correspond to the macroscopic objects we are aware of,
such as tables and chairs and so on, which have definite positions and do not enter into
superpositions of states with macroscopically separate locations. 

In \citet{GhirardiGrassiBenatti1995}, the ontology of CSL is expressed in terms of Bell's concept of
a ``beable,'' which is a thing that exists. \citet{GhirardiGrassiBenatti1995} say that the beables of CSL are
the wave function and the accessible values of the average mass density distribution.

One criticism I have of the interpretation of CSL as presented by \citet{Ghirardi1997} is
the lack of clarity in what is meant by an accessible property. In the definition of an
accessible property Ghirardi refers to an ``objectively possessed or accessible'' property which
suggests that ``accessible'' is synonymous with ``objectively possessed.'' Objectively
possessed is normally taken to mean possessed independently of human thought, and one
would think that one could replace the phrase ``objectively possessed'' with ``possessed in
reality.'' Thus a property that is nonaccessible at time $t$ would seem from this definition to be
not objectively possessed at time $t$ and thus not possessed in reality at that time. Yet, as noted
above, he says that the possible nonaccessibility of the mass density distribution does not
entail that it is not real. There seems to be some tension between these positions. 
Furthermore, it would seem that on their interpretation we could not say that an individual
microscopic system has the property of being in a superposition of two widely separate
eigenstates, because this ``property'' does not meet their criterion of being accessible, and
Ghirardi says he wishes to ``limit property attribution to $\mathbf{M}(\bx)$ itself.'' Yet he describes
microscopic systems as having properties such as these when establishing that the mass
density is not accessible for these systems. Thus there is a risk of a self-contradiction.

It is a curious feature of their interpretation that they accept a realist interpretation of
the wave function (``the wave function is everything'') but offer no picture of the properties of
the wave function on the microscopic scale other than to say that it is the source of ``potential properties.''

Before further discussing the interpretation of GRW/CSL, I will provide my
interpretation of CCQM.

Consider a wave function of N particles:

\begin{align}
\psi (\bx_{1} ,\ldots, \bx_{N} , t)&=f(\bx_{1} ,\ldots, \bx_{N} , t)e^{i\theta (\bx_{1} ,\ldots, \bx_{N} , t)} .
\end{align}
According to the standard interpretation of the wave function, the quantity $|\psi(\bx_{1},\ldots, \bx_{N}, t)|^{2}$ is
a probability density: $|\psi(\bx_{1},\ldots, \bx_{N}, t)|^{2}d\bx_{1}d\bx_{2}\ldots{}d\bx_{N}$ is the probability of finding, on
simultaneous measurement of the positions of each of the $N$ particles at time t{}, particle 1
within volume $d\bx_{1}$ of $\bx_{1}$, particle 2 within $d\bx_{2}$ of $\bx_{2}$, particle 3 within $d\bx_{3}$ of $\bx_{3}$, and so on. 
This interpretation is adequate if one is only interested in predicting the results of
experiments, but the question is whether the wave function can be given a more direct realist
interpretation. On my interpretation of CCQM, the wave function itself can be interpreted
realistically as a wave (or field) in configuration space, a complex valued wave in a 3$N$
dimensional configuration space. The magnitude of the wave at point $\bx_{1},\ldots, \bx_{N}$ is given by
$\f{}(\bx_{1},\ldots, \bx_{N})$ and its phase is given by $\theta(\bx_{1},\ldots, \bx_{N})$. (There is also a spin vector of $2^{N}$ components
multiplying the wave function. This can be alternatively written as a $2^{N}$ component wave
function. I interpret this as a vector field in configuration space, characterized by $2^{N}$ complex
magnitudes at each point in configuration space. These magnitudes will in general vary with
the location in configuration space.) Of course, the wave function can also be represented in
momentum space, and in other bases as well, although in a discrete representation the
transformation from one basis to another may not preserve all information precisely. 
However, as discussed in chapter \ref{space}, it is in configuration space that causation is local,
before collapse occurs. Furthermore it is in configuration space that the wave function
localizes when the collapse occurs, and it is in configuration space that the laws of CCQM are
formulated. It is configuration space that is divided into discrete cells, and the relative
volume which triggers collapse is measured in configuration space. Due to the privileged role
of configuration space in the formal and dynamical laws of CCQM, one might claim that this
representation gives the most fundamental description of reality, and that the other
representations merely supervene on the configuration space one. This is indeed the position
I hold.

I concur with \citet{Bell1990} that the magnitude of the (continuous) wave function
squared can be interpreted as the ``density of stuff'' of which the world is made. And this is a
density in a $3N$ dimensional configuration space. At least I concur with this interpretation up
to a certain point. This description is certainly useful in contrasting it with the merely
instrumental interpretation of the wave function's square modulus as a ``probability density.'' 
However, as I interpret it, this ``stuff'' of which the world is made does not merely have a
density in configuration space; it also has a magnitude and a phase, so it can be described as a
wave or complex valued field in configuration space. (In fact there are $2^{N}$ complex
components at each point in configuration space.) Thus I support a direct realist
interpretation of the wave function itself, not just the modulus square of the wave function.

Another quantity of interest is the distribution of each particle $k$ in three-dimensional
position space, given that this particle is represented by an $N$ particle wave function:
\begin{align}
g(\bx_{k})&= \int | \psi (\bx_{1} ,\ldots, \bx_{N}) |^{2} d\bx_{1} \ldots{}d \bx_{k-1} d \bx_{k+1} \ldots{} d \bx_{N} .
\end{align}
According to the standard probabilistic interpretation of the wave function $g(\bx_{k})$ is the
probability density of particle $k$ in position space: $g(\bx_{k})d\bx_{k}$ is the probability of finding on
measurement particle $k$ within volume $d\bx_{k}$ in position space. I suggest that as well as having
this probabilistic interpretation, the quantity $g(\bx_{k})$ can be interpreted realistically as the
density of the particle $k$ in position space: it is a ``projection'' of the wave function from
configuration space into position space.

Consider now the volume of each particle $k$ in position space. In the case of a discrete
wave function, the volume in position space of a single particle $k$ can be defined as the
volume in position space for which the quantity $g(\bx_{k})$ is nonzero. (In the discrete case, the
integral in the definition of $g(\bx_{k})$ will be replaced by a sum.) In the case of a continuous wave
function, a corresponding discrete wave function can be defined, as described earlier, and the
volume of particle $k$ can be defined in the same way: the volume in position space for which
the quantity $g(\bx_{k})$ derived from the discrete wave function is nonzero. I interpret this volume
in position space corresponding to particle $k$ as the volume occupied by that particle in
position space. (The relative volume in position space of each single particle will be the number
of cells in position space occupied in this way.)

As discussed earlier, the dynamics of CCQM gives rise, for strongly interacting
systems at the macroscopic level, to systems described by separate wave functions, each
confined to a localized region, so that $g(\bx_{k})$ will be well localized in position space for the
particles in these systems. This corresponds to the world as we are aware of it. Thus we are
able to provide an account of the world compatible with our experience, to provide an
account of psychophysical parallelism, and to ``close the circle.'' The reason why we ``see''
position space rather than configuration space, despite the fact that quantum systems are
represented in configuration space, was discussed in chapter \ref{space}: I argue that the reason is
that, at the macroscopic level of things we are directly aware of, causation is approximately
local in position space.

On the other hand, for weakly interacting systems the particles are very free to spread
out in position space. Thus there are again in a sense ``two levels of reality'': the quantum
realm where systems are free to spread out in position space, and the ``classical realm'' where
systems are confined in position space.

It is easy to see why the interpretation of $g(\bx_{k})$ as the density of a particle in position
space will not give a picture of reality compatible with experience in the case of GRW/CSL. 
For in their model, if we presume that the wave function of the universe is appropriately
symmetrized under the exchange of every identical particle, \textit{then every identical particle in
the universe must have exactly the same position probability distribution}. In other words, the
distribution $g(\bx_{k})$ of each identical particle must extend over the entire universe. This is not
just because the wave function is continuous in the GRW/CSL model and so can never fall to
zero magnitude. Even if the wave function were discrete, the distribution $g(\bx_{k})$ of each
particle of type $k$ would extend over the entire universe, other than regions in which there are
no $k$ particles, due to the presumed symmetry character of the wave function. This is why the
supporters of GRW/CSL must adopt a number density or mass density interpretation of the
wave function in order to obtain a picture of reality compatible with our experience. It seems
to me to be an advantage of CCQM that a more straightforward interpretation of the wave
function is possible than for GRW/CSL. In addition, the interpretation of CCQM for
microscopic systems provides a picture of reality that transforms smoothly into the picture of
reality for macroscopic systems. On the mass density interpretation of GRW/CSL on the
other hand, there is a distinct jump between the descriptions of reality provided for the
microscopic and macroscopic levels. No picture at all is provided of reality on the
microscopic scale (in terms of categorical properties). The only description of reality
provided for the microscopic level is that there exist in that realm ``potential properties.'' A
different picture altogether emerges at the macroscopic level, where a realist picture is given
in terms of categorical rather than dispositional properties.

Defenders of the spontaneous localization theories reject Bell's interpretation of the
(modulus square of the) wave function as describing the density of stuff in configuration
space \citep{Ghirardi1997, GhirardiGrassiBenatti1995}. It is not entirely clear to me why they do so. It
seems to me that they could use the criterion of accessibility to identify properties that will
correspond closely to properties that we observe on the macroscopic scale, but not regard the
criterion as a criterion of objectivity or reality. If they did, then they could use the
accessibility criterion to give an account of psychophysical parallelism, and at the same time
accept the direct realist interpretation of the wave function. One reason they may reject this
interpretation may be that the wave function can be represented in many different bases as
well as the position basis, and to assume a direct realist interpretation of the wave function in
the position basis is not justified. However, both GRW and CSL privilege the position
representation in the formalism of the theory, so an interpretation that privileges the view of
reality presented in the position basis would not seem unreasonable. 

Another reason that they may reject this interpretation could be because they realize
that in the GRW/CSL models of the universe there can only be a single wave function for the
entire universe, given the fact that the universe is the only truly isolated system. If there is
only a single wave function for the entire universe in a GRW/CSL model, the number of
dimensions of the configuration space would be $3N$, where $N$ is the number of particles in the
universe (or at least the number of fermions in the universe). This number $N$ may be infinite,
if the universe is infinite in extent. The reluctance to accept the picture of reality of a wave
describing the density of stuff in an infinite dimensional space is understandable. Even if the
universe is finite in size, the huge number of dimensions involved may hinder the plausibility
of such an interpretation. Also they seem to be reluctant to accept the realist interpretation of
an entity not described in space or space-time. Thus the fact that the wave function (in the
position representation) is defined in configuration space may for them automatically rule out
a realist interpretation of it. This seems to me to be an overly narrow attitude to take. 

Another reason for rejecting a direct realist interpretation of the wave function may
arise from the implications of such an interpretation for Lorentz invariance. (See \citealt{Ghirardi1996}.) 
As noted in footnote \ref{footnote} on page \pageref{footnote}, the use of a configuration space wave function,
which has only one time variable and $3N$ spatial variables, requires that we choose some
privileged frame in which this time is defined, which means that the wave function cannot be
Lorentz invariant. The wish to retain Lorentz invariance appears to be one of the reasons that
\citet{Ghirardi1996} rejects a direct realist interpretation of the wave function. I think, however,
that a privileged frame must be accepted in CCQM, and most likely a privileged frame must
be accepted in any adequate account of wave function collapse. This issue of Lorentz
invariance will be discussed in the next section.

The fact that in the GRW/CSL models the wave function is continuous and can never
fall to zero gives rise to what is sometimes called the ``tail problem.'' We have seen that the
wave function is multiplied by a Gaussian when a collapse occurs, so that while a component
in a macroscopic superposition can be rapidly suppressed, it cannot be made to disappear
entirely. \citet[82]{AlbertLoewer1995} point out that on the standard textbook understanding
of quantum theory (the Copenhagen interpretation) the connection between quantum states
and physical properties is given by the following rule:

\begin{quote}
 \textit{Eigenstate-Eigenvalue Rule}: An observable (i.e.\ any genuine physical
property\footnote{I reject the notion that the only genuine physical properties are quantum mechanical observables (that is, those quantities that can be represented by Hermitian operators in quantum mechanics). The wave function itself, and its values, I take to be genuine physical properties of systems, and these are not quantum mechanical observables.}) has a well-defined value for a system S when \textit{and only when} S's
quantum state is an eigenstate of that observable.
\end{quote}
It follows from this rule, and the fact that wave functions never have zero magnitude that no
object ever has a well-defined position. \citet{GhirardiGrassiBenatti1995} avoid this problem by using the
accessibility criterion, so that although the position of an instrument needle will never be
well-defined according to this rule, its average mass distribution will be accessible. This is
because the wave function of the needle will be sufficiently confined so that the variance in
its mass density distribution will be very much less than its mean.

In discrete CCQM, where the wave function is fully discrete, the problem of tails will
not arise, since the wave function magnitude does fall to zero, so the position of particles and
instrument needles will always be confined within some finite volume. The discrete version
of CCQM is my preferred option, and the fact that the problem of tails does not arise is a
factor in its favor. However, I do not think that the tails problem poses a serious difficulty for
the continuous version of CCQM either. 

If wave functions are truly continuous, then they are spread out over all space. 
However, they have a greater probability of causing a localization of many particles where
their density is highest, and this explains why we usually observe the ``particle'' in regions
where the wave function magnitude is the greatest. Thus I think there is no need to modify
the eigenstate-eigenvalue rule: we can admit that no systems are ever in an eigenstate of the
position observable. We just have to get used to the fact that everything consists of waves of
varying intensity, and there are no genuine ``particles.'' On the other hand, if we do wish to
retain talk of localized particles, and localized macroscopic objects, we could for this purpose
modify the eigenstate-eigenvalue rule along the lines suggested by \citet*{AlbertLoewer1995}. 
They suggest that if we wish to retain our ``particle talk,'' the eigenstate-eigenvalue rule could
be relaxed in such a way that a particle can be regarded as located in some region \A{} even if its
distribution in position space is nonzero outside this region, as long as the magnitude of the
tails outside region \A{} are sufficiently small compared to the magnitude inside region \A{}. They
say that how small is ``sufficiently small'' would be a context dependent matter. If the
continuous version of CCQM were true, then there would emerge a clear choice for where to
draw the boundaries of region \A{}: we should regard a particle as being inside the region in
which the position space distribution of the particle, derived from the discrete wave function
corresponding to the actual continuous wave function, is nonzero. This discrete wave
function corresponding to the continuous wave function is determined by the size of the
reference cells and the size of the base magnitude $ \f_{0}$, as described earlier.
\vfill

\section{Relativity and quantum field theory }

\subsection{Relativity and nonlocality}
Consider now the implications of the collapse theories for relativity, and the prospects for a
relativistic generalization of CCQM. Both CCQM and GRW/CSL are nonlocal theories, due
to the collapse of the wave function they involve, since the collapse is assumed to propagate
instantaneously, or nearly instantaneously, across the large distances that wave functions can
be distributed. The nonlocality of theories is sometimes discussed in terms of the \textit{parameter
independence} and \textit{outcome independence} conditions. Consider experiments conducted on a
singlet state of two electrons. We suppose that the wave function representing these two
particles spreads out from a source and the spin of each particle is measured by detectors on
diametrically opposite sides of the source of the particles. In terms of this example,
parameter independence says that measurement outcomes at one detector must have
probability independent of the parameter settings of the other detector. Outcome
independence says that the probability of outcomes at one detector must be independent of the
outcomes at the other detector. Orthodox quantum mechanics violates outcome independence
but not parameter independence. GRW/CSL violate both outcome independence and
parameter independence, but ``nearly preserve'' parameter independence \citep{ButterfieldFlemingGhirardiGrassi1993}. 
However, the nonlocality cannot be used to send faster-than-light signals (i.e.\ information cannot be transmitted faster-than-light across space). This is because, due to the
probabilistic nature of the collapse law, the statistics of outcomes on one side, considered
independently of the other, cannot be affected by what occurs on the other side. In this way,
one can claim there can be a ``peaceful coexistence'' of GRW/CSL with relativity, if one
interprets relativity as only ruling out sending signals faster-than-light. I have not formally
examined the relationship of CCQM to the parameter independence and outcome
independence conditions. However, as the probability law for the distribution of the collapse
centres is identical to the GRW model, the only difference being the ``jump factor'' that
multiplies the wave function at collapse, and the trigger for the collapse, I would expect that
the relationship to these conditions would be identical to that of the GRW model; namely, I
would expect that CCQM would violate both parameter independence and outcome
independence, but that no faster-than-light signals could be sent according to CCQM.

Neither CCQM nor GRW/CSL are manifestly Lorentz covariant: the laws of evolution
of the wave function they involve will not be the same in all inertial reference frames. This is
not surprising, because both are non-relativistic theories. However, due to the nonlocality
involved in the collapse of the wave function in both theories, it would seem unlikely that any
relativistic generalization of these theories could be fully Lorentz invariant at the level of
individual systems. Indeed, none of the attempted relativistic generalizations of GRW/CSL
are Lorentz invariant; instead, the authors of these theories introduce the concept of
``stochastic relativistic invariance'' which is a requirement that is weaker than Lorentz
invariance for each individual process, but instead involves invariance of probabilities of
outcomes \citep[330]{Ghirardi1996}. The relativistic versions of CSL that have been developed so
far have various unsatisfactory features. I will not be examining these models in this thesis. 
Instead I will examine the prospects for a relativistic generalization of CCQM.

In CCQM, I will assume that there is a privileged reference frame in which the various
quantities of CCQM are defined, such as the de Broglie wavelength of each particle. The size
of the relative volume would be measured in this privileged frame: in frames moving relative
to the privileged frame, the relative volume will be altered by Lorentz contraction. The
Gaussian multiplying the wave function when the collapse takes place will take its standard
symmetric form in this privileged frame: in other frames the shape of the ``Gaussian'' will be
distorted. The shape and orientation of the cells in configuration space would have their
standard form in this privileged frame. Any physics that accepts that space and time are
discrete cannot be Lorentz invariant, because the orientation of the lattice of cells will not be
invariant under rotations, and a translation of a system may shift a reference point of a system
from on a lattice point to between lattice points. This lack of invariance under rotations and
translations may not produce observable effects as long as the cells are small enough.

This privileged frame, in which various quantities of CCQM are defined, could be a
universal frame, the same for all systems, such as the ``rest-frame'' of the microwave
background radiation left over from the Big Bang (the frame in which the radiation is
isotropic). Rather than supposing that there is a universal privileged reference frame, we
could suppose that the privileged frame for each wave function in the universe is determined
by some aspect of that wave function. The rest frame of the centre of mass (or centre of
energy) of the wave function concerned is one possibility. The problem that usually arises
when one tries to define an objective centre of mass for a system in quantum mechanics is
that the boundary of the ``system'' is not clear-cut. This problem does not arise in the case of
CCQM, because the boundary of the wave function objectively identifies the boundary of the
system it represents. 

If the size of the cells were proportional to the de Broglie wavelength, and the
momentum of the particles were determined relative to the centre of energy of wave functions
in this way, then the cell size would be comparatively large for particles at low temperatures. 
Thus if we wished to look for experimental violations of standard quantum theory predicted
by CCQM, one promising area would be in the realm of low-temperature physics, such as the
physics of Bose-Einstein condensates, where the de Broglie wavelength of the particles is
comparable in size to the systems themselves.

One problem that could be foreseen for a theory that took the centre of energy of the
wave function as the privileged frame is that the wave functions of high energy particles from
distant sources would no longer have small cell sizes after the collapse of the wave function had taken place. This is because, after collapse, the centre of energy of the wave function would
travel along with the ``trajectory'' of the wave function, rather than being centred at the source
of the particle, as it would be before the collapse took place, assuming that the particle was
emitted from its source as a spherically propagating wave. This sudden jump in the cell size
when collapse takes place seems counterintuitive. One remedy would be to assume that the
privileged frame is determined by the original ``center of expansion'' of the wave function
before any collapse took place, rather than the centre of energy of the wave function. 
Otherwise one could resort to the idea of a universal privileged reference frame, which is my
preferred option.

\subsection{Quantum field theory} \label{qft}
One characteristic of quantum field theory is that the number of particles that exists at any
one time is not definite. We can represent the state of a system in quantum field theory by its
Fock space expansion. In this representation the state of the system is represented by a
superposition of wave functions, each wave function representing a different number of
particles. Each wave function in this sum is multiplied by an amplitude, providing a
weighting to that wave function. (As noted in footnote \ref{footnote} on page \pageref{footnote}, we require a privileged
frame in order to regard these wave functions as configuration space wave functions.) In a
generalization of CCQM to relativistic quantum theory, we could assume that the relative
volume of the whole state is given by the sum of the relative volumes of the wave functions
in the Fock space expansion. A possible objection to this is that, in standard quantum
theory, there is potentially an infinite number of wave functions in the Fock space expansion, each
having a certain amplitude associated with it. This would mean that the relative volume
would be infinite. However, in a discrete physics the amplitude associated with each wave
function in the expansion would itself be discrete, so there must be a cutoff in the expansion
when the magnitude of the expansion amplitude fell below $ \f_{0}$, the base magnitude of this
amplitude. This would ensure that the relative volume would remain finite. Furthermore, the
limitation in the relative volume below some critical value will also limit the number of wave
function components in the Fock space expansion. 

The cutoff in the Fock space expansion would also remove the infinities that arise in
some calculations in standard relativistic quantum theory due to the infinite number of terms
that hypothetically exist in Fock space expansions such as these. This problem of infinities
thus finds a natural solution within a relativistic generalization of CCQM. The problem of
infinities can sometimes be addressed in standard quantum theory by a procedure known as
``renormalization,'' but this procedure is of debatable validity, and it may not be successful in
all cases.

Avoiding the problem of infinities is often seen as one motivation for supposing that
space is discrete in nature.\footnote{See \citet[section 1.3]{Prugovecki1995} for a short history of attempts at introducing a minimum length into quantum theory, and for references to the original publications.} For if there is some minimum cell size, this eliminates from
these expansions those terms corresponding to energies above a certain cutoff energy, because
particles with higher energy, having wavelength smaller than the minimum cell size, are
disallowed. The result is that the calculations that were formerly infinite become finite. Thus
if we were to adopt a version of discrete physics in which the cell size is independent of the
energy of particles, then the cutoff in the expansions would be affected in this fashion. 
However, I have suggested that we should suppose that the cell size may be related to the de
Broglie wavelength of the particles represented, in which case there will be no minimum cell
size and no cutoff of high energy expansion terms that result from a minimum cell size. 
However, as shown above, a cutoff of high energy terms will still arise in a discrete physics of
this kind, due to the very small amplitude of these terms, the discreteness of the expansion
amplitude, and the upper limit placed on the relative volume. Thus I suggest that the problem
of infinities can be completely avoided whichever version of CCQM is chosen. (An
intermediate version of CCQM could also be entertained, whereby the cell size is generally
dependent on the energy but that there exists a minimum cell size, perhaps characterized by
the Planck length of $10^{-33}$ cm.)

In quantum field theory it is supposed that particles can be created and annihilated, so
the number of particles that exists varies with time. The process of creation and annihilation
of particles represented by wave functions would need to be described in an extension of
CCQM to quantum field theory. In order to allow for the fluctuation in the number of
particles with time, as well as to allow for the possible occurrence of processes with very
small amplitude, it is desirable that the components in the Fock space expansion can appear
and disappear as time goes on, and that this process is a probabilistic one. The probability of
a certain component coming into, and remaining in, existence would be related to the
magnitude of the expansion amplitude that component has, or would have, in the expansion.

As an example of this kind of evolution, we could consider a case where a wave
function representing two particles is, in response to external influences, replaced by a
superposition of two wave functions, one representing two particles and the other
representing three particles. As time goes on the amplitude of the three-particle wave
function in the superposition increases at the expense of the two-particle wave function, until
the amplitude of the two-particle wave function falls to such an extent that the probability of
its disappearing grows greatly, and it subsequently ceases to exist. The system will then be
represented by a wave function of three particles.

It is characteristic of quantum field theory that interactions between particles are
mediated by ``exchange'' particles. Thus it would seem essential that when CCQM is
extended to an account of quantum field theory, an account be given of interactions in terms of
exchange particles rather than the interaction potentials of non-relativistic quantum
mechanics. This certainly does not seem difficult to accommodate by natural modifications
of CCQM. For example, one could conceive that exchange particles carrying interactions
over long distances, such as photons and gravitons, could be represented by their own
separate wave functions propagating between the interacting wave functions. However, the
extension of CCQM to quantum field theory is beyond the scope of this work.

\chapter{Quantum Mechanics and Complexity}
\label{qm_complexity}

\section{Relative volume as a measure of complexity }
As pointed out earlier, it has been suggested by \citet{Leggett1984} that there may be ``corrections
to linear quantum mechanics which are functions, in some sense or other, of the \textit{degree of
complexity} of the physical system described.'' The GRW theory can be seen as a proposed
solution to the measurement problem of the form suggested by Leggett. GRW may be seen as
proposing that the complexity of a system should be measured by the number of particles it
represents, and that the probability that the wave function will collapse is proportional to its
complexity. The number of particles in a system is a fairly crude measure of complexity. 
From this point of view alone it is worth investigating other measures of complexity, which
may produce an alternative collapse criterion, and a rival theory of collapse. In the literature
on complexity, there are many measures proposed for different purposes. No one measure
predominates. If we take the relative volume of a system to be a measure of the complexity
of that system, then CCQM can be seen to be taking up this suggestion of Leggett, in that the
trigger to the corrections to the deterministic evolution of the wave function depends on the
complexity of the system. The theory can be seen as proposing a new measure of complexity
of wave functions, and proposing a criterion of collapse based on this measure. In this section
I wish to defend relative volume as a measure of complexity of quantum systems.

Another measure of complexity that is sometimes used is \textit{entropy} so it would be
desirable if the quantum mechanical measure of complexity made some connection with
entropy. The connection between the relative volume and entropy will be investigated in the
next chapter.
\vfill

\subsection{Physical complexity}
I have defined the complexity of a system as the relative volume of its wave function in
configuration space. I now wish to motivate this as a measure of complexity in physical
terms. 

The literature on the physical complexity of systems is largely concerned with far-from-equilibrium, irreversible systems. For isolated, deterministic, time-reversible
(conservative) systems, the distinction can be made between \textit{integrable} conservative systems,
which are separable into effectively one-particle systems with independent degrees of
freedom, and \textit{non-integrable} systems that are not separable. It is noted by Nicolis and
Prigogine in their book \textit{Exploring Complexity} \citeyearpar{NicolisPrigogine1989} that in some sense the latter are more
complex than the former. In this thesis I am dealing with systems in quantum mechanics that
are taken to be conservative systems for all times other than when the wave function
collapses, since they evolve deterministically and (in formal terms) reversibly between
collapse events. (The collapse of the wave function itself is indeterministic and irreversible.) 
For these quantum systems, in standard quantum mechanics, it is systems having wave
functions that can be written as products of single particle wave functions that are the
integrable systems, because to write a wave function as a product of one-particle wave
functions is to effectively separate the system into one-particle systems. From the earlier
discussion, it is clear that in general, other things being equal, the integrable systems will
have a relatively low relative volume, in CCQM, compared to non-integrable systems,
because they will most likely be represented by separate one-particle wave functions in
CCQM. Thus the measure of complexity provided in this chapter will generally, other things
being equal, be in accord with the ``measure'' provided by the division into integrable and
non-integrable conservative systems provided by Nicolis and Prigogine.

I suggest, however, that the relative volume is a superior measure of complexity,
preferable to the mere division of systems into integrable and non-integrable ones. Pre-theoretically, one might expect that systems of a larger number of particles should be classed
as more complex than systems of a smaller number of particles, other things being equal. The
relative volume does ensure this, whereas integrability, although sometimes connected to the
number of particles, does not distinguish at all among those systems that are integrable and
those systems that are not integrable. As it has been defined above, integrability does not
come in degrees. Furthermore, the relative volume will also be relatively low for wave
functions that are products of two or more particle wave functions, not just one-particle wave
functions, and if products of one-particle wave functions are simpler than non-product wave
functions, it would seem by parity of reasoning that other types of product wave functions
should also be classed as simpler than non-product wave functions.

We earlier discussed superconducting systems. For these systems the wave function
representing a large number of electrons can be written in standard quantum mechanics as a
product of separate two-electron wave functions, so in CCQM the electron-pairs will most
likely be represented by separate wave functions. Thus the relative volume is just the sum of
relative volumes of the electron-pair wave functions. The electron-pair wave functions are
restricted in size by the boundaries of the superconductor, so their volume in space will be
small compared to $v_{c}$ and their relative volume in configuration space will also be relatively
small. Thus the total relative volume of the superconducting systems will not be sufficient to
cause collapse, no matter how many electron pairs are contained in it. If we take the relative
volume to be a measure of complexity, then these superconducting systems are much less
complex than other systems of a similar number of particles, systems that are more strongly
interacting and cannot be represented by separate wave functions.

This is a desirable result, for intuitively, and for the reasons given above,
superconducting systems should be counted as having a relatively low degree of complexity,
compared to systems of the same number of particles whose members are all interacting
strongly with each other. On the other hand, on the measure of complexity appropriate to the
GRW theory, namely the number of particles in the system, the complexity of the
superconducting systems is the same as any other system with the same number of particles,
and superconducting systems will suffer just as many spontaneous localization events as any
other of these systems. If there are any such collapses, they do not currently produce
observable results. The absence of such observable collapses does not refute GRW. Yet
CCQM offers a better explanation for the lack of observable collapses for these systems: that
their degree of complexity is insufficient to cause a collapse of their wave functions, despite
the superconducting systems containing large numbers of particles spread over macroscopic
dimensions.

\subsection{Computational measures of complexity}
Computational notions are often used in providing definitions of complexity \citep{Bennett1989a,Bennett1989b}. 
I have proposed a measure of the complexity of a quantum mechanical system as the
relative volume that the system's wave function occupies in configuration space. I argue that
two different types of computational complexity can motivate this choice. The first of these
measures could be called a measure of the ``state complexity'' of the system and the second a
measure of ``dynamic complexity'' of the system.

In general the computational complexity of a problem is the amount of computational
resources (such as memory or time) it would take to solve that problem, given some
specification of the problem and of how it is to be solved \citep{TraubWasilkowskiWozniakowski1988}. One general
measure of computational complexity that is applicable to all systems, not just quantum
systems, is the amount of memory it would take to represent the system's state in digital form,
given some method of representation. We could regard a system as being more complex the
more memory it would require to be represented. However, the method of representation in
memory states could be specific to specific sciences. If we are considering a state that is
represented by a continuous function rather than discrete magnitudes, the continuous function
must be approximated in some way by a set of discrete magnitudes prior to its digital
representation. If we wished to have a method of representation that would apply to arbitrary
quantum systems, then one simple way the wave function could be represented is for each
discrete wave function magnitude to be stored in a separate memory state. I will define the
``state complexity'' as the number of memory states required to represent the wave function in
this way. If we assume that the wave function in the position basis is used, and that one
memory state is required per (complex) magnitude, then the ``state complexity'' will just be
the number of non-zero wave function magnitudes---and this is just the relative volume in
configuration space. 

In actual fact the state complexity will only be equal to the relative volume in the
special case where the particles in the wave function have no spin. In general the particles
will have spin. If the $N$ particles represented in the wave function are all spin-half particles,
then the full wave function will be represented by a vector of $2^{N}$ complex values multiplied by
the wave function in configuration space. Thus the state complexity will be $2^{N}$ multiplied by
the relative volume. 

Note that if we begin with a continuous wave function and we wish to represent this
by a discrete wave function, while preserving the form of the wave function to within some
degree of approximation, then presuming that the discretization is performed in the position
representation, it is natural to suppose that the size of the sides of the cells in configuration
space should be related to the rate of change of the wave function with distance in
configuration space. In general the magnitude of the wave function varies more slowly than
the phase with distance, and the rate of change of phase with distance is characterized by the
wavelength of the particle concerned, so it is a natural choice to have the cell sides
proportional to the wavelength of the particle concerned. 

As we have emphasized the relative volume of a wave function will grow
exponentially with the number of particles included in a wave function. Thus the ``state
complexity,'' the number of memory states required to represent the discrete wave function,
will grow exponentially with the number of particles, at least until the critical relative volume
is reached.

The second measure of computational complexity of a system I will define to be the
amount of computation required to compute the evolution of the system's discretized state per
unit of time, using the discretized equations of motion of the science concerned. This
measure of complexity could be called the \textit{dynamic complexity} of the system. As far as I am aware, this is a new measure of computational complexity but it resembles in some ways a
measure of complexity suggested by Bennett \citep{Bennett1989b} which he calls \textit{logical depth}.\footnote{The \textit{logical depth} of an object is (to first approximation) the amount of computation, or time of computation, that would be needed to generate the object using the ``minimal program'' of that object, where the minimal program is defined in algorithmic information theory. Thus this ``dynamic'' concept of logical depth is defined in highly abstract terms. Bennett notes that his measure is not linked to particular equations of motion, or causal histories, and takes this to be an advantage partly because these may be unknown or may vary greatly from case to case. In defining the dynamic complexity I am considering a specific class of dynamic systems with unique equations of motion, so it would seem natural in this case that the measure of dynamic complexity refers to the actual equations of motion. Note that the logical depth is the amount of computation that would 
be 
needed to generate an object ``from 
scratch'' using a maximally compact description of that object, whereas the dynamic complexity is the amount of computation needed to compute the state of the system from its state one discrete unit of time in the past, using the equations of motion. Thus the dynamic complexity is closer in concept to the ``logical depth per unit time'' than to the logical depth itself.} In what
follows I will not consider what kind of machine (Turing or otherwise) by which the
computations are to be considered as performed, as I will be making order of magnitude
estimates only, and the order of magnitude I take to be independent of the type of machine, as
long as the machine is a universal one \citep[789-90]{Bennett1989a}. Note that I am not here
considering the possibility of quantum computers (computers that utilize the laws of quantum
physics). I am assuming that the computation is to be carried out by a classical universal
computer.

As an example, consider a system of $N$ particles governed by the Schr\"{o}dinger
equation. Take the Schr\"{o}dinger equation in the position representation for a system of $N$
mutually interacting particles:

\begin{align}
i\hbar{\frac{\partial }{\partial t}} \psi (\bx_{1} ,\bx_{2} ,\ldots, \bx_{N} ,t)&= \left( - \sum_{j=1}^{N}{\frac{\hbar ^{2} }{2m_{j} }} {\nabla_{j}^{2} } + V(\bx_{1} ,\bx_{2} ,\ldots, \bx_{N} ,t) \right) \psi (\bx_{1} , \bx_{2} ,\ldots, \bx_{N} ,t) .
\end{align}
Characteristic of the Schr\"{o}dinger equation in the position representation is that the time
dependence of the wave function is \textit{local} in configuration space. That is, the evolution of the
wave function at a point in configuration space depends only on the values of the wave
function at that point and at immediately neighbouring points. This is true only in the
position representation. The time-dependence of the wave function at a particular point in
configuration space on the wave function value at that point is given by the potential
$\V(\bx_{1},\bx_{2},\ldots, \bx_{N})$, which is simply a multiple of the wave function at that point. The time-dependence of the wave function on the wave function values at immediately neighbouring
points is given by the differential operators $\nabla_{j}^{2}$. In the case of the discrete wave function, each
wave function value will have a finite number of immediate neighbours in configuration
space and the differential operators will become ``difference operators'' \citep{Lee1986}. In order
for the evolution of the state according to difference equations to agree, to within some
degree of approximation, with the evolution according to differential equations, the spacing of the
cells in configuration space must be closer the faster the rate of change of the wave function
with distance in configuration space. Thus we see again that it is natural to suppose that the
size of the cell sides corresponding to each particle should be proportional to the de Broglie
wavelength of the particle.

Consider the amount of computation needed to update an arbitrary discretized wave
function a single ``unit'' of time---some short time interval $\delta$t. Making use of the local nature
of the time dependence in configuration space, the evolution of the wave function at each cell
in configuration space could be calculated separately. The new wave function magnitude at a
cell could be calculated using just the wave function magnitudes at that cell and immediately
neighbouring cells. Assuming that the number of computations required is the same for each
new wave function magnitude, the number of computations needed to update the entire wave
function one unit of time will be proportional to the number of cells in which the wave-function is non-zero, which is equal to the relative volume in configuration space. Thus the
relative volume provides an order-of-magnitude measure of the number of computations
needed to compute the evolution of the wave function per unit of time.

In fact, this measure will underestimate somewhat the number of calculations needed
to compute the evolution of a many-particle system compared to a system with fewer particles. One point to note is that a single cell in a high-dimensional configuration space
will have more ``neighbours'' than one in a lower-dimensional space, so the calculation of the
wave function at that cell will require the processing of many more wave function values. In
the discretized Schr\"{o}dinger equation, the evolution is given in part by a sum of $3N$ difference
operators, corresponding to the $3N$ directions in configuration space, so if the effect of these
operators involves taking the difference in wave function values between one cell and each
neighbouring cell in each direction, then there will be a minimum of 2 calculations needed for
each of $3N$ directions. As a result, the minimum number of calculations for this process will be
proportional to $6N$ for each point. Furthermore, we have so far neglected the spin of the
particles in working out the dynamic complexity. A system of $N$ spin-half particles is
represented by a vector of $2^{N}$ complex values multiplying the spatial wave function. There
will be a difference operator for each spin component, so this increases the number of
computations at each point by a factor of $2^{N}$. The evolution of the wave function is also
influenced by the potential term in the Schr\"{o}dinger equation. The potential multiplying the
state vector in the Schr\"{o}dinger equation will be represented by a matrix of $2^{2N}$ complex
values. The number of computations involved in working out this matrix product will be of
order $2^{2N}$, although this number may be reduced somewhat if the potential takes on a simple
form. Thus the overall number of computations involved in calculating the evolution will be
at least of order $N k^N$ times the relative volume, where $k$ is some constant. But the relative
volume itself increases exponentially with $N$, of order $v^{N}$, so the computational complexity
will rise of order $N(kv)^{N}$. 

This does not significantly alter the conclusion that the relative volume gives an order
of magnitude measure of the dynamic complexity, and that the dynamic complexity will rise
roughly exponentially with $N$, where $N$ is the number of particles represented by the wave
function.

For non-interacting $N$-particle systems, which can be represented by $N$ separate single
particle wave functions, separate single particle Schr\"{o}dinger equations can be written for each
particle and the dynamic complexity will clearly be given by the sums of the dynamic
complexities of each single-particle wave function, since the relative volume for this system
is the sum of the relative volumes for each particle.

Thus two measures of the complexity of a quantum system, the ``state complexity'' and
the ``dynamic complexity'' are both closely linked to the relative volume in configuration
space, so I argue that this volume is a good measure of complexity of a quantum mechanical
system.

\section{The computer simulation of quantum mechanics } \label{compsim}
As discussed in \S\ref{comphp}, \citet{Feynman1982} has proposed that, as a heuristic guide to the
discovery of laws of physics, we should investigate whether those physical laws could be
simulated on finite computers. He suggests that we should accept as physically possible only
those laws that could be exactly simulated this way. He suggests certain rules of simulation,
then points out that given these rules, linear quantum mechanics could not be exactly
simulated this way, even if it were fully discretized. Feynman leaves it as an open question
how quantum mechanics should be altered so as to conform to this principle. It is my
contention that the modifications to linear quantum mechanics introduced by CCQM are
natural modifications to introduce in order to work towards this goal.

Feynman first considers the question of what sort of computer we are going to use to
simulate physics. He points out that as a long as the computer is a universal one, it makes
little difference what kind of computer is chosen. This is because the number of
computations needed to carry out any arbitrary problem will be the same order of magnitude
for any universal computer. He therefore assumes that the computer concerned is to be a
universal one. (These universal computers are generally presumed to operate in accordance
with the laws of classical physics. Feynman also considers quantum computers, computers
that utilize the laws of quantum physics, but I will not be considering quantum computers
here.) Turning to the simulation of physics, Feynman points out that the physics must be
discrete if it is to be exactly simulated by a universal computer. This is the first result that can
be derived from the heuristic principle. 

Feynman stipulates the following rule of simulation: 
\begin{quote}
that in carrying out the
simulation, if we were to double the volume of space-time within which the physics were to
be simulated, or to double the number of particles simulated, then we should only require a
computer of double the size: it must not be the case that an \textit{exponentially} larger computer
would be required. \citep[469--472]{Feynman1982} 
\end{quote}
If it were the case that a proposed law of
physics would require this exponential growth in the size of the universal computer, then this
law should be regarded as not physically permissible.

\begin{quote}
The rule of simulation that I would like to have is that the number of computer
elements required to simulate a large physical system is only to be proportional
to the space-time volume of the physical system\ldots{} If doubling the volume of
space and time means I'll need an \textit{exponentially} larger computer, I consider
this against the rules (I make up the rules, I'm allowed to do that). \citep[469]{Feynman1982}
\end{quote}
Feynman points out that in quantum mechanics the state is given by the wave function and the
evolution of the wave function is given by a differential equation such as the Schr\"{o}dinger
equation. He points out that to store the wave function in computer memory, or to simulate
the evolution of a wave function of $\R$ particles in a region of $N$ points of space-time according
to the Schr\"{o}dinger equation, would not be possible according to this rule, even if quantum
theory were discrete. 
\begin{quote}
 We emphasize, if a description of an isolated part of nature with $N$ variables
requires a general function of $N$ variables and if a computer simulates this by
actually computing or storing this function then doubling the size of nature ($N
\rightarrow2N$) would require an exponentially explosive growth in the size of the
simulating computer. \citep[472]{Feynman1982}
\end{quote}
 \begin{quote}
 The full description of quantum mechanics for a large system with $\R$ particles
is given by the function $\psi(x_{1}, x_{2},\ldots, x_{R}, t)$ which we can call the amplitude to
find at $x_{1},\ldots, x_{R}$, and therefore, because it has too many variables it \textit{cannot be simulated} with a normal computer with a number of elements proportional to
$\R$ or proportional to $N$. \citep[474]{Feynman1982}
\end{quote}
Feynman has expressed a number of closely connected requirements. I will adopt two
separate heuristic principles that I think capture Feynman's intentions closely. Firstly, I will
adopt the heuristic principle that the number of computer elements (memory states) required
to represent the state of a large system of $N$ particles at a single time should preferably grow
at most linearly with $N$, and must not grow exponentially with $N$. Call this the \textit{principle of
linear state complexity}. Secondly, I will adopt the heuristic principle that the number of
computations required to calculate the time-evolution of a large system of $N$ particles, per unit
time, should preferably be proportional to the number of particles, and must not grow
exponentially with $N$. Call this the \textit{principle of linear dynamic complexity}. This second
principle is similar to Feynman's rule of simulation expressed in terms of a volume of space-time because the number of particles in a volume of space is usually roughly proportional to
the volume. Thus if the number of computations required to simulate the $N$ particles per unit
time is proportional to $N$, then the number of computer elements required to simulate a space-time volume will be roughly proportional to that volume. The heuristic principle that both the
principle of linear state complexity and the principle of linear dynamic complexity be
satisfied I will call the \textit{principle of linear complexity}. If this principle is satisfied, then
Feynman's rules of simulation will be satisfied.

I will first investigate whether a discretized linear quantum mechanics, and CCQM,
satisfy the principle of linear state complexity. Consider as Feynman does that the discrete
wave function is represented in the computer by storing each discrete wave function value in
a separate memory state. Following the terminology of the previous section, I will give the
label ``state complexity'' to the number of memory states required to represent the wave
function in this way. If we assume that the wave function in the position basis is used, and
the spin of the $N$ particles represented in the wave function can all be represented by two
components each, then the state complexity will be $2^{N}$ multiplied by the relative volume,
using the result of the previous section. I will assume that the position basis is used for the
wave function, but in general for an arbitrary wave function the state complexity in any of the
standard bases will be of the same order of magnitude as the state complexity for the position
wave function. 

As before, I will suppose that the cell sides corresponding to each particle will be
proportional to the mean de Broglie wavelength of the particle. This will result in a discrete
wave function that preserves the form of the standard continuous wave function to within
some degree of approximation. However, the dependency of the state complexity on $N$ will be
of the same order of magnitude no matter whether the size of the cells are fixed or vary with
the energy of the particles in the wave function. 

As we have emphasized, the relative volume of a wave function will grow
exponentially with the number of particles included in a wave function. For example, if the
position probability distribution of all $N$ particles in a wave function were spread over some
relative volume $v$ in position space then the total relative volume of the wave function in
configuration space will be of order $v^{N}$. Thus the state complexity, the number of memory
states required to represent the discrete wave function, will grow exponentially with the
number of particles, and if there were no critical relative volume, beyond which the relative
volume could not grow, then discrete quantum physics could not be simulated, according to
the principle of linear state complexity. This conclusion is in agreement with Feynman's. 
However, with the rules of wave function collapse provided in CCQM, the relative volume of
any wave function will always remain under the critical value $v_{c}$, so that for any collection of
$N$ particles, which may be represented by a single wave function or several separate wave
functions, the relative volume of the whole system will be at most $v_{c}N$. Thus although the
growth in the relative volume of a system of $N$ particles with $N$ is exponential as long as the
system remains represented by a single wave function with relative volume which remains
under the critical value $v_{c}$, this exponential growth will not continue once a larger system of
particles is considered. Recall that there will be some maximum number $M$ such that no
wave function will represent more than $M$ particles. Thus the state complexity of a large
system of $N$ particles will be at most $2^{M}v_{c}N$. If we consider the dependence of the state
complexity on $N$ for large values of $N$, we see that the growth in the state complexity becomes
at most linear in $N$. Thus the growth of the state complexity of a large system will be at most
linear and the state of any system will be able to be represented on a universal computer, as
long as the number of memory states available is greater than $2^{\M}v_{c}$ for each particle in the
universe. Thus CCQM satisfies the principle of linear state complexity, whereas linear
quantum mechanics cannot satisfy this principle, even if it is discretized.

Now consider the principle of linear dynamic complexity. We wish to investigate
whether a discretized linear quantum mechanics, and CCQM, satisfy the principle of linear
dynamic complexity. Like Feynman, I will suppose that time is discretized, and I will suppose
that the state of the system is computed from the state of the system at the immediately
preceding instant of time, or perhaps the preceding few instants of time. As before, I will
define the \textit{dynamic complexity} of the system as the amount of computation required to
compute the evolution of the system's discretized state per unit of time, using the equations of
motion concerned.\footnote{In complexity theory, a mathematical problem of $N$ parameters whose solution requires an amount of computation that is exponential in $N$ $(k^{N}$ for some $k$) is known as an intractable problem, since it is not practically feasible to solve these problems within a reasonable time when $N$ is large \citep{GaryJohnson1979}. A tractable problem of $N$ parameters is one with a solution that requires at most an amount of computation that is a polynomial function of $N$. Thus the number of computations must be less than $KN^{R}$, where $K$ and $\R$ are constants. The number of parameters of a problem is defined as the number of free variables needed to define the problem. Consider the problem of calculating the evolution of the discrete wave function one unit of time with time, given the initial wave function and the equations of motion. The number of computations needed to solve this problem is the 
dynamic complexity. The number of free variables needed to define this problem precisely would be approximately equal to the state complexity of the wave function, which is much larger than the number of particles that the wave function represents. Thus the requirement of linear dynamic complexity is much stronger than the requirement that the evolution problem is tractable. Thus if the principle of linear dynamic complexity is satisfied then the evolution problem will certainly be tractable. \citet{KreinovichVazquezKosheleva1991} consider the different problem of predicting in quantum mechanics future experimental results given current experimental results: they conclude that this problem in standard quantum mechanics is intractable (NP-hard).} I am assuming that the computation is to be carried out by a classical
universal computer.

As an example, consider a system of $N$ particles governed by the Schr\"{o}dinger
equation, and represented by a single wave function. Following the argument of the
calculation of the dynamic complexity from the previous section, the overall number of
computations involved in calculating the evolution will be at least of order $Nk^{N}$ times the
relative volume, where $k$ is some constant. But the relative volume itself increases
exponentially with $N$, of order $v^{N}$, so the computational complexity will rise of order $N(kv)^{N}$. 

This does not significantly alter the conclusion that the relative volume gives an order
of magnitude measure of the dynamic complexity, and that the dynamic complexity will rise
roughly exponentially with $N$, where $N$ is the number of particles represented by the wave
function. Thus the evolution of the wave function with time is not able to be simulated,
according to the principle of linear dynamic complexity, as long as the number of particles
per wave function is unlimited, as in standard linear quantum mechanics. This conclusion is
in agreement with Feynman's. However, if the relative volume is limited below some
maximum value, and the number of particles per wave function is finite, as in CCQM, then
there will be some maximum number of computations $K$ which is sure to exceed the dynamic
complexity of any wave function. Thus for any system of $N$ particles the dynamic complexity
of the wave functions will be at most $N K$, growing at most linearly with $N$. Thus the system
can be simulated, in accordance with the principle of linear dynamic complexity. Thus it
seems that CCQM satisfies the principle of linear dynamic complexity as well as the principle
of linear state complexity. 

The situation is more complicated than this, because we have not yet fully considered
the dynamic complexity of systems. We have considered the dynamic complexity of each
separate wave function, but we have not yet considered the dynamic complexity involved in
computing the interaction between separate wave functions. I have suggested that one way
in which the interaction with external wave functions could take place is by time-dependent
``external'' potentials, described in the Schr\"{o}dinger equation for each wave function. These
external potentials would vary with time in response to the evolution of other wave functions. 
We have not yet considered how the evolution of these external potentials could be
calculated. As a constraint on possible models for the evolution of these potentials, we could
make the assumption that the dynamic complexity involved in computing these potentials
must be small enough so that the dynamic complexity of any $N$ particle system can be
simulated according to the linear time-evolution principle. This is a non-trivial constraint. 
The limitation on the size of wave functions is the \textit{minimum} modification needed to standard
linear quantum mechanics if quantum mechanics is able to be simulated. Most likely other
modifications to linear quantum mechanics as well as those already introduced in CCQM
would be required if this goal were to be met. However, I argue that the modifications to
quantum mechanics introduced in CCQM represents a step in the right direction if Feynman's
constraints on physical laws are to be fully adhered to.

As noted earlier, in quantum field theory interactions between particles are mediated
by ``exchange'' particles. Since quantum field theory is a more fundamental theory than non-relativistic quantum mechanics, it would seem that a theory that made claim to represent
reality should represent interactions via exchange particles rather than the interaction
potentials of non-relativistic quantum mechanics. The interactions which are represented as
``action-at-a-distance'' in non-relativistic quantum mechanics would, under this conception, be
converted to local interactions between particles' wave functions. For example, exchange
particles carrying interactions over long distances, such as photons and gravitons, could be
represented by their own separate wave functions propagating between the interacting wave
functions. The carrying of forces by exchange particles would most likely be a blessing from
a computational point of view, since rather than needing to make the complicated calculation
of interaction potentials based on the state of systems at various distances from some given
system, the strength of a force between two objects would depend simply on how many
exchange particles happen to be exchanged between these objects. While the use of exchange
particles increases the number of particles involved, the number of computations per particle
that would need to be carried out to ``simulate'' the evolution of the world would most likely
be reduced. Thus we see again that the use of the computational heuristic can be instructive:
it suggests that rather than reality being based on action-at-a-distance forces, it should be
based on local forces carried by exchange particles, just as quantum field theory seems to
suggest occurs in nature. Note that while the interactions will be local in the sense that wave
functions will be influenced directly only by other wave functions nearby in position space,
the interactions will not be local in the sense of influences only propagating at less than the
speed of light, because the exchange particles, sometimes called virtual particles, can travel
faster than light.

\chapter{Entropy and the Arrow of Time }
\label{entropy_time}

\section{The wave function entropy \label{entropy} }
I suggest that the relative volume of a wave function in configuration space should not only
be a measure of complexity of a quantum system, but also a measure of entropy of a single
quantum system. I propose the entropy of a wave function is proportional to the logarithm of
the relative volume of the wave function:
\begin{align}
S &= k \ln v
\end{align}
where $v$ is the relative volume in configuration space, and $k$ is Boltzmann's constant. I will
refer to this measure of entropy as the \textit{wave function entropy}. The standard von Neumann
definition of entropy in quantum mechanics is given by 
\begin{align}
S &= - k \Tr(\rho\ln\rho)
\end{align}
where $\rho$ is the density matrix of the system. This measure of entropy is also sometimes
known as the quantum ``fine-grained'' entropy \citep[162]{Davies1974}. This quantity is zero if $\rho$
represents a pure state. The density matrix of any system which is represented by a single
known wave function is a pure state, so the von Neumann entropy of any wave function is
zero. (Strictly speaking, the expression ``pure state'' refers to an ensemble of identical wave
functions rather than a single wave function.) Thus the wave function entropy will have
values quite different from this quantum ``fine-grained'' entropy.

The definition of wave function entropy is reminiscent of the microcanonical
definition of ``coarse-grained'' entropy, 
\begin{align}
S &= k\ln \Omega
\end{align}
where $\Omega$ is the number of microscopic configurations compatible with a given macroscopic
state. We can also consider the relative volume $v$ to be a measure of a number of
configurations. These are configurations implicit in the wave function itself. Consider a
single cell at a point $\{\bx_{1}', \bx_{2}', \ldots, \bx_{N}'\}$ in configuration space. The magnitude of the wave
function squared at this point in configuration space gives the probability that the $N$ particles
are found on measurement with the first particle at $\bx_{1}'$, the second at $\bx_{2}'$, and so on. Thus each
cell represents a different configuration of the $N$ particles, so $v$ is the number of possible
configurations the $N$ particles can be found in upon measurement, counting one configuration
per cell. 

The number of configurations $\Omega$ figuring in the definition of coarse-grained entropy
corresponds in classical mechanics to a relative volume in phase space rather than a relative
volume in configuration space. In the case of classical physics, phase space is divided into
cells of arbitrary size in order to work out the number of microscopic states compatible with a
given macroscopic state. In the case of quantum mechanics, in systems where there is a
corresponding classical system, in the correspondence limit the number of quantum states is
given approximately by dividing the corresponding phase space into cells of size $\h^{3N}$, for an $N$
particle system, where $\h$ is Planck's constant. 

In defining the relative volume, I suggested the size of the cells in configuration space
could be dependent on the particle and its energy, and I suggested that the cell size might be
proportional to the wavelength of the particle. In terms of a discrete wave function, in order
for a spatially discrete wave function to give rise to empirical predictions indistinguishable
from a continuous wave function, closer spacing of the cells would be required the more
rapidly the wave function changes with distance. In general the phase changes more rapidly
than the magnitude with distance, and the rate of change of phase with distance is
characterized by the de Broglie wavelength of the particle. From this point of view one could
argue that in a discrete physics the length of the cell-sides in each of the dimensions
corresponding to a particle should be some fraction of the wavelength of that particle. Thus
take the length of the three sides of a cell in configuration space corresponding to a single
particle to be proportional to the mean de Broglie wavelength of the particle. The mean de
Broglie wavelength $\lambda_{i}$ of particle $i$ is given by
\begin{align}
\lambda_{i} &= \frac{\h}{\p_{i}}
\end{align}
where $\p_{i}$ is the mean magnitude of the momentum of the particle, and $\h$ is Planck's constant. 
Thus we suppose that the cell side $a_{i}$ is some constant $\beta$ multiplied by the de Broglie
wavelength: $a_{i} = \beta\lambda_{i}$, where $\beta$ is a number between 1 and 0. This means that for a single
particle wave function the volume of a cell is proportional to $\lambda^{3}$, where $\lambda$ is the mean de Broglie wavelength for that particle. 

Consider a wave function of $N$ particles, each spread over a volume $V$ in position
space. Neglecting any symmetry considerations, the relative volume in configuration space
will be of order
\begin{equation}
\frac{V^{N}\p_{1}^{3}\p_{2}^{3}\ldots{}\p_{N}^{3}}{(\beta\h)^{3N}}
\end{equation}
This relative volume in configuration space can also be regarded as a relative volume in phase
space, since it is proportional to a product of a $3N$ dimensional configuration space volume
and a $3N$ dimensional momentum space volume, divided by a factor of ($\beta\h)^{3N}$: phase space is
divided into cells some fraction of volume $\h^{3N}$. Thus as long as the cell size is taken as
proportional to the de Broglie wavelength the relative volume is equivalent to a relative
volume in phase space. This lends added credibility to the claim that the logarithm of the
relative volume is a measure of entropy.

Note that if the factor $\beta$=1, this would give closest numerical agreement with the
quantum coarse-grained entropy, since then phase space is divided into cells some of volume
$\h^{3N}$. This is certainly one option. Although from the point of view of discrete physics one
would expect the actual cell size to be a fraction of the wavelength rather than the wavelength
itself, it is reasonable that in the definition of an important physical quantity that the constants
involved be characterized by the state of the system itself. If the wave function were truly
discrete, with wavelength some fraction of the de Broglie wavelength, then in the case that
$\beta$=1 the relative volume would not be the actual number of fundamental cells occupied by the
discrete wave function, but instead a theoretical quantity based on a cell size containing some
fixed number of fundamental cells. In the case of a truly continuous wave function, since
there is no scale provided external to the quantum mechanical system, it would be desirable
for the cell size to be characterized by a length that is intrinsic to the state of the system, and the
particles in that system. The de Broglie wavelength is one quantity that is characteristic of a
particle in quantum mechanics. Furthermore, I suggest that a single de Broglie wavelength is
a natural unit of separation between parts of the wave function, and so this would be a
reasonable choice of cell size. The Compton wavelength is another length that is
characteristic of a particle, but this length is energy independent, so does not necessarily
characterize the rate of change of the wave function with distance. Furthermore, the Compton
wavelength of a zero-mass particle is infinite, so this length is not suitable as a cell size for
photon and neutrino wave functions. Thus there is some argument for taking the de Broglie
wavelength itself as the cell size, for choosing $\beta=1$, but as long as the cell size is some
fraction of the de Broglie wavelength then the relative volume will be equivalent to a volume
in phase space, and a good candidate for being a measure of entropy. For reasons of
notational simplicity, in what follows I will assume that $\beta=1$, but the entropy calculation
would be very similar whichever value were chosen, and would differ only by a factor of $\beta^{3N}$.

This raises the question whether the numerical magnitude of the wave function
entropy would be equal to that of the corresponding coarse-grained entropy, in cases where
they can both be defined. This is a question I will not consider fully in this thesis. I will
consider briefly just one example here, that of the ``classical'' ideal gas.

Consider an ``ideal'' gas in a box of volume $\V$. The molecules are assumed to be very
rarely interacting, so they can be represented by separate single-particle wave functions. At
equilibrium each wave function will have expanded to fill the entire box. I will assume in
this case that the expectation value of the spacing of the molecules is great enough so that
there will be no quantum statistical effects operating---we have a ``classical'' gas, so the wave
function of each molecule can be considered separately.

Consider the wave function of a single molecule. The wave function magnitude will
rapidly decay at the boundaries of the box, so the volume where the wave function is above $ \f_{0}$
will be approximately $\V$, the volume of the box. The relative volume of molecule $i$ will be
proportional to
\begin{align}
\frac{\V}{\lambda_{i}^{3}} &= \frac{Vp_{i}^{3}}{\h^{3}}
\end{align}
Consider two separate wave functions of distinguishable particles. The total number
of possible configurations of the particles represented by them will be the product of the
number of possible configurations for each wave function, the product of the relative
volumes, since for each configuration of the particles in one wave function there will be the
full range of possible configurations for the particles represented by the other wave function. 
Thus the entropy of two separate wave functions of distinguishable particles is the sum of the
entropies of those wave functions.\footnote{Note that I have defined the entropy of a \textit{wave function} to be the logarithm of the relative volume of that wave function. The entropy of a \textit{system} is not in general just the logarithm of the relative volume of that system, because if the system contains two or more separate wave functions, the entropy is the logarithm of the \textit{product} of the relative volumes of the wave functions, whereas the total relative volume of the system is the \textit{sum} of the relative volumes of the separate wave functions.}

For $N$ identical gas particles the total number of configurations will be proportional to
the product 
\begin{equation}
\frac{V^{N}\p_{1}^{3}\p_{2}^{3}\ldots\p_{N}^{3}}{\h^{3N}N!}
\end{equation}
This is the product of the relative volumes of each particle, divided by $N!$. The $N!$ term is
present because $N!$ of the configurations on measurement will be identical, only with the
identical particles permuted. These should not be double counted. The lack of difference
between configurations is expressed in standard quantum theory through the symmetrization
of the ``product'' wave function of the whole gas that can be constructed from the individual
wave functions of the gas particles. 

Thus we see again the entropy is the logarithm of a volume in a $6N$ dimensional phase
space, divided by the volume of cells of size $\h^{3N}$, although in this case a factor of $N!$ that has
been included as a result of supposing that all the particles are identical.

Suppose now that the gas is in equilibrium with a heat bath of temperature $T$. Under
these conditions the mean energy of the particles in the gas is equal to $\varepsilon = \frac{3k T}{2}$. Making the
approximation that all the wave functions have the same energy $\varepsilon$, and using $\varepsilon = \frac{\p^{2}}{2\m}$, the
total number of configurations of the particles in the gas is of order
\begin{equation}
\frac{V^{N}(3\m k T)^{\frac{3N}{2}}}{\h^{3N}N!}
\end{equation}
If the volume of the box is increased from $V_{i}$ to $V_{f}$, keeping the gas at a constant
temperature, then the wave functions of each gas molecule will expand to fill the box, and the
entropy of the gas will increase as
\begin{equation}
k N\ln\left(\frac{\V_{f}}{\V_{i}}\right)
\end{equation}
And if the temperature (mean kinetic energy) of the particles is increased from $T_{i}$ to $T_{f}$, then
the entropy will increase as
\begin{equation}
kN\ln\left(\frac{T_{f}}{T_{i}}\right)^{\frac{3}{2}}
\end{equation}
This is in agreement with the coarse-grained entropy. This agreement with the coarse-grained
entropy in this single case is encouraging.

I have so far considered only the contribution to the entropy from the translational
degrees of freedom of the particles---I have not considered the contributions from the internal
degrees of freedom. One would have to consider the wave functions of the particles in the
molecules of the gas to do this, and the contribution to the entropy from the spin of the
particles. In this way a wave function entropy could be defined that included contributions
from the spin as well as the position wave function. We have seen that the wave function
entropy is characterized by the number of magnitudes that characterize the discretized wave
function in the position representation. The spin state of a particle can be characterized by
two magnitudes in any spin representation, so the contribution to the entropy from the spin
will increase the overall entropy of a single-particle system by a factor of two, and the entropy
of an $N$-particle system by a factor of $2^{N}$. The idea of collapse being triggered by the wave
function growing to a critical relative volume could be replaced by the idea that the wave
function collapses upon reaching a certain critical wave function entropy.

\section{The evolution of the wave function entropy with time }
Quite apart from any numerical considerations, the most important property an entropy
concept must have is that the entropy of isolated systems will for the most part increase in
magnitude with time. The relative volume in configuration space has the potential to fulfil
this requirement because wave functions will always, or virtually always, expand in volume
in configuration space rather than contract, other than when the wave functions suddenly
collapse after reaching the critical volume. 

The question arises why the relative volume should virtually always increase in time,
displaying an irreversible behaviour when the equation that governs the evolution of the wave
function is reversible in time. The answer is that although the Schr\"{o}dinger equation itself is
reversible in time, the evolution of the wave function is not always governed by this equation---the collapse of the wave function introduces a time-irreversible element, since the sudden
collapses only occur in the forwards-time direction. These collapses reduce the volume
instantaneously, or virtually instantaneously, and it is the gradual increase in the relative
volume of wave functions that occurs for the great majority of the time.

Consider the evolution of wave functions from the time of the Big Bang onwards. 
According to CCQM, an instant after the universe began the universe consisted of a collection
of wave functions, all these wave functions having relative volume at or below the critical
value. If at the time of the Big Bang the universe began as a single wave function, or a
collection of wave functions, some of which had relative volume greater than the critical
value, then those wave functions would have instantly collapsed, resulting in a collection of
wave functions all with relative volume less than $v_{c}$.

Some or all of these primordial wave functions may have been initially contracting in
volume while others may have been expanding. Any wave functions that were initially
contracting, and that remained isolated from other wave functions, would soon reach a
minimum relative volume and thereafter expand in volume \citep{Bradford1976}. These wave
functions, and those wave functions that were initially expanding, would soon reach the
critical volume and suffer a collapse. As long as the collapse processes were ones that merely
localized the wave functions, or split them into separate wave functions, and did not introduce
radical phase shifts in addition, then the wave functions would continue to expand after the
collapse took place. The wave functions thereafter would continue to expand, to interact and
to combine into larger wave functions, periodically collapsing and splitting into separate
wave functions, before resuming their expansion. Thus there is no need to assume any
special initial conditions at the beginning of the universe---even if many or all of the
primordial wave functions were initially contracting, the subsequent evolution of all wave
functions would be to always expand after collapsing.

It might be thought that a wave function that was initially expanding could be
reflected back onto itself by interaction with another system surrounding it, resulting in a
wave function that was contracting again. However, for most systems there will be a non-zero transmitted wave as well as a reflected wave, so that the total volume of the wave
function will continue to expand after the reflection takes place. Furthermore, when the
expanding free wave function interacts with other systems around it the wave function will
form a combined wave function with those wave functions it interacts with, increasing the
total relative volume. Thus even if the surrounding system presents a sufficient barrier to
reflect the particle totally, so that there is no wave function magnitude above the threshold
value $ \f_{0}$ shortly beyond the barrier, the increase in relative volume caused by a many-particle
wave function forming will more than compensate for any reduction caused by the probability
distribution of the original free-particle contracting on itself again. Eventually the reflected
wave will decouple from the larger system it interacted with, but that decoupling will only
occur when the critical volume is reached and the wave function collapses. After that time, if
reflected by just the right shaped barrier, the single-particle wave function may decrease in
volume for a short time (while it remains free of interaction from other particles), but such a
perfect geometric arrangement will be very rarely met.

As it happens, space itself has been expanding since the beginning of this universe, but
that is not an essential condition for the fact to hold that wave functions always expand---this will
continue to hold if the expansion of the universe eventually reverses, and it would have held even
if the universe had always been contracting from some initially large position-space volume,
as long as the same law of collapse were to hold.

There is a paper by \citet*{MisraPrigogineCourbage1983} ``Lyapounov variable -
entropy and measurement in quantum mechanics'' that proves that there can be no dynamical
variable of a system in standard linear quantum mechanics that monotonically increases with
time. I claim the relative volume of a wave function monotonically increases with time. At
least I claim that it increases with time between each collapse of the wave function, before the
critical volume is reached, except perhaps in some rare cases, and it is during this time
between collapses that the wave function obeys the evolution law of standard linear quantum
mechanics (other than when it combines with other wave functions). The relative volume
does not run foul of the proof because it is not a dynamical variable as \citet{MisraPrigogineCourbage1983}
define it: it is not a function of the $\bx$ and $\bp$ operators. 

In the same article, \citet{MisraPrigogineCourbage1983} consider variables (superoperators) defined for
the Liouvillian formulation of quantum dynamics, which is a formulation that deals with
ensembles of similar systems, and show that variables $\M$ with the required property do exist
for certain systems, namely systems not confined to some volume---the Hamiltonian must
have a continuous and unbounded spectrum. They call these operators ``entropy
superoperators.'' They note that a system of a single free particle in an infinite volume admits
of an entropy superoperator, because the Hamiltonian $\frac{\bp^{2}}{2\m}$ has a continuous spectrum. \citet[690]{MisraPrigogineCourbage1983} remark:

\begin{quotation}
\noindent But one does not expect any irreversibility for this system. The spurious
``irreversibility'' associated with the evolution of the free particle in the infinite
volume is just a reflection of the fact that initially localized wave packets
spread out and get rarefied with the particle, eventually ``escaping to infinity.''

Naturally, one would like to distinguish such spurious cases from more
physical irreversibility by requiring $\M$ to satisfy supplementary physically
motivated conditions.
\end{quotation}
My attitude to this is very different---in my view, this is not at all a ``spurious case.'' In my view, all irreversibility has its origin in such spreading of wave packets, together with
instantaneous collapses, both in position space, for single particle systems, and configuration
space, in many particle systems. It is true that the equations of motion that produce this
spreading are formally reversible and that it is the collapse law that is formally irreversible. If
one reverses time then one would find that localized wave functions expand instantaneously,
whereas in the forward time direction wave functions localize instantaneously---the collapse
law is time irreversible. However, it is also true that if one reversed time one would see wave
functions gradually localizing after instantaneous expansions, whereas in the forward time
direction we see wave functions gradually expanding after instantaneous localizations. Once
a wave function is localized, it will always spread gradually according to the equations of
motion. Thus the fact that we see wave functions expanding gradually in the forward time
direction indicates that the sudden collapses occur in the forward time direction. We should
not be embarrassed to describe the gradual spreading of wave functions as irreversible,
because these gradual expansions occur only in the forward time direction, with few
exceptions, and the gradual contractions of wave functions are virtually never seen in the
forward time direction. Thus we can pick out the forward time direction as the direction in
which these gradual expansions occur, as well as the instantaneous collapses. It is true that
under very rare conditions geometrical circumstances are met where a wave function may
gradually contract for a short time, but these cases are so rare as to be insignificant in number
compared to the ubiquitous gradual expansions after instantaneous collapse. 

\citet{MisraPrigogineCourbage1983} say that in the Liouvillian formulation of standard linear quantum
dynamics, they can only find ``entropy superoperators'' for systems in an infinite volume. 
This is because for systems confined to a finite volume, having a finite energy spectrum, the
state of the system will eventually return to any state that it was in previously, after the
Poincar\'{e} recurrence time. For any system of a large number of particles this recurrence time
will be extremely long, much longer than the age of the universe. In CCQM, the collapses
would probably prevent the return to the original state, but even if they did not, before the
Poincar\'{e} recurrence time could elapse the system would undergo a huge number of collapses
of the wave function, followed by gradual expansions. This reinforces the point I just made
that the gradual contractions of the wave function will be extremely rare compared to the
gradual expansions after instantaneous collapse. Thus there is no need to restrict the relative
volume as a measure of entropy in CCQM to systems that can expand in an infinite volume. 
The wave function entropy will gradually increase between collapses for all systems, with
rare exceptions, and to allow rare exceptions is only appropriate for a quantity that is defined
in quantum theory, which is a probabilistic theory. 

Thus I argue that because the relative volume has the property of almost always
increasing when the system is governed by the deterministic equations of motion, it has one
of the hallmarks of the entropy concept (when governed by these equations), and from this
point of view alone it deserves to be labelled as an entropy of some kind. The label I have
chosen is the ``wave function entropy.'' In the next section, the relationship of the wave
function entropy to other measures of entropy will be examined.

\section{The relationship between the wave function entropy and other entropy measures}
Before comparing the wave function entropy with other measures of entropy, I will examine
some of the many philosophical problems with the ``coarse-grained'' definition of entropy. As
noted earlier, the microcanonical definition of ``coarse-grained'' entropy is the logarithm of the
number of possible microscopic states compatible with a certain macrostate. Thus the
definition of this physical property refers to an ensemble of possible states. Possible states,
being merely possible, cannot be causally efficacious in themselves, so it is undesirable that the
definition of the property of an actual physical system makes reference to them. The
reference to this ensemble has the appearance of a subjective element, reflecting our lack of
knowledge of the actual physical state of the system, apart from the macrostate. Secondly, the
definition makes essential reference to a macrostate, whereas one would hope that if quantum
mechanics is complete, all properties that play a role in physical theory would be reducible to
properties of microscopic states. Thirdly, for the purpose of defining the ensemble, the energy
of the state cannot be specified precisely but instead an arbitrary range of energies around a
given energy must be specified in order to give a sensible result. If the energy range were
zero, then the entropy would be either zero or merely the energy redundancy of a single
quantum state. The value of the entropy is relatively insensitive to the choice of energy range,
within certain limits, but there is no principled way to choose a particular energy range. The
microscopic state itself does not dictate this choice, so this again seems a subjective factor,
reflecting our ignorance about the energy of the system. 

The von Neumann definition of quantum ``fine-grained'' entropy has also been
criticized by \citet[187--188]{Espagnat1976} on the grounds that it is subjective, since the
``mixtures'' used in its definition are usually interpreted as merely reflecting our ignorance of
what pure state the system is in. In cases where the definition is applied to what d'Espagnat
calls ``improper'' mixtures, the identification of the mixture depends on selectively ignoring
part of the full pure state, which can also be seen as a subjective element in the definition. 

For reasons such as these, some people, including \citet[4732]{Zurek1989}, have suggested
that we should attempt to find a ``new definition of entropy which does not use this ensemble
strategy and which can be applied to the individual microstates of the system.'' In other
words, we wish to find a new definition of entropy which is a property of the actual state of
the system rather than a property of an ensemble. The wave function entropy satisfies this
requirement. The definition of wave function entropy includes no reference to macrostates or
to an ensemble of possible states; it is a property of the one actual quantum system that the
entropy is assigned to. It is a fully objective property of the system. No arbitrary energy
range need be referred to.

\citet[4731]{Zurek1989} points out that it has generally been thought impossible to define
the entropy of a single microscopic state of a system. Rather, one had to consider an
equilibrium ensemble of similar states that share the same macroscopic properties, as
described above. One reason that it may have been thought impossible to define the entropy
of a single wave function may have been the proof of \citet{MisraPrigogineCourbage1983} that no dynamic
variable of a single wave function can monotonically increase with time while having its
evolution governed by the Schr\"{o}dinger equation. As I argued earlier, the wave function
entropy is able to evade this proof, since it does not fit their definition of a ``dynamical
variable,'' and because the collapse of the wave function occurs every so often, interrupting
the evolution according to the Schr\"{o}dinger equation and reducing the wave function entropy. 
The wave function entropy can always increase while the system is governed by the
Schr\"{o}dinger equation because the wave function entropy is continually ``reset'' to a lower
value by the collapse of the wave function.

On top of these conceptual problems with the definition of entropy in statistical
mechanics, various questionable assumptions are traditionally made in order to obtain the
correct dynamical evolution of the entropy, in agreement with the entropy of
thermodynamics. Accounting for the time irreversible character of thermodynamics, the
increase of entropy with time, has been a difficulty in statistical mechanics, because an
underlying reversible physics has generally been assumed, either Newtonian mechanics, in the
case of classical statistical mechanics, or linear quantum mechanics, in the case of quantum
statistical mechanics. A well-known criticism of the von Neumann definition of entropy is
that if an isolated system is assumed to evolve according to standard linear quantum
mechanics, the von Neumann entropy of the system will remain constant, whereas the
thermodynamic entropy of the system will increase in general. 

Arguments have been put forward in order to attempt to derive the increase in entropy
with time from this underlying reversible physics, but not surprisingly, these arguments are
very suspect. Consider one of these arguments. \citet*{Albert1994} considers an example in which
two macroscopic bodies of differing temperature are brought into thermal contact with each
other, and after ten minutes, the temperature-difference between those two bodies decreases. 
Albert discusses the explanations of that decrease provided by statistical mechanics.

The first statistical mechanical explanation he discusses runs as follows. The initial
state, described in macroscopic terms, is compatible with a huge number of possible
microstates; and the overwhelming majority of those compatible microstates are ones which
would evolve, over the next ten minutes, towards states in which the temperature-difference
between those two bodies is smaller. Albert calls these microscopic states the \textit{normal} states,
and the microscopic states which would evolve towards states in which the temperature-difference is larger he calls the \textit{abnormal} states. The crudest form of the explanation begins
from the fact that we don't know which one of the microscopic states obtains, and says that
since we do not know which microscopic state obtains we should assign each state an equal
probability of obtaining. Therefore we should believe that it is overwhelmingly likely that the
microscopic state of the system is a normal state, so we should expect that the temperatures of
the two bodies should approach one another over the next ten minutes. Albert points out that
this argument is unsatisfactory because it pretends to establish something about what
microstate the system is likely to be in from our ignorance of what microscopic state the
system is in. 

A more sophisticated explanation relies on the notion of perturbations. We can define
a metric on the set \{\C\} of states compatible with the original macroscopic state. In the
classical case the metric is the Euclidean metric on phase space and in the quantum-mechanical case it is the Hilbert-space metric generated by the absolute square of the inner
product. Using this metric we can capture what it is for two microstates to be microscopically
similar, and what it is for two microstates to be within one another's microscopic
neighborhoods. The normal microstates vastly outnumber abnormal microstates in every
microscopic neighborhood of \{\C\}, even within the microscopic neighborhoods of every one
of the abnormal microstates in \{\C\}. This means that the property of being a normal state is
extremely stable under small perturbations of the state, and the property of being an abnormal
state is extremely unstable under small perturbations of the state. Thus if the two bodies that
make up the state were frequently, microscopically and randomly perturbed, then the
temperatures of those two bodies would be overwhelmingly likely to approach one another no
matter which one of the microstates in \{\C\} the system was in. If such perturbations did take place then we could objectively explain why the temperatures of
the bodies approach one another.

Consider also an example provided by \citet{Penrose1990}. He asks us to consider all the
microscopic states of a gas in a box, each represented in classical physics by a point in phase
space. Divide phase space up into regions corresponding to states that are identical from the
macroscopic point of view. The points in one particular compartment represent physical
systems that are deemed to be identical with regard to macroscopically observable features. 
Such a division of the phase space into compartments is a ``coarse-graining'' of the phase
space. Most of the phase space will correspond to states in which the gas is very uniformly
distributed in the box and in thermal equilibrium. Suppose the system begins at some time in
a special macroscopic state, where all the gas is gathered in one corner of the box. There will
be many microscopic configurations in the compartment of the phase space corresponding to
this situation, each macroscopically indistinguishable, but the volume of this compartment of
the phase space will be very much smaller than the compartment corresponding to thermal
equilibrium. 

Suppose we assume that the gas suffers random perturbations, so that the point in
phase space wanders randomly in phase space; the point will most likely enter a region of
much larger phase space volume, corresponding to the macroscopic states where the gas is
spread out over a larger volume within the box. The phase space point will keep entering
larger and larger compartments in phase space, corresponding to the gas being spread further
and further out in the box, where the volume of each subsequent compartment will be
enormously larger than the preceding one. Thus, as long as the motion through phase space is
random, the point will soon enter the largest volume, corresponding to thermal equilibrium,
and it is virtually assured that the state will not return to any of the smaller volumes in any
reasonable time. (The system would eventually return to the original state, but the time of
return, the Poincar\'{e} recurrence time, for any typical example would be far longer than
the age of the universe, so this possibility can be ignored.) Thus we see that the entropy of
the system, which is the logarithm of the volume of the appropriate compartment in phase
space (multiplied by Boltzmann's constant), will, with very high probability, increase steadily
until equilibrium is reached.

\citet{Albert1994} makes the point that for explanations such as these to work, the
perturbations must be genuinely random. In the literature it is often suggested that these
perturbations can be seen as arising from the interactions of the system with its environment. 
Albert argues that whatever perturbations arise from interactions with the environment, if we
assume the equations of motion of the system and its environment are deterministic, then
these interactions will be ``random'' only in the sense that nobody happens to know what those
perturbations are. Without altering the deterministic equations of motion of linear quantum
mechanics to introduce a genuine random element into physics, the objectivity that is claimed
for this explanation will not obtain, he argues.

Albert suggests that the ``jumps'' in the GRW theory are precisely the sorts of
perturbations we need. He suggests that as a consequence of the spontaneous
localizations that occur in the GRW theory, every single one of the microstates in \{\C\} will be
overwhelmingly likely, on the GRW theory, to evolve over the next ten minutes into states in
which the temperature-difference between the two bodies is smaller. He says that because of
the radical instability of the abnormal microstates, any of a wide selection of GRW-like
perturbations would be likely to be sufficient, as long as the perturbations were genuinely
random, as well as frequent and microscopic. All of these requirements hold for the GRW
theory, he says. 

Albert points out that it has often been hoped that the time-asymmetry that would be
likely to arise from a stochastic modification of quantum mechanics would explain the time-asymmetry of thermodynamics. However, we cannot assume that the time-asymmetry of a
modified quantum mechanics would be sufficient to explain this: this must be demonstrated
for the proposed modification of linear quantum mechanics. Albert claims that the GRW
theory introduces the right sort of time-asymmetry into nature, the sort that can genuinely
underwrite the time-asymmetry in thermodynamics. He suggests that other collapse theories,
such as the one defended by \citet{Penrose1990}, may not be able to explain the increase in
entropy in cases such as this one.

I see a difficulty with Albert's explanation. \citet[408--410]{Penrose1990} points out that
there is another problem with the kind of explanation that relies on random perturbations of
the microscopic state, a problem not mentioned by Albert. Recall that within the
neighborhood of every microscopic state the majority of states are normal states, so that
whatever microscopic state the system is in, the system under the influence of perturbations is
likely to end up in a normal state, in which the temperature-difference between the bodies
reduces and the entropy increases with time. The problem that Penrose points out is that we
could have equally applied this very same argument in the reverse direction in time. Consider
again two bodies of different temperature in thermal contact, and now ask what are the states
of the system that are most likely to have preceded this. Applying the same argument it
would seem that under the influence of perturbations the system is most likely to end up in a
normal state, in which the temperature difference between the bodies reduces, but this time in
the reverse direction in time. Thus the argument suggests that the temperature difference is
likely to have been smaller in the past and then increased with time. This process, with the
warmer body growing spontaneously warmer, and the colder body colder, is never actually
seen. This time the argument from random perturbations has given us the wrong answer. 

Consider also Penrose's example. As before, consider the state of the gas gathered in
one corner of the box, and consider what its prior states are likely to have been. If we
supposed that the state suffered random perturbations, then we would expect that as we trace
the motion backwards in time, the phase space point would soon be found within phase
space volumes of much larger volume, corresponding to the gas being more spread out. Back
further in time it would be found in the largest volume, corresponding to thermal equilibrium. 
This time we seem to have deduced that given that some gas is gathered in one corner of the
box, the most likely history of the gas prior to this is of the gas beginning in thermal
equilibrium then concentrating itself into the corner of the box. Entropy would spontaneously
decrease throughout this process. This process is never seen: entropy always increases in our
universe. Again the argument from random perturbations has given us the wrong answer. As
Penrose points out, the most likely state prior to the gas being gathered in a corner of the box
is not the gas at equilibrium but the gas being released from behind a partition in the corner of
the box, and the gas spreading out from there.\footnote{Penrose agrees that a theory of wave function collapse could be of benefit in introducing an irreversible element into quantum mechanics and explaining the increase of entropy with time. He also agrees that a new measure of entropy is required for a proper treatment of entropy in quantum mechanics. He suggests that this entropy should be defined in a ``new'' kind of space, differing from the Hilbert space used in the von Neumann definition of entropy and the phase space used in the classical definition of entropy: ``My own opinion is that for the correct theory neither Hilbert space nor classical phase space would be appropriate, but one would have to use some hitherto undiscovered type of mathematical space which is intermediate between the two'' \citep[474]{Penrose1990}. Configuration space, used in the definition of wave function entropy, fits the description of being intermediate between Hilbert space and classical phase space. The state vector 
of 
quantum mechanics is defined in Hilbert space, but the representation of that vector in the position basis is defined in configuration space. Furthermore, as long as the configuration space is divided into cells with side-length proportional to the mean de Broglie wavelength of the particle concerned, a volume in configuration space is equivalent to a volume in phase space. Thus the new measure of entropy I have introduced seems to fit Penrose's requirements. However, Penrose calls for this new measure primarily to help solve a particular problem connecting entropy in quantum theory with the entropy of black holes, and I have not yet tried seriously to tackle this problem to see if the new measure of entropy can help solve it.}

The argument from perturbations as presented so far is unsuccessful, due to the time
symmetry implicit in the reasoning, and the time asymmetry in the intended conclusion. 
Albert points out that the GRW theory is time-asymmetric, but he does not explicitly make
use of this time-asymmetry in his argument from perturbations. Thus we must examine the
GRW collapses more carefully to see if Albert's argument can succeed. The type of
perturbations it introduces into quantum mechanics are not symmetric in time. In the forward
time direction, the perturbations are random localizations of individual particles: the position
probability distribution of a single particle is suddenly localized, and other particles may be
affected as well. In the backwards time direction the GRW perturbations will look like
sudden delocalizations of the particle: its probability distribution will suddenly expand. Can
this time asymmetry in the form of the perturbations rescue Albert's argument? It is hard to
see how it could, since the ``jump'' is a perturbation in both time directions. The most
promising approach would be to suggest that while the perturbations are sufficiently random
in the forward time direction, they are not sufficiently random in the backwards time
direction. Note that the GRW perturbations are not completely random in the forward time
direction, in the sense that they are most likely to occur where the strength of the wave
function before collapse is greatest, but as the time and place of their occurrences are
otherwise random, they may be speculated to be sufficiently random for the purpose Albert
envisions for them: to generate a ``sufficiently random'' walk in phase space (in the classical
description). In the backwards time direction, where and when the sudden expansions occur,
and to what state the expansions result, are unlikely to be determined uniquely by the state of
things ``before'' the expansions take place. Nevertheless, we may speculate that under some
measure of randomness the backwards-in-time perturbations are not sufficiently random. 
However, I cannot see how this kind of argument could \textit{explain} why the backwards-in-time
perturbations should result in a system in a normal state almost invariably ending up in an
abnormal state, despite the fact that the overwhelming number of states in its neighbourhood
are normal states. Thus I cannot see how an argument from perturbations can \textit{explain} why if
we look in the reverse time direction the warmer body gets warmer with time and the cooler
body gets cooler, the gas becomes more localized, or more generally why the entropy in the
past is lower than it is in the present.

There is another problem with Albert's argument that arises if we try to apply it to the
example of the gas in the box. Consider the gas in the corner of the box again. We would
observe the gas to spread away from the corner even if there were only a few thousand, or a
few hundred gas molecules. Indeed, even if there were only a single molecule in the box, and
its wave function were initially physically confined in the corner, then the wave function
would invariably spread away from the corner. Yet each particle will suffer a GRW collapse
on average only once every $10^{9}$ years! We would need to have $10^{16}$ gas particles to be
confident of the gas experiencing one collapse per second, and the effect of that collapse
would be to localize a single particle: the distribution of the remaining particles would be
affected very little. Thus it seems that the GRW perturbations are neither sufficiently
frequent nor of the right kind to explain, via the argument from perturbations, why the gas
expands to fill the box in the forward time direction. It seems to me that this expansion
would proceed without any perturbations of any kind, so we should not look to a continual
supply of perturbations to explain the increase in entropy in the forward time direction, at
least not in this case.

Note that this criticism does not seem to apply to the example that \citet*{Albert1994}
gives, since as he describes it one only needs a single perturbation to the system and thereafter
the deterministic equations of motion will ensure that the entropy increases. Thus one does
not need a continual supply of perturbations, only a few over the ten minute period, say. 
Furthermore, the huge number of particles involved in the two bodies in thermal contact
would ensure that there would easily be a sufficient number of GRW perturbations to cause
the entropy to continue to increase in the forward time direction. 

It is possible that if we looked more closely at the GRW theory and the example of the
gas in the box, then the argument from perturbations may be successful, but the argument
would be an indirect one.\footnote{There exists a general proof of the increase with time of the von Neumann entropy in the GRW model. \citet{Benatti1988} proves that the von Neumann entropy diverges at $t = +\infty$ if the density matrix of the system evolves according to the GRW model. Just how this proof relates to individual cases of the examples under consideration is not clear. The density matrix of a system characterizes a collection of possible states, one of which corresponds to the actual state, and it is not necessarily clear that one can say anything about how a particular system will behave from a statement of how an ensemble of possible states will behave. It is the question of how an ensemble of possible states can be objectively chosen and how and why the behaviour of this ensemble should tell us anything about the behaviour of a particular system that is the main problem that I am trying to address in this section.} On the other hand, if we assume CCQM and concentrate on the
evolution of wave functions in CCQM, then we can explain directly why the gas always
expands to fill the box in the forward time direction. The reason that the gas expands is that
wave functions virtually always gradually expand in the forward time direction, the
distribution in intensity of the wave function becoming more spread out and more uniform
with time. Thus we should expect the wave functions of the gas particles to expand to fill the
box. After the wave functions fill the box and interact with particles in the walls of the box
we would expect some collapses of the wave function to occur. However, since the particle
interacts weakly with the walls of the box, we would not generally expect that the gas particle
wave functions would collapse to a size much smaller than the box. Instead, the wave
functions of the gas particles would become energy eigenstates of the particles-in-the-box
system and the gas would remain in equilibrium, uniformly filling the box. 

Now consider the explanation of the behaviour of the gas in the backwards time
direction, starting with the gas in the corner of the box. In the backwards time direction,
wave functions gradually contract (with the exception of wave functions in energy
eigenstates, which remain static in size), so we can expect that at earlier times the gas will be
even more concentrated in the corner of the box. Thus this argument gives us directly the
correct result in both the forwards and backwards time directions. 

A similar explanation for the expansion of the gas may be possible if the GRW model
of collapse is assumed, since wave functions will expand in the forward time direction after a
GRW ``jump'' occurs. Thus, in the backwards time direction, there will be many wave
functions that contract gradually in size. However, the argument is not as straightforward as
in the case of CCQM because the spontaneous localizations occur so rarely to each particle,
and have limited effects on other particles. On the other hand, the argument for the increase
in entropy with time in CCQM, which is relatively straightforward in the case of the
expanding gas, is not directly applicable to Albert's example of the bodies in thermal contact. 
It is possible that an argument from perturbations could be utilized to demonstrate the
increase of the ensemble entropy, similar to Albert's argument using the perturbations
introduced by GRW, since the perturbations introduced by CCQM are fairly similar to those
introduced by GRW and even more frequent. The probability distribution of the centre of
collapse in CCQM is governed by a similar law to the probability distribution of the centre of
collapse in GRW, so it could be argued, along similar lines to the argument I suggested in the
case of GRW, that the perturbations will be sufficiently random in the forward time direction,
but not in the backwards time direction. However, merely showing that an ensemble entropy
will increase with time, in accordance with the thermodynamic entropy, would be insufficient,
in my view, to remove all the philosophical problems of statistical mechanics, since there
would remain the question of the objectivity of the concept of the ensemble entropy defined
in statistical mechanics. Thus I will return to the question of the status of the ensemble
entropy and the relation of the wave function entropy to the ensemble entropy and the
thermodynamic entropy.

What is the relationship between the wave function entropy and the ensemble entropy
of statistical mechanics? The most radical program would be to derive the thermodynamic
entropy directly from the wave function entropy, and to discard the ensemble entropy defined
in statistical mechanics altogether. This is in some ways an attractive idea. The entropy of
thermodynamics is taken to be a property of a single macroscopic system. The wave function
entropy is also taken to be a property of a single physical system: the actual collection of
particles that make up that macroscopic system. It would be desirable if the thermodynamic
entropy could be derived from the properties of the microscopic entities that actually make up
that macroscopic system, rather than referring to ensembles of similar systems, as is done in
statistical mechanics. Indeed, I postulate that in every case there is \textit{some} objective property of
the actual system (that preferably has a similar value to the thermodynamic entropy) that
explains why that system has the thermodynamic entropy it does have and why that entropy
evolves the way it does evolve. This wave function entropy can play the role of this property
in the example of the ``classical'' gas given above, because the thermodynamic entropy is only
defined up to a constant, so the size of the cells is not crucial in deriving the value of the
entropy. Furthermore, we have seen that the wave function entropy has the property of
always, or virtually always, increasing in value in time, prior to any collapse occurring, due to
the spread of the wave function. Thus this property of the single wave function can explain
why the thermodynamic entropy increases in value in time, prior to wave function collapses. 

Directly deriving the value of the thermodynamic entropy and the way it evolves with
time seems to work out well in the case of the ``classical'' ideal gas, where the collapse of the
wave function does not play a significant role. In this case, the wave function entropy has the
same value as the thermodynamic entropy (up to a constant factor), but we will see that in
situations where wave function collapses significantly affect the wave function entropy of a
system, the wave function entropy will not have the same value as the thermodynamic
entropy. Furthermore, while the wave function entropy will suddenly decrease when the
wave function collapses, the classical thermodynamic entropy will in general continue to
increase after wave function collapse. Thus it will not be possible to derive the
thermodynamic entropy directly from the wave function entropy in general. However, I hope
that for these systems where wave function collapses occur, the wave function entropy can
help provide objective grounding for the coarse-grained (ensemble) entropy of statistical
mechanics, and the thermodynamic entropy can thus be ``derived'' indirectly from the wave
function entropy in this way.

There is one weak sense in which the quantum coarse-grained entropy might be
grounded in the wave function entropy. Consider the example of the ideal gas in a box at
equilibrium, where we assume that the wave functions of the gas particles have expanded to
fill the box. The wave function entropy increases as the volume and temperature increase
because the higher the volume and temperature the more wavelengths it takes to span the
dimensions of the box. The quantum coarse-grained entropy increases when more
wavelengths are required to span the dimensions of the box because for a given range of
energy, the number of possible states that are eigenstates of energy will be greater. This is
because it will take a very small shift in the energy to shift from one eigenstate to the next
highest eigenstate, for which one more wavelength fits exactly into the box. Here, it is the
relative volume that is the objective property of the wave function of the system that most
directly explains why the quantum coarse-grained entropy has the value it has. Thus the wave
function entropy is a plausible objective property of the wave function to provide objective
``grounding'' for the quantum coarse-grained entropy, at least in this case. In this case the
wave function entropy and the coarse-grained entropy also both have a similar numerical
value, or are proportional in size, depending on what fraction of the de Broglie wavelength
the cell size is. In other cases, where wave function collapse comes into consideration, this
close relationship between the two types of entropy will not hold. Furthermore, the
``grounding'' provided in this case is fairly weak, merely explaining the numerical value of the
ensemble entropy, given the definition of that entropy. The type of grounding I am looking
for is providing objective grounding for the choice of the ensemble that is used, not just
grounding for the numerical size of the ensemble.

Let us now examine in a more general way the relationship between the quantum
wave function entropy and the classical coarse-grained entropy. As mentioned earlier, for a
wave function of $N$ particles, each cell in configuration space represents a possible
configuration of the $N$ particles on ``measurement.'' (If a system consists in several separate
wave functions, the total number of possible configurations for the system will be the product
of the number of configurations of the separate wave functions, unless the system contains
identical particles, in which case care must be taken not to double count the possible
configurations involving identical particles.) Thus there is implicit within the actual quantum
state an ensemble of possible states: each cell represents one configuration in that ensemble. 
In this case the ensemble does not arise because of any lack of knowledge of the system, but
rather the ensemble is objectively implicit in the actual wave function. The elements of the
ensemble represent \textit{potentialities} of the quantum system, potentialities for localized wave
functions to arise in certain configurations if and when the wave function undergoes a
collapse process, or a series of collapse processes. Before any collapse takes place, these
potentialities remain potentialities, and the value of the wave function entropy will closely
match the corresponding classical entropy, in the cases where the classical coarse-grained
entropy match the traditional quantum coarse-grained entropy, such as for the ``classical''
ideal gas. When a collapse occurs, the magnitude of the wave function entropy will no longer
correspond in value to the corresponding coarse-grained entropy, because the collapse will
cause the wave function entropy to instantaneously contract, whereas the coarse-grained
entropy will remain the same. 

However, if we consider the evolution of an ensemble of identical wave functions,
when these collapse processes occur, causing localizations of wave functions, the statistics of
the distribution of the centres of localization in configuration space will match the
configuration space probability distribution of the original wave function. The ``ensemble of
possible states'' implicit in the wave function will be ``transformed'' at the time of collapse to
an ensemble of states in the more usual sense: a collection of possible states that share some
properties with the actual state, and have the property of being states that the system ``could
have been'' in, in some sense of ``could have been.''\footnote{The process of measurement, or wave function collapse, is often described as a transformation from a ``pure state'' (a definite wave function or, more correctly, an ensemble of identical wave functions) to a ``mixture'', which is an ensemble of different possible states. I have not taken this approach, preferring to discuss wave function collapse in terms of the evolution of individual wave functions, but the ensemble description can be recovered from the description in terms of individual wave functions.} This ensemble is the ensemble of
possible states the wave function could have collapsed to, given the wave function before
collapse. It is in this way that one could hope that objective grounding could be found for the
ensemble approach of statistical mechanics. The ensemble chosen would not be a subjective
one, based on a lack of knowledge of the actual state, but an objective one, based on the
causal history of the system; based on the properties of some original wave function which
existed prior to any collapses occurring, and from which the quantum state of the system
evolved.

As an example, consider again the case of the ideal gas in a box, and consider
expanding the walls of the box to astronomical proportions. Then as before the wave
functions of the gas particles would expand, but their volumes would eventually reach the
critical value and collapse, reducing to the collapse volume. After this the volume would
resume growing, then collapse again after reaching the critical volume, and so forth. On the
other hand, the classical coarse-grained entropy would continue to grow relentlessly as the
volume of the box increased, given time for the system to reach equilibrium. 

Consider the wave function of a single particle in the gas which has just reached the
critical relative volume, but has not yet collapsed. If we assume the mechanism of collapse
adapted from GRW, the wave function before collapse gives, approximately, the probability
distribution of the centres of collapse. Each cell contained within the boundaries of the wave
function is a possible place that the Gaussian jump factor could be centred on, so we can
identify an ensemble of possible systems implicit in the wave function: the ensemble of
possible post-collapse wave functions centred at each of these cells. Immediately after the
collapse occurs, this ensemble of possible states is the ensemble of collapse wave functions
centred at cells where collapse \textit{could have been} centred. We could define an ensemble
entropy after collapse using this ensemble of possible states: the logarithm of the number of
states in this ensemble. This entropy would be equal in value to the wave function entropy
before collapse, since the wave function entropy is the logarithm of the number of cells
contained within the wave function, and a collapse could be centred at each of these cells. 
Thus at least we have identified some ensemble entropy that can find objective grounding in a
wave function entropy, even though after the collapse takes place the wave function entropy
will no longer have the same value as the ensemble entropy. This ensemble entropy, after the
collapse takes place, can be viewed as subjective from one point of view, since this ensemble
is the appropriate one to choose if one lacks knowledge of where the wave function collapse
occurred, but it is objective from another point of view, since the ensemble can be picked out
by reference to the original wave function. As time goes on one could take as the reference
ensemble the ensemble of collapse wave functions centred on cells of the original wave
function the way it \textit{would have} evolved if there had been no collapse. 

The ensemble entropy defined by this ensemble would continue to increase for all
time, until the original wave function would have filled the entire astronomically sized box. 
The process we have just gone through for a single wave function could be repeated for the
wave function of every particle in the gas, building up an ensemble of possible states for the
whole gas, enabling an entropy for the whole gas to be defined in terms of the actual wave
functions that existed at some time before any collapses took place. 

More generally, for any particular system, an ensemble of possible states could be
defined in terms of some prior wave function or collection of prior wave functions that made
up that system at some earlier time; namely, the ensemble of possible after-collapse wave
functions centred on the cells of the prior wave functions, the way those prior wave functions
would have evolved if there had been no collapses of the prior wave functions. Theoretically,
one could take the system of interest to be the entire universe, and these prior reference wave
functions to be the wave functions that existed at the beginning of the universe. In this way
an ``objective'' ensemble of possible states could be identified for the universe at any time. The
logarithm of the number of states in this ensemble is the quantity I will tentatively identify
with the objective ensemble entropy of the entire universe. This ensemble entropy will
increase for the entire history of the universe. For any particular system of interest, the
reference wave functions could be taken to be the wave functions of that system that existed
just prior to the period under consideration, or the wave functions that existed at the ``origin''
of that system, although it may not be easy to identify the true ``origin'' of any particular
system.

One problem with this suggestion is that the entropies of thermodynamics and
statistical mechanics are normally taken to be quantities that are independent of the causal
history of a system's state, dependent only on the state itself, whereas the ensemble entropy I
have identified seems to be critically dependent on the causal history of the system's state. 
One could attempt to rectify this by defining the ensemble entropy of a state in terms of a pre-collapse wave function, or collection of wave functions, that would ``typically'' precede the
state in question, existing at the ``typical origin'' of the system, or some other wave function,
or collection of wave functions, related to the typical past of this system or similar systems. 
Just how this could be done is not clear, and it is not clear how ``objective'' this entropy would
be, quite apart from the problem of showing that the ensemble entropy that resulted would
have a value similar to the entropy of statistical mechanics.

We have seen that the wave function entropy reduces after wave function collapse. 
For most of the time, the relative volume ranges between two large values, the collapse
volume and $v_{c}$. These volumes do not differ greatly in the model of collapse proposed -
perhaps by a factor of 1000 or so. The entropy, being the logarithm of the relative volume,
will vary by even less. In fact, one could imagine developing a continuous collapse model in
which the relative volume remains $v_{c}$ at all times after, only reducing below this value when
wave functions separate. Either way, if we presume that the average number of particles per
wave function in the universe does not change with time, and that the number of particles in
the universe is approximately a conserved quantity, then the total wave function entropy of
the universe is an approximately conserved quantity.\footnote{The number of particles in the universe is not necessarily approximately conserved with time. The number of baryons may be approximately conserved, except for those that disappear in black holes. However, there are many more photons in the universe than baryons, so what happens to the number of particles in the universe will depend crucially on what happens to the number of photons as the universe expands and cools. This might be expected to increase with time, as high energy photons emitted from energetic sources become absorbed and re-emitted in the form of a higher number of lower energy photons. If this is right, then the wave function entropy of the universe will increase with time, as the increase in wave function entropy derived from the growth in the number of particles will more than make up for the loss of entropy due to the drop in the average energy of the photons. (A similar process is responsible for the increase in coarse-grained 
entropy that occurs when high energy photons are absorbed by the Earth from the Sun, and re-emitted in the form of a larger number of lower energy photons.) Thus it appears that the wave function entropy of the universe will continue to increase for all time after all.} This contrasts with the classical
coarse-grained entropy of the universe which is presumed always to increase with time. I
suggest that this increase may be objective from one point of view, because it reflects how the
wave functions within a system continue to disperse away from each other even after they
individually collapse, and the choice of ensemble may find objective grounding in the wave
function before collapse. The coarse-grained entropy may be viewed as subjective from
another point of view, because it could be regarded as a measure of entropy that is appropriate
to a situation of lack of knowledge of wave function collapse. 

We have seen that prior to collapse, the wave function entropy has approximately the
same magnitude as the quantum coarse-grained entropy, at least in the one case so far
examined. After collapse, the values of the wave function entropy and the coarse-grained
entropy will diverge, but I have provided some grounds for suggesting that the value of the
coarse-grained entropy may nevertheless be able to find objective grounding in the wave
function entropy. This is speculative, and the arguments of this section are incomplete and in
need of further investigation. But I hope that my discussion so far has served to suggest that
the concept of wave function entropy may be a fruitful one in probing the connections
between quantum mechanics, thermodynamics, statistical mechanics, irreversible processes
and the arrow of time.

\citeindexfalse 

\backmatter

\bibliographystyle{plainnat} 
\bibliography{MartinPhD}

\printindex

\end{document}